\documentclass{article} 
\usepackage[cp1251]{inputenc}                            
\usepackage{amsmath,amsfonts,amssymb,amscd,euscript}
\usepackage[russian,english]{babel}
\usepackage{fancyhdr}
\usepackage{graphicx}
\usepackage{psfrag}
\usepackage{cite}
\makeatletter

\def\thispagecolontitul{%
\thispagestyle{fancy}
\fancyfoot{}
\fancyhead{}
\lhead{\scriptsize {SPACE, TIME AND FUNDAMENTAL INTERACTIONS}}
\rhead{\scriptsize {2019, Issue\,4}}
\renewcommand{\headrulewidth}{0.4pt}}

\def\received{\vspace{3ex} \hfill Received \ \datereceive \par \vspace{5ex} }

\def\annotationandkeywordseng{\noindent {\small \referateng \par } \vspace{8pt}
\noindent {\small {\bf {Keywords}}: \keywordseng}
\par \vspace{5pt}
\noindent {\small {\bf {PACS}}: \PACS }
\par \vspace{10pt}}

\@addtoreset{equation}{section}
\renewcommand{\section}{\@startsection{section}{1}{0pt}{1.3ex
plus 1ex minus .1ex}{1.3ex plus .1ex}{\bf\,}}
\newcommand{\point}{\hspace*{-4mm}{\bf.}\;}
\newcommand{\sect}[1]{\begin{flushleft}%
\protect{\section{\point#1}}\end{flushleft}}

\renewcommand{\@begintheorem}[2]{\begin{trivlist}
\item[\hspace{\labelsep}{\bf \mbox{~~~}#1\ #2.}]}
\renewcommand{\@opargbegintheorem}[3]{\begin{trivlist}
\item[\hspace{\labelsep}{\bf \mbox{~~~}#1\ #2 {\rm (#3).}}]}
\renewcommand{\@endtheorem}{\end{trivlist}}

\renewcommand{\@evenfoot}{}
\renewcommand{\@oddfoot}{}

\renewcommand*{\@biblabel}[1]{#1.\hfill}

\newcommand*{\CSep}{.\ }
\renewcommand{\@makecaption}[2]{%
  \vskip\abovecaptionskip
  \sbox\@tempboxa{{\bf #1\CSep}{#2}}%
  \ifdim \wd\@tempboxa >\hsize
  \begin{center}%
    {\footnotesize{\bf #1\CSep}{#2\par}}%
  \end{center}%
  \else
    \global \@minipagefalse
    \hb@xt@\hsize{\hfil\box\@tempboxa\hfil}%
  \fi
  \vskip\belowcaptionskip%
}

\renewcommand{\@evenhead}{\raisebox{0pt}[\headheight][0pt]{\vbox{\hbox to\textwidth{\thepage \strut \hfil
\text{\scriptsize {\authorseng}}} \hrule }}}

\renewcommand{\@oddhead}{\raisebox{0pt}[\headheight][0pt]{\vbox{\hbox to\textwidth{  \strut
\text{\articleshortname} \hfil \thepage} \hrule }}}

\usepackage{standalone}
\usepackage{bm}
\usepackage{empheq}
\usepackage{epsfig}
\usepackage{epstopdf}
\usepackage{mathtools}
\usepackage{citesort}

\newcommand{\authorseng}{A.\,N.~Petrov,  J.\,B.~Pitts }

\newcommand{\articleshortname}{\scriptsize {Field-theoretic Approach in Metric Theories }}
\newcommand{\UDK}{530.12 + 531.51} 
\newcommand{\PACS}{04.20-q 04.20.Cv  04.20.Fy  04.20.Jb 04.50.Kd} 

\newcommand{\referateng}{The representation of General Relativity (GR) and other metric theories of gravity in field-theoretic form on a background is reviewed. The gravitational field potential (metric perturbation) and other physical fields are propagated in an auxiliary background spacetime, which may be curved and may lack symmetries. Such a reformulation of a metric theory is exact and generally equivalent to its initial formulation in the standard geometrical form. The formalism  is Lagrangian-based, in that the equations for the propagating fields are obtained by varying the related Lagrangian, as are the background field equations. A new sketch of how to include spinor fields is included.

 Conserved quantities are obtained by applying the Noether theorem to the Lagrangian as well. Conserved currents
are expressed through divergences of antisymmetric tensor densities
(superpotentials), connecting local perturbations with quasi-local conserved quantities.  The gauge dependence due to the background metric  is
studied, reflecting the so-called  non-localizability  of gravitational energy in exact mathematical expressions formally, an infinity of localized energy distributions that, combined with the material energy, satisfy the continuity equation. The exact expressions can be related to pure GR pseudotensors (especially Papapetrou's) employing the matrix $diag(-1,1,1,1)$, as Nester \emph{et al.} consider on independent grounds.

The field-theoretic formalism admits  two partially overlapping uses. The first one is practical applications of pure GR, where the background presents merely a useful fiction. The second one is foundational considerations in which a background notion of causality, $\eta$-causality, is useful for  making sense of equal-time or space-like commutation relations, in which case the background metric \emph{via} inequalities has qualitative but not strict quantitative physical meaning.

The Schwarzschild solution is the main object for demonstration of the power of the method.  Various possibilities for calculating the mass of the Schwarzschild black hole using surface integration of superpotentials are given. Presenting the Schwarzschild solution as a field configuration on a  Minkowski background, we describe a curved spacetime from spatial infinity to the horizon and even to the true singularity, which is represented in consistently as a point particle using the Dirac $\delta$-function. Trajectories of test particles in the Schwarzschild geometry are gauge-dependent in that even breakdowns at the horizon can be suppressed (or generated) by naive  gauge transformations. This fact illustrates the auxiliary nature of the background metric and the need for some notion of maximal extension---much as with coordinate transformations in geometric GR.  A continuous collapse to a point-like state modelled by the Dirac $\delta$-function in the framework of the field-theoretic method is presented.

The field-theoretic method is generalized to arbitrary metric theories in arbitrary $D$ dimensions. The results are developed in the framework of Lovelock gravity and applied to calculate masses of Schwarzschild-like black holes. Future applications are discussed. The formalism also makes it natural to consider adding a graviton mass.  The works of Babak and Grishchuk, which are partly numerical and hence nonperturbative, are reviewed, shedding light on the traditional questions of a (dis)continuous massless limit for massive pure spin-2 and the (in)stability of a \emph{classical} theory including massive spin-2 and spin-0 gravitons.
}

\newcommand{\keywordseng}{conservation laws, general relativity, modified metric theories}

\newcommand{\datereceive}{88.88.2019} 

\newcommand{\mailone}{E-mail: alex.petrov55@gmail.com}
\newcommand{\mailtwo}{E-mail: jamesbrianpitts@gmail.com}

\newcommand{\contactinformationenglish}{Petrov Alexander Nikolaevich,
Doctor Habilitation of Physical and Mathematical Sciences; Leading Scientific Researcher, Sternberg Astronomical  Institute, M.\,V.\, Lomonosov Moscow State University, Universitetskii Prospekt 13, Moscow 119992, Russia. \\
\mailone\ \\ [5pt]
James Brian Pitts,
Ph.D. in Physics,  Ph.D. in History and Philosophy of Science;  Faculty of Philosophy, University of Cambridge, Sidgwick Avenue,
Cambridge, CB3 9DA, United Kingdom.\\
\mailtwo\ \\ [5pt]
}

\def\txt{\textstyle}

\def\sst{\scriptscriptstyle}

\def\be{\begin{equation}}
\def\ee{\end{equation}}
\def\bea{\begin{eqnarray}}
\def\eea{\end{eqnarray}}
\def\beaN{\begin{eqnarray*}}
\def\eeaN{\end{eqnarray*}}
\def\ed{\end{document}}
\def\bit{\begin{itemize}}
\def\eit{\end{itemize}}

\def\bn{\bar\nabla}

\def\sig{\sigma}

\def\lam{\lambda}

\def\Del{\Delta}
\def\del{\delta}
\def\k{\kappa}

\def\alf{\alpha}

\def\di{\partial}

\def\Lix{\pounds_\xi}

\def\half{{\textstyle{1 \over 2}}}
\def\~{\tilde}
\def\lag{{\cal L}}
\def\m{\label}
\def\l{\left}
\def\r{\right}

\def\goto{\rightarrow}
\def\Bar{\overline}
\def\const{\rm const}

\newcommand{\gog}{\mathfrak{g}}
\newcommand{\goh}{\mathfrak{h}}

\begin{document}

\thispagecolontitul

\hbox{UDK \UDK}  \vspace{20pt}

\begin{center}
{\bf { \textit { \authorseng}} \\[2ex]
{THE FIELD-THEORETIC APPROACH IN GENERAL RELATIVITY AND OTHER METRIC THEORIES. A REVIEW}
}

\par
\vspace{10pt}
\end{center}

\annotationandkeywordseng



\begin{flushleft}
{\bf{Introduction}}
\end{flushleft}



In the framework of the field-theoretic approach, a metric theory is represented in the form where gravitational field (metric perturbation) together with other physical fields are propagated in a specified  (curved or flat and not necessarily symmetric) background space-time. Such a reformulation of a metric theory is equivalent to its initial geometrical formulation, apart, perhaps, from global considerations, though the freedom to make the background metric not too different from the effective metric provides the freedom to minimize global issues.
Review materials related to the field-theoretic method have been published previously  \cite{Petrov_2008,Petrov_2008_a,Petrov+_2017}. 
However, the approach continues to be developed; see, for example, a description of a continues collapse to a point in the framework of the field-theoretic approach \cite{Petrov_2018} and the field-theoretic formalism in Einstein-Gauss-Bonnet gravity \cite{Petrov_2009} and in  Lovelock gravity of an arbitrary order \cite{Petrov_2019}. Important properties of the method gradually become clearer. To further popularize the field-theoretic formalism, the present paper gives a review of its development and  current applications.

The mature form of special relativity explained why all the attempts to detect the luminiferous {\em`ether'} failed. The ether as a {\em`true physical background'} was discarded as physically idle \cite{Kostro}. Instead  the notion of Minkowski space has been introduced:  a flat $4$-dimensional space-time in which  physical bodies, particles and fields propagate, evolve and interact.  Could gravity be included within the framework of special relativity?
Einstein's attempts were unsuccessful \cite{NortonNordstrom}, though his criticism of the approach was not  correct \cite{GiuliniScalar}.  Gunnar Nordstr\"{o}m later largely succeeded with a relativistic scalar gravity theory with some help from Einstein; as Einstein and Fokker showed, Nordstr\"{o}m's theory was in effect that of a curved, conformally flat space-time \cite{EinsteinFokker}.  The Minkowski volumes are clothed with the gravitational potential so that (one might say)  volumes are universally distorted by gravity.  Using universal coupling of gravitational and non-gravitational energy-momentum (or rather its trace), one can derive the Nordstr\"{o}m-Einstein-Fokker theory and various massive scalar graviton theories
 \cite{Kraichnan,FreundNambu,DeserHalpern,PittsScalar}, according to which free gravity (that is, disregarding the nonlinearities describing self-interaction) satisfies the Klein-Gordon equation.  Massive scalar gravity, though fully developed very belatedly or `postmaturely' (to borrow a word \cite{StachelNewstein}), has clear non-relativistic antecedents in the 19th century work of Carl Neumann and Hugo von Seeliger and the first half of Einstein's $\Lambda$ paper before the false analogy to the cosmological constant $\Lambda$ is introduced \cite{FMS,Schucking,LambdaMPIWG}. Given the smooth massless graviton limit, massive scalar gravity---actually there are infinitely many such theories---approximates the Nordstr\"{o}m-Einstein-Fokker theory arbitrarily well, exemplifying permanent underdetermination from approximate but arbitrarily close empirical equivalence \cite{UnderdeterminationPhoton}.  Of course the nonzero graviton mass has no evident empirical support, while the scalar character makes massive scalar gravities just as refuted by the 1919 light-bending observations as is the Nordstr\"{o}m-Einstein-Fokker theory.  Massive scalar gravities, like their massless relative \cite{MisnerScalar}, remain a useful conceptual testbed for the foundations of space-time, such as helping to adjudicate whether space-time geometry explains the Euler-Lagrange equations or \emph{vice versa} \cite{BrownPhysicalRelativity,ScalarGravityPhil,BrownFest} and even assessing the non-viability of Kantian views about space and time  given modern science \cite{KantParticle}.

Given a symmetric rank $2$ tensor gravitational potential (as the bending of light evidently requires), attempts to develop viable theories that do not have ``ghosts'' (negative-energy degrees of freedom and hence presumably instability, especially under quantization) lead to gauge freedom (at any rate for the kinetic term), and the introduction sources leads to universal coupling and hence to the coalescence of the graviton potential and background metric into an effective curved metric \cite{Kraichnan,NSSexlLinear,VanN,Deser,DaviesFang,PittsSchive2001a}; the flat background is unobservable, except perhaps \emph{via} an extra scalar degree of freedom  \cite{PittsSchive2001a} or a graviton mass term.  Graviton mass terms, however, seemed to have two key devils in the details in the early 1970s \cite{DeserMass}:  either a scalar ghost (but see \cite{deRhamGabadadze,HassanRosen,MaheshwariIdentity,MassiveGravity3}) or a discontinuous massless limit \cite{Zakharov,vDVmass1,vDVmass2,Iwasaki} (but see \cite{Vainshtein,Vainshtein2}).
 Thus it is either impossible or at least difficult to construct a viable theory of gravity that bends light and does not closely resemble GR, even if one avoids any \emph{a priori} appeal to curved space-time, the equivalence principle, general covariance, generalized relativity of motion, Mach's principle, \emph{etc.}
It is worth noting that the particle physics approach to GR bears a strong resemblance (albeit considerably improved) \cite{EinsteinEnergyStability,ConverseHilbertian} to Einstein's recently reappreciated ``physical strategy'' that he pursued in the first half of the 1910s alongside his subsequently endorsed ``mathematical strategy'' \cite{NortonField,RennSauer,Janssen,BradingConserve,RennDwarfEmergence,Renn,JanssenRenn,RennSauerPathways}.  (Einstein later retold his own history in a way that suppressed the physical strategy and credited success to the mathematical strategy, at least partly in order to justify his decreasingly appreciated unified field theory quest \cite{vanDongenBook}.)
Indeed one can see that the particle physics ``spin 2'' derivation just sketched is powered by Noether's converse Hilbertian assertion  that the energy-momentum for gravity consists in a piece vanishing using the field equations and a piece with automatically vanishing coordinate divergence \cite{Noether,ConverseHilbertian}.
 Hence most or all roads, wherever they start, lead to  or at least near to GR.
As a result, effectively space-time becomes a dynamic structure with metric components as dynamical variables. Thus in GR, space-time, while continuing to be an arena for evolution of non-gravitational physical fields, gives gravity an exceptional position and  cannot be interpreted as a `background' in the sense of a fixed structure.

 Nevertheless, many problems in GR,  both  theoretical  and as applied, require  considering perturbations, including metric perturbations, in a given (fixed) space-time. In such an interpretation, the fixed (not necessarily symmetric) space-time is to be a solution to the Einstein equations and plays the role of a {\em background}. One  could list some examples. Einstein himself, just after constructing GR, studied weak gravitational waves as metric perturbations propagating in Minkowski space. Later gravitational waves and other perturbations have been considered on backgrounds of the Friedmann solution and other cosmological solutions. Isolated gravitating systems both at spatial and at light-like infinities are considered as perturbed systems on backgrounds of Minkowski space, anti-de Sitter space, or another background geometry. The stability of many GR solutions, including but not limited to black hole solutions, is examined by studying the evolution of metric perturbations on the background of these solutions. B. DeWitt's background field method of quantization of gravity makes systematic use of a (curved) background \cite{DeWitt-book}.

However, many of the aforementioned studies (excluding DeWitt's) are carried out under restrictions, which are determined by the problem under consideration. Frequently a linear approximation is considered without taking into account backreaction. Thus, one has to make a separate study because the background is changed by the backreaction; see the pioneering work on this relation \cite{Isaacson}. This topic has been developed in Efroimsky's works \cite{Efroimsky_1992,Efroimsky_1994}, where weak gravitation waves in vacuum and in media in a cosmological context are studied. Attention is paid to the role of nonlinearity provided by the  energy-momentum of metric perturbations while taking into account the low-frequency cut-off. In \cite{Svitek+}, the Efroimsky approach is corrected and developed. In recent  years (see, for example, \cite{Efroimsky+,Efroimsky++} and references therein) the Efroimsky method has gained popularity, but it also becomes quite specific.

Frequently only  flat or strongly simplified curved backgrounds are considered, so it is not clear how to develop a theory of perturbations if a background becomes  more complicated, general and non-symmetric. Frequently additional assumptions are used; then it is not always clear how the results depend on these  assumptions, {\em etc}.
Keeping in mind all the above, one concludes that a generalized and unique description of perturbed systems in GR on a given background is necessary. The main requirement is to be that such a description of perturbations in GR has to be equivalent to GR itself. (There could in some cases be global issues.  However, the admission of an arbitrary background, rather than a one-size-fits-all approach with, say, a Minkowski background, minimizes the expectation of significant global differences between the effective and background metrics.) This description has to be in the form of a field theory, where fields present exact (without approximations) perturbations in a fixed background space-time. We call it a {\em field-theoretic presentation of GR}. Its desired properties are:
 \bit
 \item Unlike the standard geometrical presentation where a space-time is a dynamic structure, in the framework of the  field-theoretic approach the field
configuration consists of dynamic fields (which represent perturbations) propagating in a `fixed' background
space-time. This background can be curved and hence in some sense dynamical, but is specified somewhat independently of the effective metric.

\item The field-theoretic approach is to be Lagrangian-based. This means that a) a Lagrangian density  is defined for the field configuration; b) field equations are defined by varying the action; c) conserved quantities such energy and momentum (and their densities)  are defined by applying Noether's theorem to the symmetries of  the action. 

 \item Equations and conserved quantities are to be covariant under coordinate transformations. This property, being desirable in itself (although  partially offset by a new gauge dependence), gives the possibility to study the field configurations on arbitrary curved backgrounds which are themselves  solutions to the GR field equations.

\item Gauge (non-coordinate) transformations are to be defined explicitly with well described properties convenient in applications.

\item Because the field configuration is to be exact (without approximations), gauge transformations and conserved quantities are to be exact as well. All of these features give the  possibility to construct approximations of all important expressions up to an arbitrary order more naturally and easily.
 \eit

Let us discuss in a more detail a problem of defining energy and other conserved quantities in GR.
As well known, in many theoretical studies and in applications, notions of conserved quantities, like energy, momentum, angular momentum and their densities, play essential role. However, conserved quantities are usually said to be ``not localizable'' in GR  \cite[p. 467]{MTW}.  Sticking closer to the facts rather than interpretations, one can say that it is impossible to construct  densities of conserved quantities in GR in a unique way.
If one introduces by hand as usual the  tacit assumption that there should be a unique density of gravitational energy-momentum, then one infers that the local descriptions lack physical meaning and hence that gravitational energy-momentum is not localizable.  If, on the other hand, one simply takes the mathematics at face value \cite{BergmannConservation,EnergyGravity}, then one notices that any time-like vector field locally takes the form $(1,0,0,0)$ and so is formally a rigid time translation.  Hence GR has  uncountably infinitely many formally rigid symmetries of the action (\emph{not} symmetries of the geometry, an issue of no direct relevance to Noether's first theorem).  Noether's first theorem associates to each formally rigid symmetry of the action a conserved current and \emph{vice versa} \cite{Noether,KosmannSchwarzbachNoether}.  Thus one can  infer that there are infinitely many conserved energies and momenta \cite{EnergyGravity}.  Such a result is not very familiar and is somewhat inconvenient for accountants, but seems otherwise plausible enough; gravitational energy, instead of being non-localizable, is infinitely plural and thus has no single objective localized $10$- or $16$-component description. Within the field formulation, one can arrive at such an inference by taking the gravitational energy-momentum tensor in \emph{all gauges}.  Within pure geometric GR with no background, the analogous entity is a pseudotensor in    \emph{all coordinate systems}, potentially with some dependence on an auxiliary matrix $diag(-1,1,1,1)$ as a reference configuration \cite{Papapetrou,NesterQuasiPseudo,ChangNesterChen}. Pseudotensors, besides being supposedly coordinate-dependent in a vicious way (which the interpretation in terms of infinitely many energies suggests is a virtue rather than a vice), are also worrisomely nonunique; Nester and collaborators claim to find physical meaning in this multiplicity in terms of differing boundary conditions.  On the other hand, the energy-momentum expression employed in this paper in section 4  seems especially virtuous.  Either view could address the nonuniqueness problem.  While the supposed nonlocalizability issue is logically independent of the field formulation,  the field formulation was in fact apparently the first occasion of the proposal of taking the formal infinity of gravitational energies seriously \cite{EnergyGravity}.  Also the field formulation might more readily suggest (gauge) transformations connected to the identity rather than large (coordinate) transformations under which gravitational energy localizations behave badly \cite{BauerEnergy}.  Thus the field formulation may be of some heuristic relevance in relation to the localization question \cite{EnergyGravity}.

This problem of defining localized energy-momentum (energies-momenta?) arises because GR is a geometrical theory in which space-time has a double role:  as an arena on which physical fields evolve and a dynamical object. This double role of a space-time follows from the equivalence principle (see, for example, \cite{MTW}).  Some take the view that a definition of conserved quantities in GR is meaningless except in special cases.  There seems to be increasing interest, however, in revisiting the question in the last two decades or so.   In spite of the pseudotensoriality of the Noether-based conserved quantities in GR, gravitational interaction gives a contribution to  {\em total} conserved quantities of gravitating systems \cite{MTW}.  Here are some familiar examples. To describe a binary system, one has to include a notion of gravitational energy as a binding energy  \cite{MTW}. Concerning gravitational waves, first,  a bounded domain of empty space filled by gravitational waves has to have a total positive energy, see \cite{MTW}; second, observations of double pulsars show that the orbit's axis becomes smaller by gravitational radiation because gravitational waves carry away positive energy, see, for example, \cite{WNT_2010}; third, the recent direct detection of gravitational waves, see, for example, \cite{GW150914,GW151226,GW-A-LIGO}, tells the same story. Thus, energy and other conserved quantities in GR are naturally construed as physically  real.  Thus the formal description of such quantities and their spatio-temporal localization merits continued investigation, especially once the traditional objections from  pseudotensoriality and nonuniqueness have come to seem less decisive.  The problem of conserved quantities  can be considered conveniently in the framework of the field-theoretic approach in GR.

There is another topic concerning energy-momentum and its relations to space-time:  in recent decades  one associates energy-momentum and angular momentum with a finite spatial domain (whether or not gravitational radiation is present)  and its  boundary. Such conserved quantities are called  {\em quasi-local}; see the nice review by Szabados \cite{Szabados04} and references therein. Such a treatment can be useful, for example, in studies of cosmological problems, where more frequently local properties of perturbations are examined. Thus a  study of the connection of local characteristics with quasi-local quantities could be very fruitful.  Pseudotensors naturally yield quasi-local quantities  \cite{NesterQuasiPseudo,ChangNesterChen} and depend on coordinates only on the boundary, not in the interior. In the field-theoretic formulation, one would expect a tensorial but gauge-dependent energy-momentum distribution to yield quasi-local conserved quantities that are tensorial but depend on the gauge at the boundary.

Thus, for the purposes of studying conserved quantities in the field-theoretic formulation of GR, it is desirable:
\bit

\item to have consistent definitions of conserved quantities,

\item to derive their properties that can be useful in applications,

\item to give a mathematical (exact and concrete) derivation of the so-called non-localization (\emph{i.e.}, the lack of a unique localization due to gauge dependence), and

\item to connect local and quasi-local quantities.

\eit

The field-theoretic formulation is intended to be equivalent to the  geometrical formulation, so some version of the spirit of  GR {\it as
a geometrical theory} has to be preserved in the field-theoretic formulation as well.  Thus, \emph{e.g.}, elevating some gauge or coordinate system to a physical law is not appropriate.
 In the geometrical formulation,  there is no background,  so of course none  can be observed. The field-theoretic formulation thus has to have the same empirical property of no \emph{observable} background. If one studies  the movement of test particles and light rays, one cannot connect it (quantitatively) with the geometry of the background space-time. The metric perturbations considered as a gravitational field on a given background play a role of a refractive medium so that the background is screened (perhaps one should say clothed) totally.  An  analogous interpretation, for example, can be found in the paper \cite{Nandi_1995}. This means that the background space-time in the framework of the field-theoretic formulation can be interpreted as an auxiliary and nonphysical (fictitious) concept, at least in its precise quantitative properties.   

In the last two decades much attention has been paid to modifications of GR. An arbitrary field theory can be represented in the field-theoretic form, as will be shown in  sections 9 and 10. A special place among modified theories is taken by metric theories including theories in more than $4$ dimensions. Among them $f(R)$ theories \cite{Sotiriou_2010}, quadratic-in-curvature theories, and Einstein-Gauss-Bonnet theory and its generalization for arbitrary order (Lovelock theory \cite{Lovelock_1971}), are among the most  popular. Because the equivalence principle can be considered as the basis for an arbitrary metric theory, such theories must  have  field-theoretic representations similar to that of  GR. Thus, the field-theoretic approach to other metric theories can be desirable for  the aforementioned reasons.

 Below we will demonstrate important and interesting properties of the field-theoretic approach by  presenting very well known solutions in the form of exact field configurations on a given backgrounds. We will also pay  attention to exotic applications to consider the possibility of a  physical interpretation of some \emph{qualitative} features of the background. For example, we derive the Schwarzschild black hole solution as a field configuration on a Minkowski background up to the horizon, beyond the horizon, and even all the way to the true singularity.  This exhibition demonstrates the power of the method and its less-explored possibilities as well. Concerning natural applications of the field-theoretic approach, we refer the reader to the original papers, 1) where cosmological perturbations on the Friedmann-Lema\^{i}tre-Robertson-Walker (FLRW) backgrounds are studied \cite{PK1,PK2,PK3,PK4}, 2) where asymptotically flat space-times are examined at spatial infinity \cite{Petrov_1995,Petrov_1997,Baskaran_Lau_Petrov_2003}, \emph{etc}. Finally, we show some ways for a development of the method and its new applications.

The paper is organized as follows:

In section 1,
we give the mathematical
foundation of the field-theoretic formulation of GR in detail. This gives a
basis for the remainder of the article. This section also sketches how to include spinor fields in the field formulation, a topic rarely considered.

In section 2, we study the gauge invariance properties of the Lagrangian, field equations and conserved quantities in the framework of the field-theoretic approach to GR.

In section 3, we review various possibilities for how the field-theoretic representation of GR can be constructed starting from a fixed background space-time. We demonstrate also that ultimately such a background metric is not observable and plays a merely auxiliary role.

In section 4,  using the results of previous  works
we present conservation laws in the field-theoretic
formulation of GR. The conserved currents are constructed on the
basis of a symmetric energy-momentum tensor and express localized
 conserved quantities. At the same time the
currents are derived as divergences of antisymmetric tensor
densities (superpotentials), integration of which just leads to
surface integrals, which are quasi-local conserved quantities.

In section 5, we present various possibilities to calculate the mass of the Schwarzschild black hole with the use of the surface integration of superpotentials defining the gravitational charge.

In section 6, we presenting the Schwarzschild solution as a field configuration on a Minkowski background, including not only the horizon but also the true singularity. It is represented as a point particle \emph{via} a Dirac $\delta$-function.

In section 7,  the example of trajectories of test particles in the Schwarzschild geometry illustrates  that the background space-time in the field-theoretic formalism is an auxiliary structure. Trajectories are gauge dependent in the sense that even break-downs at the horizon can be suppressed or generated by the gauge transformations. A natural conclusion is that some notion of maximal extension is required, much as one requires in pure geometric GR.

In section 8, we present a continuous collapse to a point-like state modelled by the Dirac $\delta$-function in the framework of the field-theoretic method.

In section 9, we present a field-theoretic treatment of an arbitrary $D$-dimensional metric theory of space-time and gravity.

In section 10, currents and superpotentials are obtained for the more specific example of the Lovelock class of theories.

In section 11, we find conserved quantities in the Lovelock theories.

In section 12, we study the mass of the Schwarzschild-like black hole and consider future applications of the method.

In section 13, there is an explicit discussion of some massive gravity theories due to Babak and Grishchuk. Their work includes numerical simulations and hence is nonperturbative.  In these and most other massive gravity theories, the background metric is now physically real and indirectly observable due to the graviton mass term; while rods and clocks do not conform to the background metric, it plays an essential role in the field equations. The question whether negative-energy degrees of freedom are bad already in classical field theory, or become bad only when quantization is entertained, is addressed. The technique of nonlinear group realizations, which Ogievetsky and Polubarinov invented and used to derive graviton mass terms using arbitrary powers of the metric, is also relevant to spinors.

In appendix  A, we derive expressions that follow in an arbitrary field theory after application of Noether's theorem.

\bigskip
\noindent{\bf Notations}:
\bit
\item $\psi^A,~{P}_B, \ldots$ - sets of tensor densities of arbitrary ranks and weights
with the collective indices $A,~ B,\ldots$ in a compressed notation;

\item $\bar {\psi^A}$ - the ``bar'' above $\psi^A$ means a background value of $\psi^A$;

 \item ${\bm t}_\sig{}^\mu, {\bm m}_\sig{}^{\mu\nu}, \ldots$ - notations in calligraphic boldface for small letters, if they represent quantities of mathematic weight $+1$.  For example, ${\bm t}_\sig{}^\mu$ could be a density  expressed with the use of the tensor ${t}_\sig{}^\mu$:  ${\bm t}_\sig{}^\mu = \sqrt{-\bar g}{t}_\sig{}^\mu$, or ${\bm t}_\sig{}^\mu$  could be a density itself, {\em etc}.  Some authors define density weight with the opposite sign;

\item ${\cal L},~ {\cal U}_\sig{}^{\mu}, \ldots$ - the capital calligraphic letters
denote geometric quantities of weight $+1$ analogously to previous item;

\item ${\xi^\alf}$ and ${\bar\xi^\alf}$ - arbitrary displacement vectors and Killing vectors, respectively, in a space-time;

\item  $g_{\mu\nu}$ and $\bar g_{\mu\nu}$ - the dynamical (also called effective or physical) and background metrics;

\item $g ={\det}\,g_{\mu\nu}$ and $\bar g =\det\,\bar g_{\mu\nu}$ - the determinants of the dynamical and background metrics;

\item the indices of tensor fields on the physical quantities or  background quantities  are lowered and raised with the use of $g_{\alf\beta}$ or $\bar g_{\alf\beta}$ and their inverses, respectively;

\item $R^\rho{}_{\alf\sig\beta}$ and $\bar R^\rho{}_{\alf\sig\beta}$, $R_{\alf\beta}$ and $\bar R_{\alf\beta}$, and $R$ and $\bar R$ - the Riemann and Ricci tensors and the Ricci scalars for the  physical and background metrics, respectively;

\item ${\di{\psi^A} }/{\di x^\alf}= \di_\alf \psi^A = {\psi}^{A}{}_{,\alf}$ -  the partial derivative;

\item $\nabla_\alf {\psi}^A$ and $\bar\nabla_\alf {\psi}^A$ -  the covariant derivatives of ${\psi}^A$ compatible with $g_{\mu\nu}$ and with $\bar g_{\mu\nu}$, respectively;

\item the Lagrange derivative of the quantity ${\psi}^A={\psi}^A(q^B;q^B{}_{,\alf};q^B{}_{,\alf\beta})$ is
 $$\displaystyle\frac{\del{\psi}^A}{\del q^B} = \frac{\di{\psi}^A}{\di q^B} - \di_\alf\l(\frac{\di{\psi}^A}{\di q^B{}_{,\alf}}\r) + \di_\alf\di_\beta\l(\frac{\di{\psi}^A}{\di q^B{}_{,\alf\beta}}\r)\,;$$

\item $\l.{\psi}^A\r|^\alf_\beta$ is a permutation linear operator depending on the transformation properties of ${\psi}^A$, for example, for the tensor density $\psi^A = {\bm t}_\sig{}^\mu$ one has $\l. {\bm t}_\sig{}^\mu \r|^\gamma_\beta = - {\bm t}_\sig{}^\mu \delta^\gamma_\beta + {\bm t}_\sig{}^\gamma \delta^\mu_\beta  - {\bm t}_\beta{}^\mu \delta^\gamma_\sig$ (with the `extra' density weight-dependent term that appears in Lie derivatives and covariant derivatives \cite{Schouten,Anderson,Israel});

\item $\Lix {\psi}^A = -\xi^\alf \psi^A{}_{,\alf} + \xi^\beta{}_{,\alf}\l.{\psi}^A\r|^\alf_\beta$ - the Lie derivative of the quantity ${\psi}^A$ along $\xi^\alf$ defined here. Note that many authors define the Lie derivative with the opposite sign, see, {\em e.g.}, \cite{Schouten}.

 \eit


\vspace{0.3cm}
\sect{The mathematical basis for the field-theoretic formulation of GR}
\m{FTF-GR}

Perturbations and conservation laws for them have been  studied in GR for many years. What is the simplest and most usual way?  The linear terms in metric perturbations are placed on the left-hand side, whereas all nonlinear terms are moved to the right-hand side and, together with the matter energy-momentum tensor, are treated as a total (effective) energy-momentum ${\bm t}^{tot}_{\mu\nu}$, see, for example, the book \cite{Weinberg-book}. (Linearity \emph{vs.} nonlinearity is not altogether invariant under field redefinitions, on which more below.) Perturbations are considered on arbitrary backgrounds (flat or curved), and  equations for them are in the form of the field theory. However, such equations are derived `by hand'. This line of research has been pursued since the  1940s \cite{Papapetrou}. Deser in \cite{Deser}, generalizing earlier works (see, for example, \cite{Kraichnan,[23],[22]}),  suggested a Lagrangian-based  presentation where perturbations propagate (at least formally) in Minkowski space-time, yielding an exact (without expansions and approximations) Lagrangian theory of tensor field with self-interaction. In this framework, the effective energy-momentum ${\bm t}^{tot}_{\mu\nu}$ is obtained by variation of an action with respect to a background metric. Subsequently  \cite{GPP,Grishchuk92},  the Lagrangian-based theory for perturbations in GR on arbitrary curved backgrounds has been developed. A related bibliography of earlier works particularly can be found in \cite{Deser,GPP,Grishchuk92,PittsSchive2001a}.  There are also similarities to  DeWitt's background field formalism for quantum gravity (\emph{e.g.}, \cite{DeWitt-book}).

In this section, building on the results of the papers \cite{GPP,[15]} (see also chapter 2 in the book \cite{Petrov+_2017}), we give the mathematical formalism for the Lagrangian based field-theoretic presentation of GR with an arbitrary curved background space-time. We show  the connection with the usual geometrical formulation of GR explicitly.

Let us start with the Einstein-Hilbert action in the usual form:
 \be S = {1 \over c} \int d^4x {\cal L}_{\sst EH}
\equiv
 -{1 \over {2\kappa c}} \int d^4x {\cal R}(g_{\mu\nu})
+ {1 \over c} \int d^4x {\cal L}^M (\Phi^A,~g_{\mu\nu})
\m{(a2.1)}
 \ee
where ${\cal L}^M (\Phi^A,~g_{\mu\nu})$ is a Lagrangian of matter fields $\Phi^A$ consisting of  a set of tensor densities\footnote{At the end of this section we present a new proposal how spinors/fermions can be included in the field-theoretic formalism.} of an arbitrary order and rank with the collective index `$^A$' ; ${\cal L}^M $ depends on derivatives of variables up second order.  Here  $\kappa = 8 \pi G c^{-4}.$
One could of course add an arbitrary divergence to this Lagrangian density.  Considering components of the inverse metric density $\gog^{\mu\nu} = \sqrt{-g}g^{\mu\nu}$, after variation of (\ref{(a2.1)}) we derive the  Einstein equations together with the matter ones in the form:
 \bea
 &&{{\del {\lag}_{\sst EH}} \over {\del \gog^{\mu\nu}}} =
 -{1 \over {2\k }}
 {{\del \cal R} \over {\del \gog^{\mu\nu}}} +
 {{\del {\lag}^M} \over {\del \gog^{\mu\nu}}} = 0\, ,
\m{(a2.2)}\\
 &&{{\del {\lag}_{\sst EH}} \over {\del \Phi^A}} =
 {{\del \lag^M} \over {\del \Phi^A}} = 0\, .
\m{(a2.3)}
 \eea
 The equation (\ref{(a2.2)}) can be rewritten in the  form:
 \be
 R_{\mu\nu} = \kappa (T_{\mu\nu} - \half g_{\mu\nu}T)\, .
\m{(a2.2+)}
\ee

Now let us make  decompositions of the inverse metric density and the matter fields into dynamical perturbations and background values:
 \bea
 \gog^{\mu\nu} & \equiv &\bar {\gog}^{\mu\nu} + {\goh}^{\mu\nu}\, ,\m{(a2.4)}\\
\Phi^A & \equiv & \bar {\Phi}^A + \phi^A\,  \m{(a2.4m)}
 \eea
where the metric and matter perturbations $\goh^{\mu\nu}$ and
$\phi^A$ are defined with respect to a background solution $\bar{\gog}^{\mu\nu}$ and
$\bar\Phi^A$.  This decomposition is, of course, not unique, though for some purposes it is optimal, such as for writing relativistic wave equations.  Field (re)definitions  play an important role in various contexts including showing the equivalence of various massless formulations \cite{B-Deser} and for deriving inequivalent massive gravities \cite{Kraichnan,OP,MassiveGravity1,MassiveGravity2,MassiveGravity3,PittsScalar}. The possibility of field redefinitions and their (ir)relevance for definitions of energy-momentum is discussed elsewhere \cite{[15],Petrov_2008_a}.

 A notion of the background solution can be generalized to the notion of a background system. The latter
is described by the action:
 \be
 \bar S =
{1 \over c} \int{ d^4x \bar{\lag}_{\sst EH}} \equiv
 -{1 \over {2\k c}} \int {d^4x \bar{\cal R}}
+ {1 \over c} \int {d^4x \bar{\lag}^M}\,
\m{(a2.8)}
 \ee
depending on the variables $\bar{\gog}^{\mu\nu}$ and
$\bar\Phi^A$. These quantities satisfy the corresponding background Einstein
equations:
 \be
 -{1 \over {2\k }}  {{\del \bar{\cal R}} \over {\del \bar{\gog}^{\mu\nu}}} +
 {{\del \bar{\lag}^M} \over {\del \bar{\gog}^{\mu\nu}}} = 0,
 \qquad  {{\del \bar{\lag}^M} \over {\del \bar{\Phi}^A}} = 0\, .
\m{(a2.6)}
\ee

The strategy of the field-theoretic method is based on the treating perturbations $\goh^{\mu\nu}$ and $\phi^A$ as {\em independent} dynamic variables. Then a perturbed system has to be described by a corresponding Lagrangian. Now one substitutes the decompositions (\ref{(a2.4)}) into the Lagrangian of
the action (\ref{(a2.1)}) and subtracts zeroth-order and linear terms in $\goh^{\mu\nu}$ and  $\phi^A$ from  the functional expansion:
%
 \be
 {\lag}^{\rm dyn}  =
 {\lag}_{\sst EH}(\bar g+ \goh,\,\bar \Phi+\phi) -
 \goh^{\mu\nu}
 {{\del \bar{\lag}_{\sst EH}} \over {\del \bar{\gog}^{\mu\nu}}} -
\phi^A {{\delta \bar {\lag}_{\sst EH}}\over{\delta \bar{\Phi^A}}} -
 \bar{\lag}_{\sst EH} + {\rm div}
=  -{1\over{2\k}}\lag^g + \lag^m\,. \m{(2.10)}
 \ee
Because in (\ref{(2.10)}) perturbations $\goh^{\mu\nu}$ and $\phi^A$ are treated as dynamical variables, we call $ {\lag}^{\rm dyn}$  the {\it dynamical
Lagrangian} in the terminology of \cite{[15]}. A divergence is included for the sake of generality.

In (\ref{(2.10)}), the zeroth order term is the background Lagrangian from (\ref{(a2.8)}), whereas the linear term is proportional to
the left-hand side of the background equations (\ref{(a2.6)}).
For finding the field equations for the perturbation, it makes no difference whether one uses the background equations before or after varying the perturbation.
However, one should not to use the background equations in ${\lag}^{\rm dyn}$  before
its variation with respect to a background metric to find the energy-momentum tensor. The reason is that zeroth order  and linear terms in the perturbations cancel parts of  ${\lag}_{\sst EH}(\bar g + \goh,\,\bar \Phi+\phi)$ in (\ref{(2.10)}) and thus help to give a reasonable energy-momentum tensor.

The other important property of the dynamical Lagrangian is related to a role of the divergence in (\ref{(2.10)}). First, let us define  an important tensor, due originally to Levi-Civita, from the difference of two connections \cite[p. 221]{LeviCivita}:
\be
\Del^\alpha{}_{\mu\nu} \equiv \Gamma^\alpha{}_{\mu\nu} - \bar
{\Gamma}^\alpha{}_{\mu\nu} = \half g^{\alf\rho}\l( {\bar\nabla}_\mu g_{\rho\nu}
+ {\bar\nabla}_\nu g_{\rho\mu} - {\bar\nabla}_\rho g_{\mu\nu}\r)\,  \m{DeltaDef}
 \ee
that is a difference between dynamic and background Christoffel symbols, and which is linear in $\goh^{\mu\nu}$ to leading order. Then, if one chooses  a divergence ${\rm div} = \di_\alf {\bm k}^\alf $ with the vector density
 \be
 {\bm k}^\alf \equiv \gog^{\alpha\nu}\Del^\mu{}_{\mu\nu} - \gog^{\mu\nu}
\Del^\alpha{}_{\mu\nu}\, ,\m{k-KBL}
 \ee
then a pure gravitational part in the Lagrangian (\ref{(2.10)}) acquires the form:
 \bea
\lag^{g}& =&
(\bar {\gog}^{\mu\nu} + \goh^{\mu\nu}) R_{\mu\nu}(\goh +\gog) - \goh^{\mu\nu}
\bar{R}_{\mu\nu} - \bar{\gog}^{\mu\nu}\bar R_{\mu\nu} +\di_\mu {\bm k}^\mu \nonumber \\&=& -(\Delta^\rho{}_{\mu\nu} -
\Delta^\sig{}_{\mu\sig}\delta^\rho_\nu)\bn_\rho \goh^{\mu\nu} +
(\bar{\gog}^{\mu\nu} + \goh^{\mu\nu})
\l(\Delta^\rho{}_{\mu\nu}\Delta^\sig{}_{\rho\sig}
-\Delta^\rho{}_{\mu\sig}\Delta^\sig{}_{\rho\nu}\r)\,. \m {(a2.16)}
 \eea
 Note that it depends on only the first derivatives of the gravitational variables $\goh^{\mu\nu}$, which is very economical for boundary conditions under variation, whereas the pure gravitational part in (\ref{(2.10)}) with ${\rm div} = 0$ has second derivatives of  $\goh^{\mu\nu}$. Notice that in the case of a flat background, the pure gravitational part in (\ref{(2.10)}) becomes Deser's  \cite{Deser}, whereas the Lagrangian (\ref{(a2.16)}) becomes Rosen's covariant
Lagrangian \cite{[2],[2]_a}. The matter part of the dynamical Lagrangian (\ref{(2.10)}) is
 \be
 \lag^m  =
{\lag}^M\l(\bar g+ \goh, \,\bar {\Phi} + \phi\r)  - \goh^{\mu\nu} {{\delta \bar {\lag}^M}\over{\delta \bar {\gog}^{\mu\nu}}}- \phi^A {{\delta \bar {\lag}^M}\over{\delta
\bar\Phi^A}} - \bar{\lag}^M\, .
\m{(a2.15)}
 \ee
 %


Now, let us turn to deriving field equations. The variation of action with the Lagrangian ${\lag}^{\rm dyn}$ with
respect to the gravitational dynamic variables $\goh^{\alf\beta}$ and contraction with
\be
\frac{1}{\sqrt{-\bar g}}\frac{\di\bar {\gog}^{\alf\beta}}{\di\bar g^{\mu\nu}} = \del^\alf_\mu\del^\beta_\nu -
\half \bar g^{\alf\beta}\bar g_{\mu\nu}
\m{g-di_g}
\ee
lead to the field equations in the form
 \be
 {\cal G}^L_{\mu\nu} + {\bm \Phi}^L_{\mu\nu} = \k\l({\bm t}^g_{\mu\nu} +  {\bm t}^m_{\mu\nu}\r)
\equiv \k{\bm t}^{\rm tot}_{\mu\nu}\, , \m {(a2.17)}
 \ee
where the left hand side is linear in $\goh^{\mu\nu}$ and $\phi^A$ and consists
of both the pure gravitational and matter parts:
 \bea
 {\cal G}^L_{\mu\nu}(\goh) & \equiv &
{\delta \over {\delta\bar{g}^{\mu\nu}}} \goh^{\rho\sig}
{{\delta\bar{\cal R}}\over{\delta \bar{\gog}^{\rho\sig}}}\nonumber \\ &\equiv &
\half \l({\bn_\rho}{\bn^\rho}\goh_{\mu\nu} + {\bar
g_{\mu\nu}}{\bn_\rho}{\bn_\sig}\goh^{\rho\sig} - {\bn_\rho}{\bn_\nu}\goh_{\mu}^{~\rho} - {\bn_\rho}{\bn_\mu}\goh_{\nu}^{~\rho}\r) , \m {(a2.18)}\\
  {\bm \Phi}^L_{\mu\nu}(\goh, \phi) &\equiv & -2\k {\delta \over {\delta\bar
g^{\mu\nu}}} \l(\goh^{\rho\sig} {{\delta\bar{\lag}^M}\over{\delta
\bar{\gog}^{\rho\sig}}} + \phi^A {{\delta
{{\bar{\lag}}^M}}\over{\delta {\bar\Phi^A}}}\r)\, . \m{(a2.19)}
 \eea
Indices on $\goh^{\mu\nu}$ are moved with the background metric.
The right hand side of (\ref{(a2.17)}) is the total symmetric (metric)
energy-momentum tensor density obtained by the variation with respect to the background metric:
 \be
 {\bm t}^{\rm tot}_{\mu\nu} \equiv
2{{\delta{\lag}^{\rm dyn}}\over{\delta \bar g^{\mu\nu}}} \equiv
2{{\delta}\over {\delta \bar g^{\mu\nu}}}\l(-{1\over{2\k}}\lag^g +
\lag^m  \r) \equiv {\bm t}^g_{\mu\nu} +  {\bm t}^m_{\mu\nu}.
\m{(2.20)}
 \ee

First, because under variation a divergence does not contribute to the field equations,  we note that all the expressions in (\ref{(a2.17)})-(\ref{(2.20)}) do not depend on $\rm div$ in (\ref{(2.10)}). Second, by the construction of the dynamical Lagrangian ${\lag}^{\rm dyn}$ in (\ref{(2.10)}), we note that the energy-momentum ${\bm t}^{\rm tot}_{\mu\nu}$ is not less than quadratic in $\goh^{\mu\nu}$ and $\phi^A$. Thus, the pure
gravitational part is
 \be {\bm t}^g_{\mu\nu} =
{1 \over \k} \l[%
\sqrt{-\bar g}\l(-\del^\rho_\mu \del^\sig_\nu +\half \bar g_{\mu\nu} \bar
g^{\rho\sig}\r)\l(\Del^\alf{}_{\rho\sig}\Del^\beta{}_{\alf\beta} -
\Del^\alf{}_{\rho\beta}\Del^\beta{}_{\alf\sig}\r) + \bn_\tau
{\cal Q}^\tau{}_{\mu\nu}\r]\, . \m{(2.20')}
 \ee
 with
 \bea
 2{\cal Q}^\tau_{\mu\nu}& \equiv &  -\bar g_{\mu\nu} \goh^{\alf\beta}\Del^\tau{}_{\alf\beta}+ \goh_{\mu\nu} \Del^\tau{}_{\alf\beta}\bar g^{\alf\beta}- \goh^\tau_{\mu}
\Del^\alf{}_{\nu\alf}- \goh^\tau_{\nu} \Del^\alf{}_{\mu\alf}\nonumber\\
& +& \goh^{\beta\tau}\l( \Del^\alf{}_{\mu\beta}\bar g_{\alf\nu} + \Del^\alf{}_{\nu\beta}\bar g_{\alf\mu}\r)
 +  \goh^{\beta}_{\mu}\l( \Del^\tau{}_{\nu\beta}-
 \Del^\alf{}_{\beta\rho}\bar g^{\rho\tau}\bar g_{\alf\nu}\r)
\nonumber\\ &+&\goh^{\beta}_{\nu}\l( \Del^\tau{}_{\mu\beta}-
 \Del^\alf{}_{\beta\rho}\bar g^{\rho\tau}\bar g_{\alf\mu}\r)\, .
\m{(2.20'')}
 \eea
The matter part is expressed through the usual material
energy-momentum tensor $ T_{\mu\nu}$ defined in the Einstein equations (\ref{(a2.2+)}) by the expression:
 \bea
{\bm t}^m_{\mu\nu}& = &\sqrt{-\bar g}\l[ \l(\del^\rho_\mu
\del^\sig_\nu -\half \bar g_{\mu\nu} \bar g^{\rho\sig}\r)
\l(T_{\rho\sig} - {\half}g_{\rho\sig}T_{\pi\lam}g^{\pi\lam}\r)
 -\bar{T}_{\mu\nu}\r]\nonumber \\
&{}& -\,2{\delta \over {\delta\bar g^{\mu\nu}}} \l(\goh^{\rho\sig}
 {{\delta\bar{\lag}^M}\over{\delta \bar{\gog}^{\rho\sig}}} + \phi^A
{{\delta {\bar{\lag}^M}}\over{\delta {\bar\Phi^A}}}\r)\, .
\m{(2.20+)}
 \eea

Let us compare the energy-momentum definition in (\ref{(2.20)}) with an attempt to define a symmetrical energy-momentum using  $\delta \lag^E/ \delta
g^{\mu\nu}$ in the usual description of GR. The latter is not reasonable  because it vanishes if the Einstein equations themselves (\ref{(a2.2)}) hold. On the other hand, ${\bm t}^{\rm tot}_{\mu\nu}$ defined in
(\ref{(2.20)}) does not vanish on the field equations (\ref{(a2.17)}). A formal reason is that in the Lagrangian
(\ref{(2.10)}) the linear terms are subtracted off in the Lagrangian (\ref{(2.10)}) and these terms affect the energy-momentum tensor.

Let us simplify the field equations (\ref{(a2.17)}). By  the definitions (\ref{(a2.19)}) and (\ref{(2.20+)}), they  can be rewritten in the form:
 \be
 {\cal G}^L_{\mu\nu} = \k\l( {\bm t}^g_{\mu\nu} +  \delta {\bm t}^M_{\mu\nu} \r) = \k{\bm t}^{\rm eff}_{\mu\nu}\,.
 \m{B30}
 \ee
Such a form could be obtained if the construction of equations in the field-theoretic form is provided `by hand' as  is discussed in the Introduction. The left hand side is linear in metric perturbations, whereas other terms are moved to the right hand side and united in ${\bm t}^{\rm eff}_{\mu\nu}$. Now we will show that (\ref{(a2.17)}), the same (\ref{B30}), are equivalent to (\ref{(a2.2)}).
Considering $\delta{\bm t}^M_{\mu\nu}$ formally we remark that it is the first line in (\ref{(2.20+)}) without the second line, which includes
even linear perturbations in the dynamic fields.  Note especially that ${\bm t}^{\rm eff}_{\mu\nu}$ does not follow from any Lagrangian, and this is disadvantageous compared to  ${\bm t}^{\rm tot}_{\mu\nu}$, which follows from the Lagrangian (\ref{(2.10)}).

At last, we demonstrate the equivalence of the field theoretical equations (\ref{(a2.17)}) with the Einstein equations (\ref{(a2.2)}).
Let us transfer ${\bm t}^{\rm tot}_{\mu\nu}$ to the left hand side of (\ref{(a2.17)}) and use the definitions (\ref{(a2.18)}),
(\ref{(a2.19)}) and (\ref{(2.20)}) with (\ref{(2.10)}). As a result one has
 \bea
 &{}& {\cal G}^L_{\mu\nu} + {\bm \Phi}^L_{\mu\nu}
- \k{\bm t}^{\rm tot}_{\mu\nu}\nonumber\\ & \equiv & - 2\k {{\di \bar
{\gog}^{\rho\sig}}\over {\di \bar{g}^{\mu\nu}}} {{\delta}\over
{\delta {\goh^{\rho\sig}}}} \l[ -{1 \over {2\k }} {\cal R}\l(\bar{\gog}+ \goh\r) +  {\lag}^M
\l(\bar{\Phi} + \phi;~\Bar{\gog} + \goh\r)\r] \nonumber \\
&{}&+~ 2\k{{\delta}\over {\delta \bar{g}^{\mu\nu}}} \l( -{1 \over
{2\k }} \bar{\cal R} +  \bar{\lag}^M \r). \m {(2.17')}
 \eea
One easily recognizes that the second line is proportional to the operator in the Einstein equations (\ref{(a2.2)}), whereas the third line is proportional to the operator of the
background equations in (\ref{(a2.6)}). Thus we conclude that the equations (\ref{(a2.17)}) are equivalent to the Einstein equations (\ref{(a2.2)}) assuming that the background equations in (\ref{(a2.6)}) hold.  Thus we have the Einstein equations in the field-theoretic form.

The above field-theoretic reformulation of GR permits one easily to construct algorithm for production of an approximate scheme up to an arbitrary order in perturbations. Assume that ${\lag}_{\sst EH}$ is smooth enough. Then the Lagrangian $\lag_{\sst EH}(\bar \gog + \goh,\, \bar \Phi + \phi)$ can be presented in the form:  
 \bea
&{}&{\lag}_{\sst EH}  = \bar {\lag}_{\sst EH} + \goh^{\rho\sig}
{{\delta\bar{\lag}_{\sst EH}}\over{\delta \bar{\gog}^{\rho\sig}}} + \phi^B
{{\delta {\bar{\lag}_{\sst EH}}}\over{\delta {\bar\Phi^B}}}  \nonumber\\
&+& \frac{1}{2!} \goh^{\alf\beta} {{\delta} \over{\delta \bar
{\gog}^{\alf\beta}}} \goh^{\rho\sig} {\delta  {{{\bar{\lag}_{\sst EH}
}}}\over{\delta \bar { \gog}^{\rho\sig}}}+ \goh^{\rho\sig}
{{\delta }\over{\delta \bar { \gog}^{\rho\sig}}}\phi^A {{\delta
{{\bar{\lag}_{\sst EH}}}}\over{\delta \bar\Phi^A}} + \frac{1}{2!}\phi^B
{{\delta }\over{\delta \bar\Phi^B}}\phi^A {{\delta {{\bar{\lag}_{\sst EH}}}}
\over{\delta \bar\Phi^A}} +      \ldots + {\rm div}\, .
 \m{b-b33}
 \eea
The expansion is the Lagrangian derivative form of the  functional expansion employed in DeWitt's background-field formalism \cite{DeWitt-book}.
 The main property, which has been used for presentation (\ref{b-b33}) is that the Lagrangian derivatives commute up to a divergence, such as
 \bea
 \goh^{\rho\sig}
{{\delta }\over{\delta \bar { \gog}^{\rho\sig}}}\phi^A {{\delta
{{\bar\lag_{\sst EH}}}}\over{\delta \bar\Phi^A}} = \phi^A {{\delta
{}}\over{\delta \bar\Phi^A}}\goh^{\rho\sig}
{{\delta {\bar\lag_{\sst EH}} }\over{\delta \bar { \gog}^{\rho\sig}}} + {\rm div}\,.
\m{b-b33_a}
 \eea

On substituting (\ref{b-b33}) into the dynamical Lagrangian (\ref{(2.10)}), one finds that it is not less than quadratic in the dynamical variables and has the form:
 \bea
{\lag}^{\rm dyn} & =& \frac{1}{2!} \goh^{\alf\beta} {{\delta}
\over{\delta \bar {\gog}^{\alf\beta}}} \goh^{\rho\sig} {\delta
{{{\bar{\lag}_{\sst EH}}}}\over{\delta \bar { \gog}^{\rho\sig}}}+ \goh^{\rho\sig} {{\delta }\over{\delta \bar { \gog}^{\rho\sig}}}\phi^A {{\delta {{\bar{\lag}_{\sst EH}}}}\over{\delta
\bar\Phi^A}} + \frac{1}{2!}\phi^B {{\delta }\over{\delta
\bar\Phi^B}}\phi^A {{\delta  {{\bar{\lag}_{\sst EH}}}}\over{\delta
\bar\Phi^A}}\nonumber\\ &+&
\frac{1}{3!}\goh^{\mu\nu} {{\delta} \over{\delta \bar {\gog}^{\mu\nu}}} \goh^{\alf\beta} {{\delta} \over{\delta \bar {\gog}^{\alf\beta}}} \goh^{\rho\sig} {{\delta
{{\bar{\lag}_{\sst EH}}}}\over{\delta \bar { \gog}^{\rho\sig}}} + \frac{1}{2!}\goh^{\mu\nu} {{\delta} \over{\delta \bar {\gog}^{\mu\nu}}} \goh^{\alf\beta} {{\delta} \over{\delta \bar {\gog}^{\alf\beta}}}\phi^A {{\delta  {{\bar{\lag}_{\sst EH}}}}\over{\delta\bar\Phi^A}}\nonumber\\ &+& \frac{1}{2!}\goh^{\mu\nu} {{\delta} \over{\delta \bar {\gog}^{\mu\nu}}} \phi^B {{\delta }\over{\delta \bar\Phi^B}}\phi^A {{\delta  {{\bar{\lag}_{\sst EH}}}}\over{\delta
\bar\Phi^A}} + \frac{1}{3!}\phi^C {{\delta }\over{\delta \bar\Phi^C}}\phi^B {{\delta }\over{\delta
\bar\Phi^B}}\phi^A {{\delta  {{\bar{\lag}_{\sst EH}}}}\over{\delta \bar\Phi^A}}+\ldots  + {\rm div}\, .
 \m{b-b34}   
 \eea
 The remarkable structure of (\ref{b-b34}) permits one to represent the variation with respect to  dynamical variables, $\goh^{\mu\nu}$, in the form:
 \be
 \frac{\delta {\lag}^{\rm dyn}}{\delta \goh^{\mu\nu}}= {{\delta}
\over{\delta \bar {\gog}^{\mu\nu}}} \l(\goh^{\rho\sig} {\delta
{{{\bar\lag_{\sst EH}}}}\over{\delta \bar { \gog}^{\rho\sig}}} + \phi^A {{\delta
{\bar\lag_{\sst EH}}}\over{\delta {\bar\Phi^A}}}\r)+ \frac{\delta {\lag}^{\rm dyn}}{\delta \bar{\gog}^{\mu\nu}} = 0.
 \m{b-b35}
 \ee
One immediately recognizes the equations (\ref{(a2.17)}), which are the gravitational equations in the framework of the field-theoretic formulation of GR.

The dynamical Lagrangian in the form (\ref{b-b34}) gives the possibility to construct field equations and the energy-momentum tensor up to a necessary order of approximation. Thus, the quadratic approximation of (\ref{b-b34}) leads to linear equations
 \be
-\frac{1}{2\k}\frac{\di g^{\rho\sig}}{\di \gog^{\mu\nu}}\l({\cal G}^L_{\rho\sig}(\goh)+
{\bm \Phi}^L_{\rho\sig}(\goh, \phi)\r)\equiv  {{\delta}
\over{\delta \bar {\gog}^{\mu\nu}}} \l(\goh^{\rho\sig} {\delta
{{{\bar\lag_{\sst EH}}}}\over{\delta \bar { \gog}^{\rho\sig}}}  + \phi^A {{\delta
{\bar\lag_{\sst EH}}}\over{\delta {\bar\Phi^A}}}\r)= 0\, ,
\m{b-b37}
 \ee
and to the quadratic energy-momentum:
 \bea
 {\bm t}^{\rm tot}_{\mu\nu}  = {2}\frac{\delta}{\delta
\Bar g^{\mu\nu}}\l(\frac{1}{2!} \goh^{\alf\beta} {{\delta}
\over{\delta \bar {\gog}^{\alf\beta}}} \goh^{\rho\sig} {\delta
{{{\bar{\lag}_{\sst EH}}}}\over{\delta \bar { \gog}^{\rho\sig}}}+
\goh^{\rho\sig} {{\delta }\over{\delta \bar { \gog}^{\rho\sig}}}\phi^A {{\delta {{\bar{\lag}_{\sst EH}}}}\over{\delta
\bar\Phi^A}} + \frac{1}{2!}\phi^B {{\delta }\over{\delta
\bar\Phi^B}}\phi^A {{\delta  {{\bar{\lag}_{\sst EH}}}}\over{\delta
\bar\Phi^A}}\r) \, .
 \m{b-b39}
 \eea
The next approximations of (\ref{b-b34}) give the possibility to construct the field equations and the energy-momentum  in next approximations as well. Thus  the presentation of the dynamical Lagrangian  in the form (\ref{b-b34}) presents a concrete algorithm for constructing an approximate scheme.

 We note that the above formalism has been used to develop a so-called nonlinear quantum mechanics with non-classical gravitational self-interaction \cite{Popova_P_1993_a,Popova_P_1993_b}; it has been  applied  to study problems regarding the early universe \cite{Petrov_P_1994_a,Petrov_P_1994_b}. Besides, the formalism has been used to generalize possibilities to construct variants of unimodular gravity \cite{Petrov_1991}. 

Finally, let us suggest a new proposal how spinors/fermions can be included into the above consideration. Including spinorial/fermionic matter poses additional challenges, some of which have been mentioned previously \cite{MassiveGravity3}.  Two possibilities are  either a bi-tetrad  formalism (\emph{e.g.}, \cite{MassiveGravity2})  or  a nonlinear metric-dependent spinor formalism   distinguishing spinors with a transformation law dependent on the background metric and spinors with a transformation law dependent on the total effective metric (using the Ogievetsky-Polubarinov formalism \cite{OPspinor,OP,PittsSpinor} twice) and then attempting to define a spinor perturbation.  In the bi-tetrad case, the gravitational perturbation (the difference between the tetrads), when suitably combined with the background tetrad, forms a locally Lorentz-invariant rank $2$ tensor gravitational potential.  This tensor must  be symmetric in order to avoid introducing a new antisymmetric gravitational potential into the theory with no analog in geometrical GR.  The local Lorentz gauge freedom can be fixed by making the background tetrad symmetric, thus yielding the Ogievetsky-Polubarinov nonlinear spinor formalism in terms of the background metric.  In the formalism with two kinds of Ogievetsky-Polubarinov metric-dependent nonlinear spinors relating to the two metrics, one faces the problem of attempting to subtract spinors defined relative to two different (more specifically, not conformally related) metrics.  Defining the spinor perturbation thus calls for making a bimetric field-dependent local Lorentz transformation on the full spinor that was initially defined in relation to the full metric.  Then  the full spinor, the background spinor, and consequently the spinor perturbation are all defined relative to the background metric. Again one has the Ogievetsky-Polubarinov nonlinear spinor formalism relative to the background metric.  Thus presumably these two approaches are equivalent.  This sketch indicates that the background metric formalism discussed in this review also admits spinors with no essential difficulty. However, one does not expect fermionic matter as such to be relevant macroscopically for the most common  astrophysical and cosmological applications (even if neutrinos are a dark matter candidate),  so a sketch suffices for present purposes. These approaches also indicate possibilities for spin $2$ derivations including fermions (a subject frequently neglected in that decades-long tradition) and are likely to provide an alternative to Shirafuji's conclusion \cite{Shirafuji} that spin $2$ derivations involving spinors require a physically meaningful antisymmetric gravitational potential.


\vspace{0.3cm}
\sect{Gauge invariance properties}
\m{gauge}

We identify as ``gauge transformations'' those transformations that act on the dynamical variables only; they  do not act  on either the coordinates and on the background (fixed) quantities. Properties of the field-theoretic formulation of GR under gauge
transformations follow from the usual covariance of GR in the geometrical formulation. Here, we follow the presentation in \cite{GPP,[15]}).  We also  note that the gauge transformation properties are used for construction of the field-theoretic formulation of GR like a typical gauge theory; see  \cite{[16]}).

Let us consider any arbitrary solution to GR in two different coordinate systems: $\gog^{\mu\nu}(x)$ and $\gog'^{\mu\nu}(x')$.
Coordinate systems $\{x\}$ and $\{x'\}$, are connected by the coordinate transformation  $x' = x'(x)$. Now let us apply a decomposition of the type (\ref{(a2.4)}) in both the cases:
\bea
\gog^{\mu\nu}(x)&=& \bar{\gog}^{\mu\nu}(x)+\goh^{\mu\nu}(x), \m{h+h}\\
\gog'^{\mu\nu}(x')&=& \bar{\gog}^{\mu\nu}(x')+\goh'^{\mu\nu}(x')
\m{h+h_prime}
\eea
with the specification of the {\em same} functional form of the background metric density $\bar{\gog}^{\mu\nu}$:  that is, $\bar{\gog}^{\mu\nu}(x)$  depends mathematically on its (unprimed) coordinates in the same way as $ \bar{\gog}^{\mu\nu}(x')$ depends on its (primed) coordinates (though the same coordinate values in the two coordinate systems of course pick out distinct space-time points).  
This means that for the same geometrical solution, perturbations $\goh'^{\mu\nu}$ and $\goh^{\mu\nu}$ are defined in two different ways.

Now, let us express $\goh'^{\mu\nu}$ in terms of $\goh^{\mu\nu}$. For the solution in the {\em primed} coordinates from the points with coordinate values $x'$, we go to the points with coordinate values $x$ using the transformation functions $x' = x'(x)$.  Then, the equality (\ref{h+h_prime}) transforms to
\be
\gog'^{\mu\nu}(x)= \bar{\gog}^{\mu\nu}(x)+\goh'^{\mu\nu}(x)\,.
\m{h+h+}
\ee
Now, comparing (\ref{h+h+}) with (\ref{h+h}), we  turn to the right hand sides and {\em do not touch} the first terms. Thus, transformations are related to the perturbations (dynamical variables) $\goh^{\mu\nu}$  and $\goh'^{\mu\nu}$ only. The same procedure has to be implemented for the matter variables.
Now we are in a position to connect $\goh^{\mu\nu}$  and $\goh'^{\mu\nu}$ as well as $\phi^{A}$  and $\phi'^{A}$ by gauge (not coordinate) transformations in explicit form.  The coordinate transformation $x' = x'(x)$ can be represented in the form:
\be
x^{\prime\alf} = x^{\alf} + \xi^\alf + { 1\over {2!}}\xi^\beta
\xi^\alf_{~,\beta}+ { 1\over {3!}}\xi^\rho\l(\xi^\beta
\xi^\alf_{~,\beta}\r)_{,\rho} + \ldots\,
\m{trans}
\ee
with the displacement vector $\xi^\mu$. Knowing the connection between $\gog^{\mu\nu}(x)$ and $\gog'^{\mu\nu}(x)$ on the left hand sides of (\ref{h+h}) and (\ref{h+h_prime}), we infer the transformations of the right hand sides and transfer  all the changes to $\goh^{\mu\nu}$  and $\goh'^{\mu\nu}$. Finally, assuming that $\xi^\mu$ is sufficiently smooth, we obtain  the gauge transformations in the field-theoretic formulation of GR \cite{GPP,[15]} (recalling the definition of the Lie derivative above):
 \bea
 {\goh}'^{\mu\nu} &=& \goh^{\mu\nu} + \sum^{\infty}_{k = 1}{1\over{k!}}~ \hbox{$\pounds$}_\xi^k \l(\bar {\gog}^{\mu\nu} + \goh^{\mu\nu}\r)\,, \m{(5.9)}\\
{\phi}'^A &=& \phi^A + \sum^\infty_{k =
1}{1\over{k!}}~\hbox{$\pounds$}_\xi^k \l(\bar\Phi^A+\phi^A\r)\, .
\m{(5.9)+}
 \eea
 Indeed, they  affect neither the coordinates nor the background quantities.  As one recalls from Einstein's point-coincidence argument, in GR space-time points are physically individuated empirically by the observable events that happen there \cite{HowardPointCoincidence}; thus the physical meaning of these gauge transformations is not immediate. One sees that if one adds the background quantities to each side, then these gauge transformations leave the background metric and matter fields alone while altering the total curved metric and total matter fields just as coordinate transformations do.

Now, it is important to show why transformations  (\ref{(5.9)}) and (\ref{(5.9)+}) are called gauge transformations. Then, it is necessary to show how under substitution of them the dynamical Lagrangian, equations in the field-theoretic form of GR and energy-momentum complexes are changed. First, let us turn to the Lagrangian (\ref{(2.10)}) and substitute  (\ref{(5.9)}) and (\ref{(5.9)+}):
 \be
 {\lag}'^{\rm dyn}  = {\lag}^{\rm dyn}  -
(\goh'^{\mu\nu}-\goh^{\mu\nu}) {{\del \bar{\lag}_{\sst EH}} \over {\del
\bar{\gog}^{\mu\nu}}} - (\phi'^A - \phi^A){{\delta \bar
{\lag}^M}\over{\delta \bar{\Phi^A}}}+ {\rm div}\,.
\m{LagrangianGPP-gauge}
\ee
One can see that ${\lag}^{\rm dyn}$ is invariant under the transformations (\ref{(5.9)}) and (\ref{(5.9)+})
up to a divergence if the background equations (\ref{(a2.6)}) hold.
Second, turn to the equations (\ref{(a2.17)}) and consider their operator in the form (\ref{(2.17')}), which enables finding how they are changed under the substitution of (\ref{(5.9)}) and (\ref{(5.9)+}). One easily finds
  \bea
  &{}& \l[{\cal G}^L_{\mu\nu} + {\bm \Phi}^L_{\mu\nu} - \k{\bm t}^{\rm tot}_{\mu\nu} \r]' =
 \l[{\cal G}^L_{\mu\nu} + {\bm \Phi}^L_{\mu\nu} - \k{\bm t}^{\rm tot}_{\mu\nu}  \r]
\nonumber \\
&+& {{{\di \bar {\gog}^{\rho\sig}}}\over {\di \bar{g}^{\mu\nu}}}
\sum^\infty_{k=1} {1\over k!} {\Lix}^k\l[ {{\di \bar {g}^{\delta\pi}}\over {\di \bar{\gog}^{\rho\sig}}}
 \l({\cal G}^L_{\delta\pi} + {\bm \Phi}^L_{\delta\pi}
- \k{{\bm t}^{\rm tot}}_{\delta\pi} \r) -2\k {{\delta}\over {\delta \bar{\gog}^{\rho\sig}}}  {\bar{\lag}_{\sst EH}}\r].
\m{(2.17-gauge)}
 \eea
 One can see that the field-theoretic equations in GR are invariant under the transformations (\ref{(5.9)}) and (\ref{(5.9)+}) if a) they themselves  hold, and if b) the background equations (\ref{(a2.6)}) hold. Thus, the gauge transformations (\ref{LagrangianGPP-gauge}) and (\ref{(2.17-gauge)}) reflect the gauge invariance properties of the field-theoretic formulation of GR. One could also notice that the Lagrangian density changes by only a divergence.  Third, let us consider the gauge invariance properties of the energy-momentum tensor density (\ref{(2.20)}) (or
(\ref{B30})). Keeping in mind  the field-theoretic equations, one
has only
 \bea
 \k{{\bm t}}'^{\rm tot}_{\mu\nu}& = & \k{\bm t}^{\rm tot}_{\mu\nu}
  + {\cal G}^L_{\mu\nu}(\goh'-\goh)+{\bm \Phi}^L_{\mu\nu}(\goh'-\goh,\,,
  \phi'-\phi)\, ,
  \m{tei-gauge}\\
   \k{{\bm t}'}^{\rm eff}_{\mu\nu}& = & \k{\bm t}^{\rm eff}_{\mu\nu}
  + {\cal G}^L_{\mu\nu}(\goh'-\goh)\,:
  \m{tei-gaugeBG}
  \eea
that is, the energy-momentum complexes are not gauge invariant. The mathematical reason is by the presence of  second and third terms in
(\ref{LagrangianGPP-gauge}) and by a requirement that the background equations must  not be used before variation of (\ref{LagrangianGPP-gauge}) with respect to
$\bar{g}^{\mu\nu}$.

The longstanding (1910s+) problem known as the non-localizability of gravitational energy is illustrated by the {\em non-covariance} of pseudotensors and related superpotentials; see chapter 1 in the book \cite{Petrov+_2017}. A covariantization of pseudotensors and superpotentials can be achieved using  an auxiliary background metric. However, in this case, the non-localization problem transforms into an ambiguity in the choice of the background.
But such a formalism does not suggest any unique mathematical derivation for a concrete description of such an ambiguity. We close this gap here: the gauge transformations (\ref{tei-gauge}) and (\ref{tei-gaugeBG}) for the total energy-momentum and the effective energy-momentum show how the non-localization initiated by different choices of backgrounds is expressed mathematical terms. It is  one of the advantages when the field-theoretic formulation of GR is applied.

It is important to note that in the case of a {\em Ricci-flat background}, $\bar R_{\mu\nu}=0$,
one has $\Phi^L_{\mu\nu}=0$, therefore the energy-momentum  ${\bm t}^{\rm tot}_{\mu\nu}$ is not gauge invariant
up to $G^L_{\mu\nu}$, a covariant divergence. Note again that the energy-momentum ${ t}^{\rm eff}_{\mu\nu}$ is not gauge invariant up to a covariant divergence even in the case of {\em arbitrary curved backgrounds}. These facts could be important for determining gauge invariance of conserved charges because divergences just contribute  surface integrals.

It is also important to consider equations and gauge transformations in linear, quadratic and other approximations. Assume that perturbations  are small ($\goh^{\mu\nu} \ll \bar{\gog}^{\mu\nu}$, $\phi^A \ll \bar\Phi^A$), 
and so are their derivatives (low-frequency approximation). Assume also that the background equations (\ref{(a2.6)})
give a solution $\bar{\gog}^{\mu\nu} \sim f(\k)\bar\Phi^A$ with a coefficient $f(\k)$ of the order of the Einstein's constant. Then one can set ${\goh^{\mu\nu}} \sim f(\k){\phi^A}$, {\it etc}. To understand better the main properties of the approximation scheme, we derive the equations (\ref{(a2.17)}) up to second order:
 \be
G^L_{\mu\nu}(\goh) + \Phi^L_{\mu\nu}(\goh,~\phi) -
8\pi\, {}_2t^{\rm tot}_{\mu\nu}(\goh\goh,~\goh\phi,~\phi\phi) = 0\,.
 \m{b-b62}
 \ee
The perturbations can be expanded as usual, $\goh^{\mu\nu} = \goh^{\mu\nu}_{1} + \goh^{\mu\nu}_{2} + ...~$, and $\phi^{A} = \phi^{A}_{1} + \phi^{A}_{2} + ...~$. Then one can obtain a solution to the equations (\ref{(a2.17)}) step by step. Thus, to obtain the solution of (\ref{b-b62}) one has to find, firstly, $\goh_{1}$ and $\phi_{1}$ and, secondly, $\goh_{2}$ and $\phi_{2}$. Besides, assume $\xi^{\mu} = \xi_{1}^{\mu} + \xi_{2}^{\mu}+ ...$
with  $ \xi_{1}^{\mu} \sim \di_\alf \xi_{1}^{\mu}\sim \ldots \sim \goh^{\mu\nu}_{1}\sim f(\k) \phi^A_{1}$  and $\xi_{2}^{\mu} \sim \di_\alf \xi_{2}^{\mu} \sim \ldots \sim \goh^{\mu\nu}_{2}\sim f(\k) \phi^A_{2}$.

After these assumptions are made, the linear version of the equations (\ref{b-b62}) is
 \be
 G^L_{\alf\beta}(\goh_{1}) + \Phi^L_{\alf\beta}(\goh_{1},~\phi_{1}) = 0\,.
\m{b-b63}
 \ee
The linear approximation the transformations (\ref{(5.9)}) and (\ref{(5.9)+}) is
 \bea \goh'^{\mu\nu}_{1} &=&
\goh^{\mu\nu}_{1} + {{\hbox{$\pounds$}}}_{\xi_{1}} \bar{\gog}^{\mu\nu} = \goh^{\mu\nu}_{1} - \bar {\gog}^{\mu\nu}~\bn_\rho
\xi_{1}^\rho+ \sqrt{-\bar g}\l(\bn^\mu \xi_{1}^\nu +
\bn^\nu \xi_{1}^\mu\r)\, ,
\m{b-b64} \\
\phi'^{A}_{1} &=& \phi^{A}_{1} +{{\hbox{$\pounds$}}}_{\xi_{1}}
\bar{\Phi}^{A}. \m{b-b65}
 \eea
Substituting (\ref{b-b64}) and (\ref{b-b65}) into (\ref{(2.17-gauge)}) and retaining  the linear approximation\index{approximations!linear}, one has
 \bea
&{}&\l[G^L_{\mu\nu}(\goh_{1}) + \Phi^L_{\mu\nu}(\goh_{1},~\phi_{1})\r]' = \l[
G^L_{\mu\nu}(\goh_{1}) + \Phi^L_{\mu\nu}(\goh_{1},~\phi_{1})\r] \nonumber \\&+&
\l(\del^\rho_\mu\del^\sig_\mu - \half \bar
g_{\mu\nu}\bar{g}^{\rho\sig}\r) {{\hbox{$\pounds$}}}_{\xi_{1}} \l[ \bar R_{\rho\sig} -8\pi\l(\bar T_{\rho\sig} - \half \bar g_{\rho\sig}\bar T\r) \r]. \m{b-b66}
 \eea
Thus, the linear equations are gauge invariant on the background equations only; it is not necessary to require that the fields $\goh_{1}$ and $\phi_{1}$ satisfy (\ref{b-b63}). In the simplest case of the Ricci-flat background, the linear transformations have the form (\ref{b-b64}) only, without (\ref{b-b65}). Then the formula (\ref{b-b66}) transfers to the formula $G'^L_{\mu\nu} = G^L_{\mu\nu}$, which expresses the gauge invariance\index{gauge!invariance} of the linear spin-2 field that can be found in the text books. Thus (\ref{b-b66}) the generalization of the well known gauge invariance in the linear gravity.

The equations (\ref{b-b62}) rewritten in the quadratic order are %
  \be
 G^L_{\alf\beta}(\goh_{2}) + \Phi^L_{\alf\beta}(\goh_{2},~\phi_{2}) -
 8\pi\l({}_2t^{\rm g}_{\alf\beta}(\goh_{1}\goh_{1}) +
{}_2t^{\rm m}_{\alf\beta}(\goh_{1}\goh_{1}, ~\goh_{1}\phi_{1},~\phi_{1}\phi_{1})\r) = 0\, . \m{b-b67}
 \ee
The gauge transformations (\ref{(5.9)}) and (\ref{(5.9)+}) in the quadratic order are
 \bea
\goh'^{\mu\nu}_{2} &=& \goh^{\mu\nu}_{2} +
{{\hbox{$\pounds$}}}_{\xi_{2}} \bar{\gog}^{\mu\nu}+
{1\over{2!}}{{\hbox{$\pounds$}}}^2_{\xi_{1}} \bar{\gog}^{\mu\nu} +
{{\hbox{$\pounds$}}}_{\xi_{1}} \goh^{\mu\nu}_{1}
\m{b-b68} \\
\phi'^{A}_{2} &=& \phi^{A}_{2} + {{\hbox{$\pounds$}}}_{\xi_{2}}
\bar{\Phi}^{A}+ {1\over{2!}}{{\hbox{$\pounds$}}}^2_{\xi_{1}}
\bar{\Phi}^{A} + {\hbox{$\pounds$}}_{\xi_{1}} \phi^{A}_{1}\, .
\m{b-b69}
 \eea
Substitution of (\ref{b-b68}) and (\ref{b-b69}) into (\ref{(2.17-gauge)}) gives for  the quadratic approximation:
 \bea
&{}& \l[G^L_{\mu\nu}(\goh_{2}) + \Phi^L_{\mu\nu}(\goh_{2},~\phi_{2})
 - 8\pi\, {}_2t^{\rm tot}_{\mu\nu}(\goh_{1}\goh_{1}, ~\goh_{1}\phi_{1},~\phi_{1}\phi_{1})\r]'
\nonumber \\
&=&
\l[G^L_{\mu\nu}(\goh_{2}) + \Phi^L_{\mu\nu}(\goh_{2},~\phi_{2})
 - 8\pi\, {}_2t^{\rm tot}_{\mu\nu}(\goh_{1}\goh_{1}, ~\goh_{1}\phi_{1},~\phi_{1}\phi_{1})\r]
 \nonumber\\
 &+&
\frac{1}{\sqrt{-\bar g}}{{\di \bar {\gog}^{\rho\sig}}\over {\di \bar g^{\mu\nu}}}
\l({\hbox{$\pounds$}}_{\xi_{2}}  + {1\over
2!}{\hbox{$\pounds$}}^2_{\xi_{1}}\r) \l[ \bar R_{\rho\sig}
-8\pi\l(\bar T_{\rho\sig} - \half \bar g_{\rho\sig}\bar T\r) \r] +
\nonumber\\
&+& \frac{1}{\sqrt{-\bar g}}{{\di \bar {\gog}^{\rho\sig}}\over {\di \bar g^{\mu\nu}}}
{\hbox{$\pounds$}}_{\xi_{1}}\l[{\sqrt{-\bar g}} {{\di \bar g^{\delta\pi}}\over {\di
\bar{\gog}^{\rho\sig}}} \l[G^L_{\delta\pi}(\goh_{1})) +
\Phi^L_{\delta\pi} (\goh_{1},~\phi_{1})\r]\r]. \m{b-b70}
 \eea
One can see that equations (\ref{b-b67})  are gauge invariant on the background
equations (\ref{(a2.6)}) and on the linear equations (\ref{b-b63}). The procedure in the next orders is similar.

One can compare  coordinate transformations, gauge transformations, and a partially compensating combination of coordinate and gauge transformations to appreciate how the background metric is gauge dependent \cite{GPP,PittsSchive2001a}.
Coordinate transformations connected to the identity take the form
\begin{eqnarray}
 \mathfrak{g}^{\prime\mu\nu} = \sum_{k=0}^{\infty}{1\over{k!}} \pounds^k_{\xi} \mathfrak{g}^{\mu\nu}, \\
  \mathfrak{\bar g}^{\prime\mu\nu} = \sum_{k=0}^{\infty} {1\over{k!}} \pounds^k_{\xi} \mathfrak{\bar g}^{\mu\nu}, \\
 \Phi^{\prime A} = \sum_{k=0}^{\infty} {1\over{k!}} \pounds^k_{\xi} \Phi^{A}, \\
\bar \Phi^{\prime A} = \sum_{k=0}^{\infty}{1\over{k!}} \pounds^k_{\xi} \bar \Phi^{A},
\end{eqnarray}
 with the Lie derivative acting on all the field variables including the background metric and background matter.  Gauge transformations, such as were described above, can be presented as acting on the total effective curved metric and matter fields, while leaving the background metric and background matter  alone:
\begin{eqnarray}
 \mathfrak{g}^{\prime\mu\nu} = \sum_{k=0}^{\infty}{1\over{k!}} \pounds^k_{\xi} \mathfrak{g}^{\mu\nu}, \\
  \mathfrak{\bar g}^{\prime\mu\nu} =  \mathfrak{\bar g}^{\mu\nu}, \\
 \Phi^{\prime A} = \sum_{k=0}^{\infty} {1\over{k!}} \pounds^k_{\xi} \Phi^{A}, \\
\bar \Phi^{\prime A} = \bar \Phi^{A}.
\end{eqnarray}
 What happens if one combines these two transformations with equal-but-opposite descriptor vector fields?  The resulting combined transformation alters the background metric and background matter  while leaving the effective/total curved metric and effective/total matter fields alone \cite{PittsSchive2001a}:
\begin{eqnarray}
 \mathfrak{g}^{\prime\mu\nu} =  \mathfrak{g}^{\mu\nu}, \\
  \mathfrak{\bar g}^{\prime\mu\nu} = \sum_{k=0}^{\infty} {1\over{k!}} \pounds^k_{-\xi} \mathfrak{\bar g}^{\mu\nu}, \\
 \Phi^{\prime A} =  \Phi^{A}, \\
\bar \Phi^{\prime A} = \sum_{k=0}^{\infty}{1\over{k!}} \pounds^k_{-\xi} \bar \Phi^{A}.
\end{eqnarray}  Thus dependence on the background metric and/or background matter  makes an expression gauge-dependent under these transformations.  Thus some of the advantage for describing gravitational energy using a background metric and hence tensorially is offset by the  additional gauge dependence \cite{BrazilLocalize}.

To conclude the section we note that  on the basis of gauge invariance properties of the field-theoretic formalism, in recent papers \cite{PK1,PK2,PK3,PK4}  a gauge invariant theory of the cosmological perturbations has been elaborated. In order to ascertain the  gauge invariance of global conserved quantities for isolated systems, the weakest fall-off for gravitational potentials at spatial infinity has been determined \cite{Petrov_1995,Petrov_1997}.


\vspace{0.3cm}
\sect{Gravitational field on fixed backgrounds}
\m{fixed_STs}
%


What has been presented above is a construction of the field-theoretic formulation
of GR  when from the start in the framework of the geometrical formulation of
GR the decompositions ({\ref{(a2.4)}) and ({\ref{(a2.4m)}) have been provided. Then, the dynamical Lagrangian ({\ref{(2.10)}) has been suggested, and, next, all the structures of the theory have been obtained like in an arbitrary Lagrangian-based field theory. This presentation gives an evident connection of geometrical and field-theoretic representations of GR. However, in order to appreciate the  properties of the theory more clearly,  it is useful to outline other ways (not only on the basis of decompositions ({\ref{(a2.4)}) and ({\ref{(a2.4m)})) of constructing the field theory of gravity
on flat or fixed curved  backgrounds.

\medskip

\noindent {\em Firstly, we recall briefly  how to construct GR as a field theory of gravity in Minkowski space, keeping in mind that we are working in the framework of special relativity. (Also see above.)}
\medskip

To construct such a gravitational theory usually one follows natural requirements:
 \bit
 \item Such a theory has to be Lagrangian-based.
 \item All the fields including gravitational field are to be propagated in Minkowski space.
 \item The main observable tests have to be explained.
 \item In limit of weak fields and low velocities the gravitational theory under construction has to go to the Newtonian theory.
 \eit
As main candidates for gravitational fields in special relativity, researchers considered scalar, vector and the tensor fields. In scalar gravitational theories (see
\cite{MTW,[22]}), the deflection of light in the gravitational
field of the Sun is not described correctly, because a scalar theory does not bend light due to the conformally flat geometry of space-time (exactly for the massless case, or to an arbitrarily good approximation for a small graviton mass) and the conformal invariance of Maxwell's equations in $4$ space-time dimensions.  In the vector theories
(see \cite{MTW,[22],[23]}), in the case of positive
energy of gravitational waves, massive bodies are repulsive, contrary to the most basic features of gravitation. Thus, {\em pure} scalar or {\em pure} vector gravitational theories in Minkowski space are not interesting candidates for the real world. On the other hand, (symmetric rank $2$)  tensor variants of the gravitational theory
can satisfy the aforementioned requirements. Scalar and vector admixtures, nevertheless, can be considered as corrections for tensor theories, see, for example,
 \cite{[25],Horndeski,IgorVlasov}.
We consider the tensor variant only, which leads to the field-theoretic formulation of GR.

When a construction of the pure tensor theory is provided, one assumes the following. The first requirement is that in
Minkowski space-time with Cartesian  coordinates in the field equations, the source of the part
linear in gravitational variables $\goh^{\mu\nu}$  is to be the
symmetric energy-momentum tensor of the matter variables $\phi^A$. The next
requirement is the positivity of the energy of the gravitational
waves, leading to the unique quadratic Lagrangian $\lag^2_{grav}(\goh)$. The last describes the massless spin two field; a mass terms disappears if one assumes the correspondence to the Newtonian potential in the weak field
limit. (Massive spin $2$ gravity famously has multiple theoretical challenges  involving nonlinear negative-energy degrees of freedom typically \cite{TyutinMass,DeserMass} (but see \cite{MaheshwariIdentity,deRhamGabadadze,HassanRosen} and many subsequent works)  and a discontinuous massless limit under a perturbative treatment \cite{vDVmass1,vDVmass2} (but see \cite{Vainshtein,Vainshtein2,VainshteinReview} and many subsequent works; two reviews are \cite{HinterbichlerRMP,deRhamLRR}).
Thus the relativistic equations for the tensor gravitational
field acquire the form:
 \be
G^L_{\mu\nu}(\goh) = \kappa T_{\mu\nu}(\phi,\eta)\,.
 \m{GL=Tphi}
 \ee
At this step, the linear in $\goh^{\mu\nu}$ left hand side of (\ref{GL=Tphi}) is a result of varying $\lag^2_{grav}(\goh)$ with respect to
$\goh^{\mu\nu}$, whereas the right hand side of (\ref{GL=Tphi}) is the symmetric (metric) energy-momentum tensor of the matter variables $\phi^A$.
Because identically $G^L_{\mu\nu}{}^{,\nu} \equiv 0$ one has the conservation law  $T_{\mu\nu}{}^{,\nu} = 0$. The last, however, contradicts to field equations for $\phi^A$, which interact with $\goh^{\mu\nu}$, see  \cite{MTW}. To improve the equations (\ref{GL=Tphi}) one has to change the right hand side:
 \be
G^L_{\mu\nu}(\goh) = \kappa \left[T_{\mu\nu}(\phi, {\bm \eta + \goh}) +
t^{(2)}_{\mu\nu}(\goh) \right]
 \m{GL=T+quadr}
 \ee
where $t^{(2)}_{\mu\nu}(\goh)$ is the symmetric energy-momentum tensor
obtained from $\lag^2_{grav}(\goh)$ by the ordinary variation with respect to background (Minkowski) metric. After that the Lagrangian corresponding to (\ref{GL=T+quadr}) has to be  cubic, that is $\lag^2_{grav}(\goh) + \lag^3_{grav}(\goh)$,  {\it etc.} The contradiction vanishes when iterations are provided an infinite number of times:
 \be
G^L_{\mu\nu}(\goh) = \kappa \left[T_{\mu\nu}(\phi, \goh + \eta) +
\sum_{n=2}^\infty t^{(n)}_{\mu\nu}(\goh) \right] \,.
\m{GL=T+ttot}
 \ee
Notice that now $T_{\mu\nu}$ depends on $\goh^{\mu\nu}$ in the sum $ \gog^{\mu\nu} = \goh^{\mu\nu} +
{\bm \eta}^{\mu\nu}. $ Then, one can show that the equations (\ref{GL=T+ttot}) are equivalent to the Einstein equations in the usual form, see \cite{[23],OP,[26]}.

\medskip

\noindent {\em Secondly, a construction of GR in the field-theoretic form in Minkowski space is generalized to a tensor theory in a curved space-time.}

\medskip


In spite of considerable efforts, up to the beginning of the 1980s, there was not a  completed version of the field-theoretic
formulation of GR with all the properties of a field theory in an arbitrary curved background space-time. In the  works
\cite{GPP,[15],[16]}  this formulation was provided. It took inspiration from  Deser \cite{Deser}, who suggested a covariant
formulation on a flat background where the results given in the
(\ref{GL=Tphi}) - (\ref{GL=T+ttot}) have been suggested in closed
form without expansions. In \cite{GPP}, we have generalized  the Deser principle.
 It could be formulated as:
 \bit
\item {\em  In an arbitrary curved fixed space-time,
the total energy-momentum tensor of all the fields, including
gravitational one, has to be the source of the linear massless field
of  spin $2$ (tensor gravitational field) and linear
perturbations of matter fields}.
 \eit
On the basis of this principle the
formulation in \cite{GPP} presented above in sections 1 was constructed.


\medskip
\noindent {\em Thirdly, another principle for constructing the field-theoretic formulation of GR can be
formulated as a generalization of the Newtonian theory. This means a transformation from a gravistatic law to
gravidynamics (Einstein equations).}

\medskip

Steps of the construction are as follows.

 \bit

\item  To follow the relativistic requirement one has to replace the mass density $\rho$ in the
Newtonian law by the ten components of the matter energy-momentum tensor.

\item Then, the number of the gravitational potentials
should be increased from 1 component $\phi$ to 10 components $\goh^{\mu\nu}$.

\item  After that the Laplace operator in the Newtonian law should be replaced
 by the d'Alembert operator.

\item Besides, the relativistic theory of gravity should
be a theory with self-interaction, that is the gravitational field has to be a source for
itself. Then the gravitational equations become
 \be
\goh_{\mu\nu}{}^{,\alpha}{}_{,\alpha} = \kappa({\bm t}^g_{\mu\nu} +
{\bm t}^m_{\mu\nu}) \equiv
 {\kappa{{\bm t}^{\rm tot}_{\mu\nu}}}.
 \m{Newtonian-nonfull}
 \ee
\item In these equations the gauge condition $\goh^{\mu\nu}{}_{,\nu} = 0$ is already chosen. The
condition when the conservation law $({\bm t}^g_{\mu\nu} + {\bm t}^m_{\mu\nu})^{,\nu} = 0$ holds automatically leads to a necessity
to add the terms $\l[ \eta_{\mu\nu}\goh^{\alpha\beta}{}_{,\alpha,\beta} - \goh^\alpha{}_{\mu,\nu,\alpha}
 - \goh^\alpha{}_{\nu,\mu,\alpha}\r]$ to the left hand side of (\ref{Newtonian-nonfull}). Finally,
(\ref{Newtonian-nonfull}) transforms to the Einstein equations in the form (\ref{GL=T+ttot}).
 \eit
For more detail see the work \cite{Grishchuk92}.

\medskip

\noindent {\em Fourthly, because GR in the field-theoretic form has gauge freedom, it can be constructed as a gauge theory itself.}

\medskip


The gauge principle of constructing the field-theoretic formulation
of GR has been presented and analyzed in the work \cite{[16]}. A non-standard way of localization is postulated. From the very start, the existence
of a fixed background space-time (it can be even curved) with  symmetries presented by Killing vectors is assumed. It is also assumed
that initial dynamic fields in this space-time are propagated and
their action is under consideration. Then, one notices that the initial action is invariant up to a surface term under the addition of Lie derivatives (with respect to aforementioned Killing vector)
to the initial fields to themselves. Then, the Killing vector
in this transformation is replaced by an arbitrary vector. Next we
require the same invariance of a sought-for action for the same
dynamic fields under the {\it localized} transformation.  In the process
the coordinates and the background metric do not change. All of this
plays a role of a local invariance. We note that our concept of
`localization' should not be taken literally. Indeed, it turns out
that in our case a background space-time can have no symmetries. It is enough to require the aforementioned invariance of
the initial action for Lie derivatives with respect to arbitrary
vectors only. As a result of `localization' the compensating (gauge)
field appears. The requirement to have this gauge field as a universal
one with the same (as the other fields) gauge transformation law
leads to the theory with the Lagrangian (\ref{(2.10)}) and the
field equations (\ref{(a2.17)}) where the gauge field is just $\goh^{\mu\nu}$.

All the above methods begin from a concrete background. However, finally it becomes the background in  the field-theoretic formulation of GR.
Already, we have noted that the Einstein equations in field-theoretic form are equivalent to Einstein equations in the standard geometrical form. But the latter have no any background structure at all.\footnote{In the case of spinor fields, the matrix $diag(-1,1,1,1)$ appears (as do $\Gamma$ matrices).  One might call this quantity a confined object rather than an absolute object \cite{TLL,FriedmanJones} because it does not change at all order coordinate transformations. Some authors use this matrix (or the identity matrix with $x^4=ict$) as a background for a perturbative expansion of the effective curved metric  \cite{GuptaPPSL2,Huggins,Feynman63,OP,OPspinor,Veltman}.  Then the gravitational potential  has an inhomogeneous coordinate transformation law and no new gauge freedom arises.  Thus the idea of a background is more common and potentially `thinner' than one might have thought.  }  Therefore, it is quite important to clarify the role, physical or auxiliary, of the background. Indeed, after the identification (\ref{(a2.4)}) and
(\ref{(a2.4m)}) the fixed metric $\bar g_{\mu\nu} $ and the fixed
fields $\bar \Phi^A$ disappear from the equations (\ref{(a2.17)}). This means that a fixed background appearing  in the field-theoretic expressions cannot in fact be observed. Let us demonstrate it from the physical viewpoint for a Minkowski background space-time.

Consider the intermediate equations
(\ref{GL=Tphi}) of the field-theoretic formulation with  Cartesian coordinates for  the flat background. The latter cannot be observed with the use of the light signals, see the
works  \cite{Grishchuk92,[22]}). Indeed, the propagation of light in gravitational field can be interpreted as a propagation in a refractive  medium \cite{Nandi_1995}; the velocity of light changes in this medium. Besides  that, the energy of
relativistic particles depends on  the gravitational
field $\goh^{\mu\nu}$ as well, and, consequently, the  frequency of photons depends on gravity also. Then, the ratio of physically measured time to coordinate time  changes in the theory with the equation
(\ref{GL=Tphi}). Finally,  one concludes that distances are decreased and the Minkowski space-time is not observed when
the gravitational field $\goh^{\mu\nu}$ propagates.

What can  a study of propagation of  gravitational waves (not only relativistic particles) in
Minkowski space-time yield? Recall that the characteristic part of the left hand side of the equations (\ref{GL=T+ttot}) is the d'Alembert
operator, like in (\ref{Newtonian-nonfull}) for the flat background in the Lorentzian gauge. Only in the linear approximation do the
gravitational waves propagate along the null  geodesics of the
Minkowski space. But we consider equations with self-interaction, for example, see  (\ref{GL=T+ttot}), where the right hand side contains the second derivatives of $\goh^{\mu\nu}$ in the terms like $\goh^{\alf\beta}\goh^{\mu\nu}{}_{,\alf,\beta}$. As a result, the flat d'Alembert operator is modified. Then, of course, gravitational waves cannot feel a flat
background. The related idea of a geometry that is entirely masked by a universal distortion force goes back to the 19th and early 20th centuries in discussions of universal distortion forces and the question whether Euclidean geometry is privileged  \cite{Lotze,PoincareFoundations,ReichenbachSpace,KantParticle}.

To illustrate further the auxiliary character of the background space-time, it is instructive to consider how the gauge transformations influence trajectories of a test particle on this background. Again, it is enough to consider a flat background. Let us derive the related dynamical Lagrangian:
\be
\lag^{\rm dyn}= -\frac{1}{2\k}\lag^g + \lag^m\,.
\m{lag+}
\ee
 The related matter Lagrangian in the field-theoretic form has a general form: $\lag^m = {\lag}^M \l({\bm \eta} + \goh, \phi\r)$, where $\goh^{\mu\nu}$ is the gravitational field, ${\bm \eta}^{\mu\nu}$ is the background Minkowski metric density.
To define ${\lag}^M$ one has to recall the action for a free matter point in GR \cite{LL}:
\be
S^{m} = - m \int d\tau\,,
\m{b-b115_a}
\ee
 where $d\tau^2 = - ds^2 = g_{\mu\nu}dx^\mu dx^\nu$. To represent (\ref{b-b115_a}) in the field-theoretic form one has to express $g_{\mu\nu} = g_{\mu\nu}({\bm \eta}^{\alf\beta}, \goh^{\rho\sig})$ with the use of (\ref{(a2.4)}). The variation of $S^m$ with respect to the coordinates gives the equations of motion for a test particle. It is assumed that their solutions exist and are the vector components of the particle 4-velocity $u^\alf \equiv dx^\alf/d\tau$; we note that $d\tau$ depends on $g_{\mu\nu}dx^\mu dx^\nu$.

Let us present $S^m$ in a more suitable form:
\be
S^m= \int d^4x \lag^m= -\int d^4x\sqrt{-g}\rho g_{\mu\nu} u^\mu u^\nu;\qquad \rho \equiv \frac{m\delta (\vec{r}-\vec{r_0})}{\sqrt{- g^3g_{00}}}\frac{d\tau}{dt},
\m{b-b116}
\ee
where $\delta (\vec{r}-\vec{r_0})$ is the Dirac $\delta$-function, $g_{ab}$ is the spatial part of  the tensor $g_{\alf\beta}$ and $g^3 \equiv \det g_{ab}$. Thus, matter fields in (\ref{b-b116})  are $\phi^A =\{ \rho, u^\alf\}$.

Of course, the theory with the Lagrangian (\ref{lag+}) has to be gauge invariant with respect the gauge transformations (\ref{(5.9)}) and (\ref{(5.9)+}). In the case of the flat background the transformations (\ref{(5.9)}) and (\ref{(5.9)+}) for all the variables in (\ref{lag+}) are
 \bea
{\goh}'^{\mu\nu}(x) &= &\goh^{\mu\nu}(x) + \sum^{\infty}_{k = 1}{1\over{k!}}~ \hbox{$\pounds$}_\xi^k\l({\bm \eta}^{\mu\nu}(x) +
\goh^{\mu\nu}(x)\r),
\m{b-b117} \\
{\rho}'(x) &=& \rho(x) + \sum^{\infty}_{k = 1}{1\over{k!}}~ \hbox{$\pounds$}_\xi^k \rho(x), \m{b-b118}\\
{u}'^{\alf}(x) &=& {u}^{\alf}(x) + \sum^{\infty}_{k = 1}{1\over{k!}}~ \hbox{$\pounds$}_\xi^k {u}^{\alf}(x). \m{b-b119}
 \eea
Of course, both the set $\goh^{\mu\nu}(x) $, $\rho(x)$, ${u}^{\alf}(x)$ and the set $\goh'^{\mu\nu}(x) $, $\rho'(x)$, ${u}'^{\alf}(x)$ satisfy the  equations of the field-theoretic formulation of general relativity. However, in general, ${u}^{\alf}(x)$ and ${u}'^{\alf}(x)$ defines {\em different trajectories} in the same background space-time. This conclusion again stresses the fact that a background space-time has an auxiliary character.
However, in spite of backgrounds' lacking physical meaning (at least quantitatively)  in the field-theoretic formulation of General Relativity,  they  can be very useful for deriving important characteristics of various solutions, including interpretations of exact solutions in  GR and other metric theories.


\vspace{0.3cm}
\sect{Conservation laws in GR}
\m{ConservationLaws}
%


Here we follow the results and methods of the papers \cite{GPP,AbbottDeser82,Deser87,KBL,PK}. In section 2, we have connected the non-localization of the energy and other conserved quantities in GR with the gauge non-invariance of the energy-momentum tensor
of perturbations. Before applying the formalism to any concrete models, one has to fix the gauge freedom. In the usual geometrical formulation of GR, this procedure corresponds to fixing a coordinate system. By  gauge fixing one suppresses the ambiguities related to non-localization and can construct unambiguous conserved quantities related just to this gauge. Thus, in this section we construct conserved quantities and conservation
laws assuming that a gauge fixing was made.

From the start let discuss differential conservation laws on Ricci-flat (including flat) backgrounds, $\bar R_{\mu\nu} = 0$. Then, one has to take into
account $\bar{\Phi}^A \equiv 0$,  $\bar {\lag}^M \equiv 0$, ${\bm \Phi}^L_{\mu\nu}\equiv 0$ and use
\be
\frac{\del \bar{\cal R}}{\del \bar{\gog}^{\mu\nu}} = 0
\m{flat}
\ee
as the degenerated form  of the  background equations (\ref{(a2.6)}). Then, the dynamical Lagrangian  (\ref{(2.10)}) is simplified to
\be
 {\lag}^{\rm dyn}  =
 -{1 \over {2\k }}  \lag^g  + \lag^m =
 -{1 \over {2\k }}  \lag^g
+ {\lag}^M \l(\phi^A;~\bar{\gog}^{\mu\nu} + \goh^{\mu\nu}\r),
\m{(2.22)}
 \ee
 and the field equations (\ref{(a2.17)}) transform to the form of the equations (\ref{(2.17')}):
 \be {\cal G}^L_{\mu\nu} = \k\l({\bm
t}^g_{\mu\nu} +  {\bm t}^m_{\mu\nu}\r) \equiv \k{\bm t}^{\rm tot}_{\mu\nu}\, . \m {(a2.23)}
 \ee
Thus  for  Ricci-flat backgrounds the energy-momentum tensor densities ${\bm t}^{\rm tot}_{\mu\nu}$ and
${\bm t }^{\rm eff}_{\mu\nu}$ coincide. Furthermore, in the case of Ricci-flat backgrounds the left hand side of (\ref{(a2.23)}) is conserved identically,
\be
\bn^\nu{\cal G}^{L}_{\mu\nu} \equiv 0\,;
\m{nbG}
\ee
then taking the divergence of equation (\ref{(a2.23)}) leads to a differential conservation law:
 \be
 \bn^\nu{\bm t}^{\rm tot}_{\mu\nu} = 0\, .
 \m{(2.23')}
 \ee

 All the above permits us to construct differentially conserved current. Contracting ${\bm t}^{\rm tot}_{\mu\nu}$ with a Killing vector $\bar \xi^\alf$ defined in a background space-time, one obtains such a current:
 \be
 {\cal J}^\nu(\bar\xi) = {\bm t}^{{\rm tot}\nu}_\mu \bar\xi^\mu, \qquad \bn_\nu
{\cal J}^\nu \l(\bar\xi\r) \equiv \di_\nu {\cal J}^\nu \l(\bar\xi\r)= 0 \, .
\m{CCofFF}
\ee
Integration of this equality leads to a definitions of integral (not local) conserved quantities.  Consider a background 4-dimensional volume $V_4$, the
boundary of which consists of time-like `surrounding wall' $S$ and
two space-like sections: $\Sigma_0 := t_0 = \const$ and $\Sigma_1 :=
t_1 = \const$. Because the conservation law (\ref{CCofFF}) is
a scalar density under coordinate transformations, it can be integrated in a coordinate-independent way over  the
4-volume $V_4$:
\be
\int_{V_4} \di_\mu {\cal J}^\mu(\bar\xi) d^4 x
= 0\, .
\m{V_4}
\ee
 By the generalized Gauss theorem, it can be rewritten as
 \be \int_{\Sigma_1}d^3 x
{\cal J}^0(\bar\xi)  - \int_{\Sigma_0} d^3 x  {\cal J}^0(\bar\xi)
 +\oint_{S} ds_\mu {\cal J}^\mu(\bar\xi)  = 0\, ,
 \m{UseGauss}
 \ee
where $ds_\mu$ is the element of integration on $S$. If the integral  over `surrounding wall' in
(\ref{UseGauss}) becomes zero,
 \be
 \oint_{S}  \hat {\cal J}^\mu(\lam) dS_\mu = 0\,,
 \m{ointS}
 \ee
then the quantity
 \be
 {\cal P}(\bar\xi) = \int_{\Sigma}d^3x {\cal J}^0(\bar\xi)\,
 \m{(1.26)}
 \ee
is conserved  on space-like sections $\Sigma$ restricted by $\di\Sigma$, intersection
with $S$. It can be also assumed that $\di\Sigma \goto \infty$. In the case, when the condition (\ref{ointS}) does not hold,  the equation (\ref{UseGauss}) describes a
change of the quantity (\ref{(1.26)}), that is its flux through $\di\Sigma$. The differential conservation
laws (\ref{(2.23')}) and all the following constructions also apply for backgrounds that are Einstein spaces in A. Z. Petrov's
definition \cite{Petrov} with a vacuum $ \bar R_{\mu\nu} = \Lambda \bar g_{\mu\nu}, $ where $\Lambda$ is a constant (see
\cite{GPP,AbbottDeser82,Deser87}).

Below we will apply the formalism to study various solutions in GR using flat backgrounds only. Therefore, the above theoretical results are quite enough for such goals. However, for {\em arbitrary curved} backgrounds there are no conservation laws of the form  (\ref{(2.23')}). That  is because, in the general case for the linear operators in (\ref{(a2.17)}) and (\ref{B30}) one has
\be
\bn_\nu\l({\cal G}^{L\nu}_{\mu} + {\bm \Phi}_{\mu}^{L\nu}\r) \neq 0, \qquad \bn^\nu{\cal G}^{L}_{\mu\nu} \neq 0\,
\m{nbG_neq}
\ee
instead of (\ref{nbG}). The reason is that the system (\ref{(2.10)}) interacts
with a complicated background geometry determined by the background
matter fields $\bar{\Phi}^A$. Cosmological solutions, for example,  are not flat or Einstein's spaces.

Nevertheless, in spite of the inequalities  (\ref{nbG_neq}), one expects  conservation laws for arbitrary curved backgrounds and arbitrary displacement vectors $\xi^\alf$.
(This fact follows from Noether's first theorem and the fact that the \emph{laws} (not the geometry, which \emph{as such} is irrelevant to Noether's theorems) have continuous symmetries \cite{Noether,BergmannConservation,KosmannSchwarzbachNoether}.)
  We find such laws making use of the  canonical N{\oe}ther procedure developed in \cite{KBL} and applied to the Lagrangian (\ref{(2.10)}).  This technique is developed in detail in the framework of an arbitrary metric theory of the Lovelock class in section 10. At the end of this section we derive formulae of section 10 simplified to GR and necessary here.


Thus, let us derive the identity (\ref{BDi-sup}) adopted to GR:
\be
{\bm i}^{\mu} \equiv \bn_\nu  {\bm i}^{\mu\nu} \equiv \di_\nu  {\bm i}^{\mu\nu}\,.
\m{identity_GR}
\ee
Here, the current and superpotential are, respectively,
 \bea
{\bm i}^{\mu}(\xi) &\equiv &{1 \over {\k}} {\cal G}^{L\mu}_{\nu}\xi^\nu + {1 \over
 \k} \goh^{\mu\lam} \bar
 R_{\lam\nu}\xi^\nu + {\bm \zeta}^\mu(\xi) \,
 \m{B27'}\\
 {\bm i}^{\mu\nu}(\xi) &\equiv & {1 \over \k} \goh^{\rho[\mu}\bn_\rho\xi^{\nu]} + {\cal P}^{\mu\nu}{_\lambda} \xi^\lambda \equiv {1 \over \k} \l(\goh^{\rho[\mu}\bn_\rho\xi^{\nu]}+ \xi^{[\mu}\bn_\sig \goh^{\nu]\sig}- \xi^\sig \bn^{[\mu}\goh^{\nu ]}_\sig  \r)\,.
\m{alaAbbottDeser}
 \eea
 The last term in (\ref{B27'}) is
 \bea
 2\k {\bm \zeta}^\mu(\xi) &\equiv &
2\l(z^{\rho\sig}\bn_\rho \goh^\mu_\sig - \goh^{\rho\sig}\bn_\rho z^\mu_\sig\r)- \l(z_{\rho\sig}\bn^\mu \goh^{\rho\sig} - \goh^{\rho\sig}\bn^\mu
z_{\rho\sig}\r) \nonumber\\& +& \l(\goh^{\mu\nu}\bn_\nu z
- z\bn_\nu \goh^{\mu\nu}\r) \m{Zmu-new}
 \eea
where $2z_{\rho\sig} \equiv -\pounds_\xi \Bar g_{\rho\sig}$,
and, thus, disappears if $\xi^\alf = \bar\xi^\alf$, that is, if the displacement vector is a Killing vector of the background. Here  $z = z^\alf{}_\alf$ with the index moved using the background metric.

The main property demanded of  superpotentials, $\di_{\mu\nu} {\bm i}^{\mu\nu}(\xi) \equiv 0$, holds. The expression (\ref{alaAbbottDeser}) generalizes
the Papapetrou superpotential \cite{Papapetrou}, which depends on a background matrix $diag(-1,1,1,1)$ or the like. Indeed for the case of a Minkowski space-time background and rigid coordinate translations $\xi^\lam = \delta^\lam_{(\rho)}$ (in
 Cartesian coordinates), one gets
 \be
 {\bm i}^{\mu\nu}_{(\rho)} =
{\cal P}^{\mu\nu}{}_{\rho} = {1 \over 2\k}
\di_\sig\l(\delta^{\mu}_\rho \goh^{\nu\sig}-\delta_{\rho}^{\nu}
\goh^{\mu\sig}-  \bar g^{\sig\mu} \goh^{\nu}_{\rho}+\bar
g^{\sig\nu} \goh^{\mu}_{\rho}\r)\, . \m{(2.27)}
 \ee
The same superpotential (\ref{alaAbbottDeser}) was constructed in
\cite{PK} by another means, namely, by the Belinfante symmetrization
of the canonical system in \cite{KBL}.

To provide physically sensible conservation laws from the
identity (\ref{B27'}), one needs to use the field equations. We
substitute ${\cal G}^{L}_{\mu\nu}$ in the form (\ref{B30}) into the current
(\ref{B27'}) and obtain
 \be
 {\cal J}^{\mu}(\xi) \equiv {\bm \Theta}_{\nu}{}^{\mu} \xi^\nu + {\bm \zeta}^\mu(\xi) \,.
 \m{B32}
 \ee
The generalized total energy-momentum tensor density is
 \be
 {\bm \Theta}_{\nu}{}^{\mu}
  \equiv {\bm t}_{\nu}^{g}{}^{\mu} +
\delta {\bm  t}_{\nu}^{M\mu}+
 {1 \over \k} \goh^{\mu\lam} \bar R_{\lam\nu} \equiv
 {\bm t}^{{\rm eff}}_{\nu}{}^\mu +  {1 \over \k} \goh^{\mu\lam} \bar R_{\lam\nu}\,
 \m{B34}
 \ee
where the interaction with the background geometry, $\goh^{\mu\lam} \bar R_{\lam\nu}$, is taken into account. Because on the right hand side of (\ref{B32}) there is a divergence of the superpotential (\ref{alaAbbottDeser}), the current (\ref{B32}) is conserved: $\bn_\mu {\cal J}^{\mu}=\di_\mu
{\cal J}^{\mu}=0$. Thus,  ${\bm \Theta}_{\nu}{}^{\mu}$ plays the same role as ${\bm  t}^{\rm tot}_\nu{}^\mu$ in the equation (\ref{CCofFF}) on the flat background if Killing vectors exist. Thus, the current (\ref{B32}) generalizes (\ref{CCofFF}) to arbitrary backgrounds and arbitrary displacement vectors. It
can be important, for example, for models with cosmological backgrounds where not only the Killing vectors are used (see, e.g., \cite{Traschen,Traschen_1986,Uzan}).

 For a concrete solution the superpotential ${\bm i}^{\mu\nu}(\xi)$ in (\ref{alaAbbottDeser}) is rewritten in a new notation, ${\cal  J}^{\mu\nu}(\xi)$, although it has the same form. Finally, the identity (\ref{identity_GR}) acquires the form of a physically meaningful conservation law:
\be
{\cal J}^{\mu}(\xi) = {\bn_\nu} {\cal  J}^{\mu\nu}(\xi) = {\di_\nu} {\cal J}^{\mu\nu}(\xi)\,.
\m{identity_GR+}
\ee
Because the current (\ref{B32}) is conserved, the integral conserved quantity, like (\ref{(1.26)}), can be constructed.  Due to antisymmetry of the superpotential in  (\ref{identity_GR+}), this conserved quantity is expressed over a surface
integral in the form of the charge:
 \be  {\cal P}(\xi) = \int_{\Sigma}d^3x {\cal J}^0(\xi) = \oint_{\di\Sigma} d\sig_k {\cal J}^{0k}(\xi)\,,
\m{int-surface+}
 \ee
where $d\sig_k$ is the element of integration on $\di\Sigma$. It is a significant expression because it connects a quantity ${\cal P}(\xi)$
obtained by integration of {\em local} densities with a surface
integral playing a role of a {\em quasi-local} quantity (see
discussion in the Introduction).


\vspace{0.3cm}
\sect{The total mass of the Schwarzschild black hole in GR}

Already in the Introduction, we have noted that it is important to describe exact solutions in GR in terms of the field-theoretic formalism. This means that the solution is represented by the field configuration propagating on a fixed background. In the present and next sections, we concentrate on the first exact solution of GR, which is the Schwarzschild solution. It is simplest yet most relevant solution in GR, and its properties interpreted in the framework of the geometrical description are well known. In the present section (based on the results of the papers \cite{GPP,PetrovNarlikar1,FPL2005}, see also chapter 4 in the book \cite{Petrov+_2017}), we calculate the total mass of the Schwarzschild black hole presented by various field configurations connected by gauge transformations.

For the Schwarzschild solution, which  is asymptotically flat, it is quite natural to admit the flat space-time at spatial infinity as a background space-time. Therefore we choose a flat metric coinciding with the asymptotic metric as the  background metric. In spherical coordinates, the metric is
  \be
  d\bar s^2 = -c^2dt^2 + dr^2 + r^2\l(d\theta^2 +\sin^2\theta d\phi^2 \r) \,,
 \m{c-d1}
 \ee
where, as usual, the coordinates are numerated as  $x^0 = ct$, $x^1 =r$, $x^2 =\theta$ and $x^3 =\phi$. (The freedom to use Cartesian coordinates distinguishes the field-theoretic formulation even with a flat background from the use of a numerical matrix background $diag(-1,1,1,1)$.) We denote the background metric of the Minkowski space in curved coordinates  as $\bar g_{\mu\nu}=\gamma_{\mu\nu}$. Non-zero components of the  Christoffel symbols corresponding the metric (\ref{c-d1}) are
 \bea
 C^1{}_{22} &=& -r\,, ~~ C^1{}_{33} = -r\sin^2\theta\,,~~ C^2{}_{12} =  C^3{}_{13} = \frac{1}{r}\,,
  \nonumber\\
  C^2{}_{33} &=& -\sin\theta\cos\theta\,,~~ C^3{}_{23} = \cot\theta\,.
\m{c-d11}
 \eea

From the start let us consider the Schwarzschild solution in the typical Droste  coordinates:
  \be
  ds^2 = -\l(1-\frac{r_g}{r}\r)c^2dt^2 + \frac{1}{1-({r_g}/{r})}dr^2 + r^2\l(d\theta^2 +\sin^2\theta d\phi^2 \r)\,,
    \m{c-d2}
 \ee
where ${r_g} \equiv {2mG}/{c^2}$. Below we consider other presentations of the Schwarzschild solution important for our considerations. First, we  change the radial coordinate by what one might call a radial translation:
\be
r \goto r\l(1+ \frac{r_g}{4r} \r)\,.
\m{radial}
\ee
Then, the metric element (\ref{c-d2}) is represented in the so-called  isotropic coordinates \cite{LL}:
 \be
ds^2 =  -\frac{\left(1 - {{r_g}/ {4r}}\right)^2}{ \left(1 +
{{r_g}/ {4r}}\right)^2} c^2 dt^2 + \left(1 + {\frac{r_g}{4r}}\right)^4 \left[dr^2 + r^2(d \theta^2 + \sin^2 \theta d \phi^2)
\right]\, .
\m{c-d3}
 \ee
Of course, the coordinate `$r$' here is not the same as the coordinate `$r$' in (\ref{c-d2}). This is because the same background metric in the form (\ref{c-d1}) is used to derive the field configuration in both the above cases.
While the detailed quantitative features of the background metric are not observable, the background metric still helps to delimit places that actually exist.  While the world does not have edges that one could fall off, values such as $r=0$ and infinite values of the coordinates for a well-chosen background metric do have the significance of delimiting the furthest reaches of the world.  As will appear below, one aims to stuff as much of a curved metric's observable events onto the background as possible (giving a bimetric notion of maximal extension), without leaving any bare spots.  The non-internal character of gravitational gauge transformations makes such questions disanalogous to Maxwell or Yang-Mills and analogous to questions of extending space-times in geometrical GR.

Next we  only change the time coordinate
 \be
 ct \goto ct - r_g\ln\l|1 - \frac{r_g}{r}\r|\,,
 \m{c-d49}
 \ee
whereas the other coordinates $\{r,\,\theta,\, \phi \}$ are not changed. As a  result one has
 \be
 ds^2 = -\l(1 - \frac{r_g}{r}\r)c^2dt^2 +
 2\,\frac{r_g^2}{r^2}\, c\, dt\, dr
 + \l(1 + \frac{r_g}{r}\r) \l(1 + \frac{r_g^2}{r^2}\r)dr^2 +
 r^2\l(d\theta^2 + \sin^2\theta d\phi^2\r).
 \m{c-d50}
 \ee
 Of course, the coordinate `$t$' here is not the same as the coordinate `$t$' in (\ref{c-d2}). 
The important properties of this solution are that a
falling test particle approaches the horizon $r= r_g$ in finite coordinate time $t$.  Below the horizon, it  is always falling towards the
singularity, it gets arbitrarily close to it, but only hits it, at, $t= \infty$.
Finally, let us provide the time transformation for the Schwarzschild time in the form:
\be
ct \goto ct - r_g\ln\l|\frac{r}{r_g}-1\r|\,.
 \m{c-d57}
 \ee
As a result, the metric element (\ref{c-d2}) is represented as
 \be
 \hspace*{-0.5cm}ds^2 = - \l(1 - \frac{r_g}{r}\r)c^2d t^2 + 2\,\frac{r_g}{r}\, c\, d t\, dr
+ \l(1 + \frac{r_g}{r}\r)dr^2 -
 r^2\l(d\theta^2 + \sin^2\theta d\phi^2\r).
 \m{c-d55}
 \ee
Again, the coordinate `$t$' here is not the same as the coordinate `$t$' in (\ref{c-d2}). One has to remark that the metric (\ref{c-d55}) is the well known Eddington-Finkelstein (EF) metric for the Schwarzschild geometry \cite{MTW}. Only one has to make a transformation from the null coordinate $\~V$ to the time coordinate $t$: $ct = c\~V - r$.

Now, let us derive field configurations corresponding the above geometrical representations of the Schwarzschild solution. We use the decomposition (\ref{(a2.4)}) adopted to these solutions. Thus, for the background (\ref{c-d2}) one has
 \be \gog^{\mu\nu} \equiv \bar {\gog}^{\mu\nu} + \goh^{\mu\nu}= {\bm \gamma}^{\mu\nu} +\goh^{\mu\nu}  = \sqrt{-\gamma}\l(\gamma^{\mu\nu} +h^{\mu\nu} \r)\,
 \m{c-d4}
  \ee
where $\sqrt{-\gamma} = r^2\sin\theta$. The equation above defines the non-densitized (non-Gothic letter) gravitational perturbation $h^{\mu\nu},$ which is freed of the strong coordinate effects manifest in $ r^2\sin\theta$ in spherical coordinates by the use of $\sqrt{-\gamma}$ to de-densitize.

 Then, the field configurations related to (\ref{c-d2}), (\ref{c-d3}), (\ref{c-d50}) and (\ref{c-d55}), respectively, are as follows,
 \bea
h^{00} &=&  -{{r_g} \over r}~{1 \over {1 - ({r_g} / r)}}\, ,\qquad ~~~~~~h^{11} =  -{{r_g} \over r} \,;
\m{c-d5}\\
h'^{00} &=& 1-\frac {\left(1 + {{r_g}/ {4r}}\right)^7} {\l({1 -
{{r_g}/{4r}}}\r)}\, ,\qquad ~~~~~h'^{11} = h'^{22} = h'^{33} = -
\left({{r_g}\over {4r}}\right)^2\, ;
\m{c-d6}\\
h''^{00}  &=& -\l(\frac{r_g}{r}+ \frac{r_g^2}{r^2} + \frac{r_g^3}{r^3}\r)\,
 ,~\qquad h''^{01}  =   \frac{r_g^2}{r^2}\, ,\qquad
 h''^{11}  = - \frac{r_g}{r}\, ;
 \m{c-d51}\\
h'''^{00}  &=&  -\frac{r_g}{r}\,
 ,\qquad h'''^{01}  =   \frac{r_g}{r}\, ,\qquad
 h'''^{11}  =  -\frac{r_g}{r}\, .
  \m{c-d56}
 \eea
 Let us outline a connection of variables in (\ref{c-d5})-(\ref{c-d56}) to gauge transformations. The transformations (\ref{radial}),  (\ref{c-d49}) and (\ref{c-d57}) lead to the metric elements (\ref{c-d3}), (\ref{c-d50}) and (\ref{c-d55}), respectively. The last three are united in the general formula (\ref{h+h+}). Thus, $h^{\mu\nu}$, $h'^{\mu\nu}$, $h''^{\mu\nu}$ and $h'''^{\mu\nu}$ are connected by gauge transformations (\ref{(5.9)}), only in (\ref{c-d5})-(\ref{c-d56}) displacement vectors $\xi^\alf$ are not derived explicitly.

 One can see that the field configurations (\ref{c-d5})-(\ref{c-d56}) have breaks and singularities at $r=r_g$ and/or at $r=0$. The breaks at $r=r_g$ reflect the coordinate problems at the horizon in the framework of the geometrical description. Because it is not a physical singularity, it can be suppressed by a (naive) gauge transformation. Indeed, after related gauge transformations a break at $r=r_g$ in (\ref{c-d5}) and (\ref{c-d6}) is cancelled in (\ref{c-d51}) and (\ref{c-d56}). The singularity at $r=0$ corresponds to a true singularity of a black hole, so it cannot be suppressed, whether  by coordinate transformations or by  gauge transformations.

 In the present section all the field configurations (\ref{c-d5})-(\ref{c-d56}) (combined with the background metric) are asymptotically flat. What does this mean? To show this explicitly,  it is best to use Cartesian coordinates, $x^1 = x$, $x^2 = y$, $x^3 = z$, instead of spherical ones. After that the background metric element (\ref{c-d1}) goes to Minkowski form and the Christoffel symbols (\ref{c-d11}) disappear. As a result, at spatial infinity, $r\goto \infty$, all the components of the configurations (\ref{c-d5})-(\ref{c-d56}) acquire the fall-off not weaker than $h^{\mu\nu} \sim 1/r$. Then, by the conclusions presented in \cite{PetrovH}, one concludes  that configurations  (\ref{c-d5})-(\ref{c-d56}) have to give the same total mass for the related isolated system. Below, we illustrate it explicitly.

 To calculate the total mass/energy of the Schwarzschild black hole, we use the surface integral (\ref{int-surface+}) at $r\goto \infty$. The superpotential (\ref{alaAbbottDeser}) in this integral is universal and is valid for arbitrary curved backgrounds, including a flat one. In our case we have to consider covariant derivatives in (\ref{alaAbbottDeser}) defined by the Christoffel  symbols (\ref{c-d11}). To calculate the energy we choose the Killing vector of the background in the form: $\bar\xi^\alf = \{-1,0,0,0 \}$. Then for all the kinds of configurations (\ref{c-d5})-(\ref{c-d56}) the total mass is
 \be
 {\cal P}(\bar\xi) = \oint_\infty d\theta d\phi {\cal J}^{01}(\bar \xi) = mc^2\,.
\m{total_E}
 \ee

 Let us explain why one has the unique result (\ref{total_E}) for the different aforementioned configurations. Because we use the simplest (flat) background, one can use 
the current  defined in (\ref{CCofFF}). Because we consider a concrete solution to GR, the field equations (\ref{(a2.23)}) hold. Then, the total energy-momentum in (\ref{CCofFF}) can be expressed through the left hand side of the field equations:
 \be
{\bm t}^{\rm tot}_{\mu\nu} = \frac{1}{\kappa}{\cal G}^L_{\mu\nu} = \frac{1}{2\kappa}\l(\goh_{\mu\nu}{}^{;\alf}{}_{;\alf} + \gamma_{\mu\nu}
\goh^{\alf\beta}{}_{;\alf\beta} -\goh^{\alf}{}_{\mu;\nu\alf}-\goh^{\alf}{}_{\nu;\mu\alf}\r)\,,
\m{c-d39}
 \ee
 where `$_{;\alf}$' means the covariant derivative with respect to $\gamma_{\mu\nu}$. Substituting this expression into the current (\ref{CCofFF}) one can transform it into a divergence of the superpotential (\ref{alaAbbottDeser}) with the displacement vector $\xi^\alf =\bar\xi^\alf $ only. Finally, one obtains again the charge (the total energy) in the form and with the result in (\ref{total_E}).

 Now, recall that the field configurations are connected by gauge transformations. Then, for the flat background the transformation for the total energy-momentum (\ref{tei-gauge}) acquires the form:
 \be
  \k{{\bm t}}'^{\rm tot}_{\mu\nu} =  \k{\bm t}^{\rm tot}_{\mu\nu}
  + {\cal G}^L_{\mu\nu}(\goh'-\goh)\, .
  \m{tei-gauge_flat}
  \ee
 The difference ${\cal G}^L_{\mu\nu}(\goh'-\goh)$ is a double divergence, and is, of course, incorporated in the integrand in (\ref{total_E}). One can easily check that all the differences, like $\goh'^{\mu\nu}-\goh^{\mu\nu}$, $\goh''^{\mu\nu}-\goh^{\mu\nu}$, etc., do not contribute to the charge (\ref{total_E}). Thus the total energy (\ref{total_E}) is invariant with respect to gauge transformations connecting the configurations (\ref{c-d5})-(\ref{c-d56}).
One can, of course, interpret this result in terms of either a gauge-dependent localization of an objectively non-localizable quantity or as the sameness of the total amounts of distinct localized energies.

\vspace{0.3cm}


\sect{The Schwarzschild black hole in GR as a point mass}

The Schwarzschild solution, being a surprisingly a non-trivial solution, has  well-known problems with regard to its  interpretation in the geometric language. These problems have been discussed in many papers and textbooks and in most cases are resolved. Many of the problems have analogs in the field-theoretic formalism.  One of these problems is a description of the point mass in GR. Here we take inspiration from a paper by Narlikar  \cite{Narlikar}. In Newtonian gravity the problem is resolved simply:  one can describe the point mass with the Newtonian potential $\sim m/r$ everywhere, including the point $r =0$. To satisfy this, one has to assume that the mass distribution has the form $\rho(r) =m\delta(r)$, where the Dirac  $\delta$-function satisfies the ordinary Poisson equation
 \be
 \nabla^2 \left({1\over r}\right) =
 \l( \frac{d^2}{dr^2} + \frac{2}{r}\frac{d}{dr}\r){1\over r} =
- 4\pi\delta({\bf r})\, .
 \m{c-d33}
 \ee
Then, both for a regular distribution $\rho(r)$ and for a point mass
 \be
 \rho(r) = m\delta({\bf r}),
\m{c-d34+}
\ee
the total mass of the gravitating system is calculated with the use of the {\it same} integral:
 \be
 m = \int_\Sigma dx^3 {\bm\rho}({ r})\,
\m{c-d34}
\ee
with $\bm\rho({ r}) = \sqrt{-\gamma}\rho({ r})$.
Thus, the point particle located at $r=0$ is included in a unique standard way in  Newtonian gravity by making use of the $\delta$-function.

At first glance, it might seem that one can simply  use the Schwarzschild solution in order to describe a point mass in GR. However, a conceptual difficulty arises. If  one tries to consider an ideal point mass in the framework of the geometrical description, the point mass is shrouded by  its own horizon. Under the horizon the coordinate $r$ becomes a time-like coordinate and $r=0$ describes a space-like hypersurface, not a point. Therefore it is impossible to model it by $\delta$-function at $r=0$ as one might have hoped naively. However, we will  show how this problem is resolved using the field-theoretic formalism. Already we have shown that the black hole geometry can be interpreted as a reasonable field configuration even at the horizon and behind the horizon down to the true physical singularity $r=0$; see (\ref{c-d5})-(\ref{c-d56}). Then, one anticipates that the problem of a point mass can be solved by using the volume integration (\ref{(1.26)}) with the time-like Killing vector $\bar\xi^\alf = \{-1,0,0,0 \}$ defined for the coordinates in (\ref{c-d1}):
 \be
 {\cal P}(\bar\xi) = \int_{\Sigma}d^3x {\cal J}^0(\bar\xi) = \int_{\Sigma}d^3 x {\bm t}^{0\alf}_{{\rm tot}}\bar\xi_\alf = \int_{\Sigma}d^3 x \sqrt{-\gamma}{t}^{00}_{{\rm tot}}\,
   \m{c-d36}
 \ee
  over the whole Minkowski space including $r=0$ with the energy density (energy distribution) ${t}^{00}_{\rm tot}$. The problem of the point mass can be resolved if the $\delta$-function representing the singularity is included into ${t}^{00}_{\rm tot}$ in a consistent way. Thus, (\ref{c-d36}) has to generalize the Newtonian formula (\ref{c-d34}).  We develop this proposal below.

Recall that ${t}^{00}_{\rm tot}$ is changed by gauge transformations. So, we will check the configurations (\ref{c-d5})-(\ref{c-d56}) just connected by gauge transformations and define an appropriate gauge fixing. We employ the following criteria:

\bigskip
\bit

\item[(i)] Breaks in the field configurations and the energy density at each  points of the Minkowski background space-time (except at $r=0$) are inadmissible.

\item[(ii)] A point particle at rest in the whole Minkowski space-time must be represented. Therefore it must be natural  to
describe the true singularity by the worldline $r=0$.

\item[(iii)] The mass-energy should be concentrated at point $r=0$ only without distribution of energy outside, as in the Newtonian case (\ref{c-d34+}).\footnote{This condition is imposed here not because its violation is physically absurd, but because one wants to see how fully the ideal of a point mass can be realized.}

\item[(iv)] The Schwarzschild solution in appropriate
coordinates should be asymptotically flat.

\item[(v)] A falling test particle should penetrate the horizon without obstacles and reach  the point $r=0$  at a finite time relative to the Minkowski background.\footnote{One might instead consider the alternative requirement that the test particle only approach the true singularity as Minkowski background time goes to infinity. A reason for doing so will be mentioned later.  }

\item[(vi)] The requirement of `$\eta$-causality', to be explained presently, is satisfied.

\eit

The property of  `$\eta$-causality' is that  the physical light cone of the effective metric $g_{\mu\nu}$ is tangent to or inside the flat `light' (null)
cone of the flat background metric $\eta_{\mu\nu}$ at all points of the Minkowski background space-time \cite{PittsSchive2001a,Pitts_Schive_2004,Pitts_Schive_2003,FPL2005}.\footnote{ One might reasonably generalize this requirement to use for a background a maximally symmetric space-time of constant curvature (A)dS or perhaps a pure conformal geometry with vanishing Weyl curvature tensor. In all such cases a flat null cone is retained and the geometry in question is maximally symmetric. } This requirement avoids interpretive difficulties in  the field-theoretic
presentation of GR. Given this requirement, all the causally connected
events in the physical (dynamical) space-time\index{space-time!dynamic} $g_{\mu\nu}$ are acceptably related to the  causal
structure of the Minkowski space. Hence any $g$-time-like vector is $\eta$-time-like and any $g$-null vector is $\eta$-time-like or $\eta$-null.  Thus the effect of gravity is to narrow and perhaps gently tip the physical light cone relative to the background null cone, but not to make anything physical ($g$-time-like or $g$-null) `go faster than light' as defined by the background (along which, admittedly, electromagnetic radiation does not travel).  The proper relationship of the null cones is not automatically gauge invariant in the sense of the exponentiated Lie derivative formulae \cite{[22]}. Properties of  $\eta$-causality and
gauge transformations conserving it were studied in some detail
 \cite{Pitts_Schive_2004}. In effect one aims to  restrict the naive mathematical notion of gauge transformations because gauge transformations should relate physically equivalent solutions and thus should preserve the proper relation between the two null cones. Employing a new set of variables, a generalized eigenvector formalism, can be useful so that the inequalities hold automatically.  A major motivation for the $\eta$-causality criterion is to provide a justification for quantization with equal-time or space-like commutation relations \cite{NullCones,PittsSchive2001a,Pitts_Schive_2004}, which otherwise are often used in both covariant and canonical quantization programs with no evident justification.

The requirement of the $\eta$-causality can be strengthened   by the requirement of stable $\eta$-causality \cite{PittsSchive2001a,Pitts_Schive_2004,Pitts_Schive_2003}. The latter condition means that the physical light cone of $g_{\mu\nu}$ has to be {\it strictly} inside the flat light
cone of $\eta_{\mu\nu}.$ This relation could be important when quantization problems are under
consideration. Indeed, in the case of tangency, a field is on the
verge of $\eta$-causality violation  and would be pushed into violation under some infinitesimal gauge transformations \cite{Pitts_Schive_2004}.  Assuming stable $\eta$-causality, any infinitesimal gauge transformation will change an $\eta$-causal configuration into an $\eta$-causal configuration; only for finite transformations restrictions on the descriptor vector field arise.

We note that the representation of the Schwarzschild solution by the field configuration (\ref{c-d5}) does not
satisfy the requirements (i), (iii), (v) and (vi) in the above list.  The field configuration (\ref{c-d6}) has to be excluded as well because it cannot satisfy all the requirements; indeed,  the coordinates in (\ref{c-d3}) do not cover the area under the horizon, so the configuration (\ref{c-d6}) cannot describe the true singularity at all. The configuration (\ref{c-d51}) does not satisfy the requirement (iii) and a test particle cannot reach the true singularity in finite Minkowski time.

Unlike (\ref{c-d5})-(\ref{c-d51}), the field configuration (\ref{c-d56}) satisfies all the requirements in the above list.  Finally let us present the components of the total energy-momentum. After making use of the expression (\ref{c-d39})  for the
configuration (\ref{c-d56}) we obtain
\bea
 t^{\rm tot}_{00} & = & mc^2 \delta({\bf r})\, ,\nonumber\\
t^{\rm tot}_{11} & = & - mc^2\delta({\bf r})\, ,\nonumber\\
t^{\rm tot}_{AB} & = & - \half \gamma_{AB}\, mc^2\delta({\bf r}); \qquad A,\ldots = 2,3\,.
 \m{c-d58}
 \eea
Indeed, all these  energy-momentum components are concentrated {\it
only} at $r=0$. The volume integration (\ref{c-d36}) of $t^{tot}_{00}$
from (\ref{c-d58}) again gives $E = mc^2$. Recall, the surface integration with the configuration (\ref{c-d56}) gives $E = mc^2$ also. This result follows with an arbitrary radius, $r_0$, of 2-sphere in a surface integration; it is not necessary to set  $r_0 \goto \infty$. It is an exact analog for calculating
the electric charge in electrodynamics, or calculating the point mass in Newtonian gravity, as in (\ref{c-d34}) for the point mass.
The other components $t^{tot}_{11}$ and $t^{tot}_{AB}$ in
(\ref{c-d58}) are proportional to $\delta({\bf r})$ as well, thus, could describe the ``inner
radial'' and ``inner tangent'' stresses. Formally these quantities could be related  to the intrinsic properties of the point. Thus, finally one concludes that the field configuration (\ref{c-d56}) indeed represents a point-like object in Minkowski space, though it is more complicated than in  Newtonian gravity.

Finally let us remark that we are considering only the total energy-momentum. The matter source, $t^m_{\mu\nu}$, that contributes into (\ref{c-d58}) is `localized' at $r=0$ only.  One can find how it can be separated formally from the free gravitational field, see \cite{PetrovNarlikar1}. However, it is in the spirit of GR that $t^m_{\mu\nu}$ {\em cannot} be considered separately from $t^g_{\mu\nu}$.


\vspace{0.3cm}
\sect{Particle trajectories in the Schwarzschild space-time and the harmonic gauge fixing}

To find solutions to the Einstein equations, one usually makes an appropriate choice of coordinates. Harmonic coordinates are among the most popular ones. Concerning the Schwarzschild solution, Fock \cite{Fock_1959} has suggested  such harmonic coordinates. However, the latter, like the Schwarzschild coordinates, are singular at the horizon. Many coordinate systems without this defect are known but are not harmonic. Here, following \cite{PetrovH}, we  discuss coordinates that are both harmonic and regular at the horizon.

Thus, continuing to illustrate the field-theoretic method, we apply it to interpret the transition from the Fock coordinates to the new harmonic coordinates in terms of gauge transformations. In both of the gauge fixings, we consider trajectories of test particles falling into a Schwarzschild black hole. We find that trajectories in the Minkowski space are gauge dependent, as  was remarked in the Introduction and as one expects from the combined coordinate and gauge transformation that alters only the background entities. Because gauge transformations do not change the physical meaning, we see once more that the background Minkowski metric is an auxiliary structure. Thus, a breakdown in the trajectories at the horizon for the field configuration in the Fock picture is interpreted as physically unreal. Indeed, such a breakdown is averted for the field configuration corresponding to the new harmonic coordinates. These problems, of course, are resolved clearly in the framework of the usual geometrical formalism of GR. Here we illustrate the utility of the field-theoretic formalism.

We start with the Schwarzschild metric in the Fock harmonic coordinates:
 \be d s^2
=- \frac{r-\alpha}{r+\alpha}c^2 dt^2 + \frac{r+\alpha}{r-\alpha} dr^2
+ (r + \alpha)^2 (d\theta^2 + \sin^2 \theta\, d \phi^2)\,,
\m{c-d59}
 \ee
where $\alpha= r_g/2$. Going to asymptotically Cartesian coordinates in standard way, one finds that the solution (\ref{c-d59}) satisfies the harmonic (de Donder) conditions
 \be \di_\nu\l(\sqrt{-g}g^{\mu\nu}\r) = 0\,.
 \m{c-d60}
  \ee

 For the sake of simplicity, we consider a test particle falling radially into a black hole\index{black hole}. We restrict ourselves to the `parabolic orbit' case, when a particle begins its motion from the rest at  infinity $r = \infty$. Then, the equation of motion of the test particle becomes
\be ct  =  -
2\alf \l[{{ \frac{2}{3}}} \l({\frac{r +\alf}{2\alf}}\r)^{3/2}+
2\l(\frac{r +\alf}{2\alf}\r)^{1/2} +\ln \l|{\frac{r}{\alf}- 1}\r|
- 2\ln \l|{\l({\frac{r +\alf}{2\alf}}\r)^{1/2} +1}\r|\,\r]
+ {\rm const}\,  .
\m{c-d61}
\ee
The existence of the term $-2\alf \ln |r/\alf - 1 |$ leads to the situation that a particle falling to the event horizon $r = \alf$  takes infinitely long in the coordinate time $t$ that is the time of a distant observer.

The problem of breakdown at the horizon using the de Donder harmonic conditions has been resolved  \cite{PetrovH}. Finally its results lead to the transformations:
 \be c\tau =c t + 2\alf \ln
\l|\frac{r-\alf}{r+\alf}\r|\,, \qquad r=r\, ,\qquad \theta =
\theta\, ,\qquad \phi = \phi\, .
\m{c-d64}
 \ee
Applying them, one obtains the Schwarzschild solution in the form:
 \bea d s^2& =&
- \frac{r-\alpha}{r+\alpha}c^2 d\tau^2 +
2\l(\frac{2\alf}{r+\alf}\r)^2cd\tau dr +
\l[1 +
\frac{2\alf}{r+\alf} +\l(\frac{2\alf}{r+\alf} \r)^2 +
\l(\frac{2\alf}{r+\alf} \r)^3\r]dr^2 \nonumber \\&{+}& (r +
\alpha)^2 (d\theta^2 + \sin^2 \theta\, d \phi^2)\, .
\m{c-d65}
 \eea
 Note that with the use of the shift $r\goto r-\alf$ in the transformation (\ref{c-d64}) and the metric (\ref{c-d65}), they go to (\ref{c-d49}) and (\ref{c-d50}), respectively. One can check that after transferring to asymptotically Cartesian coordinates, the metric (\ref{c-d65}) satisfies (\ref{c-d60}) also. Finally, the metric coefficients in (\ref{c-d65}) are finite everywhere except of the true singularity $r = -\alf$.

The trajectory of the test particle that is on the `parabolic orbit' is given  by the equation:
 \be c\tau = -  2\alf \l[{{ \frac{2}{3}}} \l({\frac{r
+\alf}{2\alf}}\r)^{3/2}+ 2\l(\frac{r +\alf}{2\alf}\r)^{1/2} +\ln
\l|{\frac{r}{\alf}+ 1}\r|
- 2\ln \l|{\l({\frac{r +\alf}{2\alf}}\r)^{1/2} +1}\r|\,\r]
+ {\rm const}\,  .
\m{c-d66}
 \ee
Here, unlike (\ref{c-d61}), there is no divergent logarithmic term.  Hence, in the coordinate system $(\tau, r)$, a falling particle trajectory without breakdowns goes through the horizon.

The metric (\ref{c-d65}) and the structure of the light cones
\be
\l.c\frac{d\tau}{dr} \r|_1 = \frac{(r+\alf)^2 +
(2\alf)^2}{r^2-\alf^2}\, ,\qquad \l.c\frac{d\tau}{dr} \r|_2 = -
\frac{r+ 3 \alf}{r+\alf}
\m{c-d67}
\ee
show that in the domain $r < \alf$ both $r$ and
$\tau$ become space-like, as in the Finkelstein coordinates \cite{Finkelstein_1958}. This is permissible, because the metric signature in the domain  $r<\alf$ remains correct. Only  when $r<\alf$ is the description of the particle motion  somewhat unusual: evolution of the space-like coordinate $r$ is presented as $r = r(\tau)$, where $\tau$ is another space-like coordinate. The sections $\tau = \rm const$ are space-like both outside and inside the horizon. If some events belong to the surface $\tau = \rm const$, then in this sense, one can speak about their simultaneity everywhere from infinity up to the true singularity.

Now, let us consider particle trajectories in terms of gauge transformations. Because gauge transformations act on the gravitational variables together with the matter variables, they have to act on the particle trajectories as well. Therefore, trajectories in a fixed background space-time are not gauge invariant. We consider `parabolic orbits' for the Schwarzschild solutions in harmonic coordinates, both in (\ref{c-d59}) and in (\ref{c-d65}). We consider also the {\em exact} transformations without using the $\xi^\mu$-vector used to build finite transformations by exponentiation as above.

First, we construct the field configurations related to the solutions (\ref{c-d59}) and (\ref{c-d65}). For the latter we make a mapping $\tau\,  \goto t\,$. After that for each of the solutions we choose the same background metric in the form (\ref{c-d1}). However, we exclude from the consideration the domain $-\alf \le r < 0$. Doing so is permissible here because we consider the trajectories in the  neighborhood of the event horizon only. Thus, using the decomposition (\ref{c-d4}), one finds the field configuration for the solution (\ref{c-d59}):
 \be h^{00} = 1
-\frac{\l(1+\alf/r\r)^3}{1-\alf/r},\qquad h^{11} =- \frac{\alf^2}{r^2}\,,
\m{c-d68}
 \ee
 and the field configuration for the solution (\ref{c-d65}):
 \bea h'^{00}& =& 1 - \l(1+ \frac{\alf}{r}\r)^2\l[ 1+
\frac{2\alf}{1+\alf} + \l(\frac{2\alf}{1+\alf}\r)^2+
\l(\frac{2\alf}{1+\alf}\r)^3\r] , \nonumber\\
h'^{01}& =&\frac{4\alf^2}{r^2}\, ,\qquad h'^{11}
=-\frac{\alf^2}{r^2}\,.
\m{c-d69}
 \eea
As with the configurations (\ref{c-d5}) -(\ref{c-d56}), the above configurations are connected by gauge transformations (neglecting the notion of $\eta$-causality). Now, they are induced by the coordinate transformations (\ref{c-d64}).

Let us discuss properties of the configurations (\ref{c-d68}) and (\ref{c-d69}), which are quite similar. First, they do not depend on time $t$ (stationary). Second, both of them represent asymptotically flat space-time. Third, the total energy calculated for both  of the cases is $E=mc^2$. Finally,  the tensorial de Donder condition is rewritten as
\be
h^{\mu\nu}{}_{;\nu}=0\, , ~~~{\rm and}~~~ h'^{\mu\nu}{}_{;\nu}=0
\m{c-d70}
\ee
for both of the configurations in spherical coordinates. The background metric permits the use of spherical coordinates while imposing a Fock-like generalized harmonic condition.

 To describe the trajectories of test particles, one has to vary the action (\ref{b-b116}) with respect to the coordinates. One obtains the equations for 4-velocities $u^\alf$ and $u'^\alf$; formally they are the equations for the geodesics. Maintaining the restriction to `parabolic trajectories', for the configuration (\ref{c-d68})  one has:
 \be
  u^0 =
\frac{r+\alf}{r-\alf}\, ,\qquad u^1 =
-\l(\frac{2\alf}{r+\alf}\r)^{1/2} , \qquad u^2=u^3 = 0.
\m{c-d71}
\ee
 Integrating $cdt =(u^0/u^1)dr$ one obtains the equation (\ref{c-d61}). Thus, now the particle approaches the event horizon, $r= \alf$, for an infinitely long time $t$; it fails to penetrate the horizon. However, for the field configuration (\ref{c-d69}), one has
 \bea
  u'^0 &=& \frac{1}{1+
\l(\frac{2\alf}{1+\alf}\r)^{1/2}}\l[ 1+
\l(\frac{2\alf}{1+\alf}\r)^{1/2} + \frac{2\alf}{1+\alf} +
\l(\frac{2\alf}{1+\alf}\r)^{3/2}+
\l(\frac{2\alf}{1+\alf}\r)^2\r],\nonumber\\
 u'^1 &=&-\l(\frac{2\alf}{r+\alf}\r)^{1/2}, \qquad
u'^2=u'^3 = 0\, .
\m{c-d72}
 \eea
After integration of $cdt =(u'^0/u'^1)dr$ one obtains the equations (\ref{c-d66}) by changing $\tau$ for $t$. Unlike (\ref{c-d71}), now the particle approaches the event horizon and penetrate it at a time $t$.  Thus, by a gauge transformation, trajectories are saved from a `catastrophic' discontinuity at $r = r_g$.  Much as one extends the solution in geometrical GR, one should extend the field configuration so that the interior of the horizon is included.  While infinitesimal gauge transformations remain arbitrary, finite gauge transformations require suitable boundary conditions to implement a suitable notion of maximal extension.


\vspace{0.3cm}
\sect{Continuous gravitational collapse to a point mass in GR}

In this section, we recall  some  recent results \cite{Petrov_2018}.
The Schwarzschild solution represented as a point particle with a  Minkowski background is only a  static model (at least outside the horizon). A dynamical model, that is  a description of the process by which the final point mass is formed, is desirable. In the framework of the geometrical description, the gravitational collapse was studied by Oppenheimer and Snyder \cite{Oppenheimer_Snyder_1939}; see also the textbook \cite{MTW}. The interior solution presents the Friedmann solution with dust in synchronous comoving coordinates, whereas the exterior is represented by the vacuum Schwarzschild solution. However, the interior region and the exterior region are described by different coordinates. To make the nature of the matching region more perspicuous, it would be more natural to describe both of the regions in the same coordinates. Recently \cite{Kanai_Siino_Hosoya_2011} such a task has been accomplished using a {\em generalization} of the well known Painlev\'e-Gullstrand (PG) coordinates. Below we outline the results.

Painlev\'e \cite{Peinleve_1921} and Gullstrand \cite{Gullstrand_1922}  discovered their coordinates independently. An instructive derivation of the PG coordinates and their discussion can be found in \cite{Hamilton_Lisle_2008} and references therein. The original form of the Schwarzschild vacuum solution in the PG coordinates is
 \be
 ds^2 = -c^2dt_p^2 + \l(  dr  + \l({\frac{r_g}{r}}\r)^{1/2} cdt_p \r)^2 + r^2(d\theta^2 +\sin^2\theta d\phi^2)\,.
\m{PG}
 \ee
 For the sake of definiteness, we discuss in this section the type of time coordinate, for example, here, $t_p$, that is related to PG coordinates.
 Its main property is that each of the sections defined as $ct_p = \const$ presents a flat Euclidean space. Recently  interest in these coordinates has increased; many authors, retaining  this property, have generalized the PG coordinates for more complicated black holes than the Schwarzschild one; see, for example, \cite{Kanai_Siino_Hosoya_2011}. 

 To obtain the PG coordinates, one can find the transformation from the Schwarzschild coordinates in (\ref{c-d2}) in this way:
\be
cdt_p = cdt_s + \frac{(r_g/r)^{1/2}}{1 - {r_g}/{r}}dr = cdt_s + \l(\frac{1}{1 - {r_g}/{r}} - \frac{1}{1 + (r_g/r)^{1/2}} \r)dr.
\m{to_PG}
\ee
By this transformation one removes the break in the geodesic trajectories on the space-time diagram. This fact can be seen more  explicitly after analyzing the components of the $ 4$-velocity for test particles. The transformation (\ref{to_PG}) permits us to recalculate the components of $4$-velocity for test particles (\ref{c-d71}) falling radially from infinity:
 \be
  u_p^0 = 1\, ,\qquad u_p^1 =
-\l(\frac{r_g}{r}\r)^{1/2} , \qquad u^2=u^3 = 0.
\m{u_PG}
\ee

The authors of the paper \cite{Kanai_Siino_Hosoya_2011} have generalized the vacuum PG solution (\ref{PG}) to the dust case. Here we consider the model where the surface of the star is at rest at infinity. It corresponds to our assumption of parabolic orbits.  Collapse from a finite radius has been suggested in \cite{Kanai_Siino_Hosoya_2011} as well. However, we do not consider it here because conceptually it is similar, but formulae are significantly more complicated. It is assumed that the radius of the star, $R(t_p)$, monotonically decreases from $t = -\infty$ to zero as $t_p\goto 0$. Thus, the dust interior region is contracted monotonically to the true singularity. Let us list the steps provided in  \cite{Kanai_Siino_Hosoya_2011}.

First, they have assumed that the metric element has the form
 \be
 ds^2 = -c^2dt_p^2 + \l(  dr  + \sqrt{\frac{2m}{r}\frac{G}{c^2}} cdt_p \r)^2 + r^2(d\theta^2 +\sin^2\theta d\phi^2)\,,
\m{PG_in}
 \ee
where $m=m(t_p,r)$. Second, the Einstein equations permit to express the matter energy-momentum $T_\nu{}^\mu$ at the right hand side through the function $m(t_p,r)$ unknown from the start.
Third, as usual, in the dust case it is assumed that the matter energy-momentum has the form:
 \be
 T_p^{\mu\nu} = \rho c^2 u_p^\mu u_p^\nu\,,
\m{T_PG}
 \ee
 where for the 4-velocity of matter particles moving radially it is assumed that: $u_p^\mu = \{1, v(t_p,r),0,0\}$. Fourth, the requirement of the consistency of the Einstein equations permits to find $v(t_p,r)$. Thus
 \be
  u_p^0 = 1\, ,\qquad u_p^1 =
-\l(\frac{2m}{r}\frac{G}{c^2}\r)^{1/2} , \qquad u^2=u^3 = 0.
\m{u_PG_in}
\ee
Fifth, the integration of the 00-component of the Einstein equations yields the function
 \be
 m(t_p,r) = 4\pi \int^r_0 \rho(t_p,r) r^2 dr\,.
\m{m_PG}
 \ee
Sixth, after assumption that $\rho(t,r) =\phi(r)\psi(t)$ and imposing the natural conditions $\l. m\r|_{r=0} =0$ and $\l. \rho\r|_{t=0} = \infty$, the 10-component of the Einstein equations gives:
 \be
 \rho = \frac{1}{6\pi}\frac{c^2}{G}\frac{1}{(ct_p)^2} \,
\m{rho_PG}
 \ee
 for $-\infty < t \leq 0$. Thus, a combination of (\ref{m_PG}) with (\ref{rho_PG}) gives
 \be
 \sqrt{\frac{2m}{r}\frac{G}{c^2}} = \frac{2}{3}\frac{r}{|ct_p|}\,.
\m{m_PG_ct}
 \ee
 Substituting it into (\ref{PG_in}), one obtains
 \be
 ds^2 = -\l(1- \frac{4}{9}\frac{r^2}{(ct_p)^2} \r)c^2dt_p^2 - \frac{4}{3}\frac{r}{ct_p} dr cdt_p + dr^2 + r^2(d\theta^2 +\sin^2\theta d\phi^2)\,.
\m{PG_in_tp}
 \ee
 A single non-zero component of the matter energy-momentum tensor is
\be
T_{00} = \frac{c^4}{6\pi G}\frac{1}{(ct_p)^2} \,.
\m{T_00}
\ee
Contracting it with $g^{00}=-1$ that easily can be found by (\ref{PG_in_tp}), one finds the trace of the energy-momentum tensor:
\be
 T = - \frac{c^4}{6\pi G}\frac{1}{(ct_p)^2}.
\m{T}
\ee
Thus,  (\ref{rho_PG})-(\ref{T}) describe a homogeneous distribution of dust in the interior PG coordinates.

However, only the interior solution is presented above. It is more interesting to describe a collapse of a star with the radius $r = R(t_p)$, connecting the interior dust region,  $r<R(t_p)$, with the exterior vacuum region, $r>R(t_p)$, described by (\ref{PG}). Of course, the total mass $M$ of the star is to be a constant; it is calculated by
  \be
 M = \l. m(t_p,r)\r|_{r=R(t_p)} = \frac{4\pi}{3}R^3 \rho \,.
\m{M_PG}
 \ee
 We stress that the interior and exterior regions defined in the aforementioned way are smoothly matched to each other.

 It might seem that we could directly apply the above geometrical derivation to represent the continuous gravitational collapse in the framework of the field-theoretic formulation. However, the exterior metric (\ref{PG}) does not satisfy the requirement of asymptotic flatness. To satisfy this requirement for the total model presented by the exterior metric (\ref{PG}) and the interior metric (\ref{PG_in_tp}), we can apply the transformation  in both the regions,
\be
cdt_p = cdt_e + \frac{({r_g/r})^{1/2}}{1 + ({r_g/r})^{1/2}}dr\,.
\m{EF_to_PG}
\ee
This change transfers the coordinates from the PG time $t_p$ to the EF time $t_e$ in the external region initially.
Then the exterior metric (\ref{PG}) is transformed to the EF metric (\ref{c-d55}). Following the field-theoretic prescription, we make the shift $t_e \goto t$ and obtain the field configuration (\ref{c-d56}) with  $t^{tot}_{\mu\nu} =0$  at $r>R(t)$, see (\ref{c-d58}). The next step is to be the field-theoretic reformulation for the interior solution at $r<R(t)$. Because the function (\ref{EF_to_PG}) is differentiable and monotonic at $0< r \leq \infty$, a smooth matching between the exterior and interior solutions is obtained. Because the surface $R(t_p)$ goes to zero monotonically  at $t_p \goto 0$, choosing  a vanishing  constant after integration of (\ref{EF_to_PG}), one easily finds that $R(t_e) \goto 0$, when $t_e \goto t \goto 0$ with the final state (\ref{c-d58}).

 One can easily check that in the case of the generalized EF frame for the interior and exterior regions, the above requirements (i)-(vi) of section 6, including the $\eta$-causality condition, are satisfied. Now, for the sake of generality, excluding only the requirement (iii) while preserving all the other aforementioned requirements,  we will show how this picture can be generalized. To achieve this goal we use the transformation with arbitrary $f$ combined with (\ref{EF_to_PG}):
\be
cdt_p = cdt_f + \l(\frac{({r_g/r})^{1/2}}{1 + ({r_g/r})^{1/2}} - f({r_g}/{r})\r)dr= cdt_f + F({r_g}/{r})dr\,.
\m{PG_to_generic}
\ee
First, we consider  the exterior region $r>R(t)$. Then, we obtain the metric
 \bea
 ds^2 = &-& \l(1 - \frac{r_g}{r}\r)c^2dt_f^2 + 2\l[\frac{r_g}{r} + \l(1 - \frac{r_g}{r}\r)f\r] cdt_f dr\nonumber\\
&+& \l[\l(1 + \frac{r_g}{r}\r) - 2\frac{r_g}{r}f - \l(1 - \frac{r_g}{r}\r)f^2\r]  dr^2 +
 r^2 d\Omega^2 \,.
 \m{generic}
 \eea
After applying a shift $t_f \goto t$, choosing the flat background again in the form (\ref{c-d1}), and making the use of the decomposition (\ref{c-d4}), one obtains for the
field configuration corresponding to (\ref{generic})
 \bea
 h_f^{00}  &=&  - \frac{r_g}{r} + 2\frac{r_g}{r}f + \l(1 - \frac{r_g}{r}\r)f^2\,
 ,\nonumber\\
 h_f^{01}  &=&   \frac{r_g}{r} + \l(1 - \frac{r_g}{r}\r)f\, ,
 \nonumber\\
 h_f^{11}  &=&  - \frac{r_g}{r}\, .
  \m{h-generic}
 \eea
This field configuration has to satisfy the Einstein equations in the whole Minkowski space. Because it contains an arbitrary function $f=f({r_g}/{r})$, the technique has to be generalized. Formally, one can derive for an arbitrary such function $\~f=\~f({r_g}/{r})$:
 \be
 \nabla^2 \~f\l(\frac{r_g}{r}\r) = \l(\~f''\frac{r^2_g}{r^4} - 4\pi r_g \~f'\delta(\bm r)\r) ,
 \m{Delta-f}
 \ee
 where $\~f' = \di_x f(x)$. The definition (\ref{Delta-f}) has been derived by making use of the formula (\ref{c-d33}), assuming that $\~f(x)$ in enough smooth. Application of the theory of generalized functions requires careful consideration; therefore formula (\ref{Delta-f}) tends to restrict the choice of $\~f\l({r_g}/{r}\r)$. Then, calculating the components of $t^{tot}_{\mu\nu}$  making use of (\ref{c-d39}), we obtain non-vanishing components of the total energy-momentum tensor\footnote{Note that in \cite{Petrov_2018} in the formula related to (\ref{t-tot-00}) a misprint exists.}:
\bea
 t^{tot}_{00} & = & mc^2 \delta({\bm r}) + \frac{mc^2}{2}\l[\frac{\~f''}{4\pi}\frac{r_g}{r^4} - \~f'\delta({\bm r}) \r]; \nonumber\\ && \~f \equiv 2\frac{r_g}{r}f + \l(1 - \frac{r_g}{r}\r)f^2\, ,\m{t-tot-00}\\
t^{tot}_{11} & = & - mc^2\delta({\bm r})\, ,\m{t-tot-11}\\
t^{tot}_{AB} & = & - \half \gamma_{AB}\, mc^2\delta({\bm r}); \qquad A,\ldots = 2,3\,.
 \m{t-tot-AB}
 \eea
One can see that the energy-momentum is especially concentrated at $r=0$ and is expressed  making  use of the $\delta({\bm r})$-function representing the point particle, but there is also a distribution of energy outside $r=0.$ %

The requirement for a definition of the permissible weakest fall-off for asymptotically flat space-time (see  \cite{Petrov_1995,Petrov_1997} and references there in) presented by the field configuration $h_f^{\mu\nu}$ in (\ref{h-generic}) restricts the asymptotic behavior of $f$ as
\be
\l.f(r_g/r)\r|_{r\goto \infty} \sim (r_g/r)^{\alf}; \qquad \alf > 1/2.
\m{first-f}
\ee

Consider the requirement of continuity of geodesics in the vacuum region. The transformation (\ref{PG_to_generic}) permits us to recalculate
the components of the 4-velocity for test particles (\ref{u_PG}) in the generic coordinates:
 \be
  u_f^0 = 1+
\frac{r_g/r}{1+ (r_g/r)^{1/2}} - f\cdot\l(\frac{r_g}{r}\r)^{1/2} ,~~ u_f^1 =
-\l(\frac{r_g}{r}\r)^{1/2} , ~~ u^2=u^3 = 0.
\m{c-d71_b}
\ee
This gives
 \be
\frac{cdt}{dr} = \frac{u_f^0}{u_f^1} = - \l({r}/{r_g}\r)^{1/2} -
\frac{(r_g/r)^{1/2}}{1+ (r_g/r)^{1/2}} + f .
\m{du0du1}
\ee
The requirement for geodesics to be continuous after such transformations gives a restriction on $f$:
 \be
 \l|f\r| < N
 \m{f<N}
 \ee
for some finite arbitrary large positive $N$, at least for $r>0$. Indeed, failure of  (\ref{f<N}) means that $\lim_{r\goto r_0}\l|f\r| = \infty$ and indicates  a breakdown of the geodesic at $r_0$.
Besides, to have an appropriate form for the ingoing geodesics, one needs a monotonic smooth function $f$ when ${cdt}/{dr} < 0$. Then the concrete expression (\ref{du0du1}) gives
 \be
f < 1 + \frac{(r/r_g)^{1/2}}{1+ (r_g/r)^{1/2}}.
\m{du0du1<}
\ee
After integration of (\ref{du0du1}), one obtains the equation of the radial parabolic orbits on the space-time ($t\times r$) diagram:
\bea
ct &=& -
2r_g\l[{{ \frac{1}{3}}} \l({\frac{r_g}{r}}\r)^{-3/2}+
\l(\frac{r_g}{r}\r)^{-1/2}  -
\ln \l|{\l({\frac{r_g}{r}}\r)^{-1/2} +1}\r|\,\r]\nonumber\\
&& + \int^r f\l({\frac{r_g}{r^*}}\r) dr^* +{\rm const}\,  .
\m{c-d61_a}
\eea
Summarizing, we conclude that the requirement of the continuity is satisfied by the restrictions (\ref{f<N}) and (\ref{du0du1<}) for monotonic smooth $f$ at $0< r \leq \infty$.

 Let us turn to the question of  $\eta$-causality.  Deriving the light cone expressions from $ds^2 =0$ for the quite complicated form of the metric (\ref{generic}), one obtains surprisingly  simple formulae. Thus for the ingoing light ray one has
\be
\l.\frac{cdt}{dr}\r|_{1} = f - 1,
\m{generic-cone1}
\ee
whereas for the outgoing light ray it is
\be
 \l.\frac{cdt}{dr}\r|_{2} = f + \frac{1+ {r_g}/{r}}{1 - {r_g}/{r}}.
\m{generic-cone2}
\ee
The requirement of $\eta$-causality\footnote{Note that the requirement of `$\eta$-causality' can be changed by the requirement of the `stable $\eta$-causality' by exchanged $\{\leq\}$ and $\{\geq\}$ by $\{<\}$ and $\{>\}$ in (\ref{generic-cone_x1}) and (\ref{generic-cone_x2}). } for (\ref{generic-cone1}) and (\ref{generic-cone2}) can be realized as
\bea
&&f - 1 \leq  -1,\m{generic-cone_x1}\\
&&f + \frac{1+ {r_g}/{r}}{1 - {r_g}/{r}} \geq  1.
\m{generic-cone_x2}
\eea
The restriction (\ref{generic-cone_x1}) gives $f\leq 0$  everywhere ($0\leq r \leq \infty$). Then if we impose the requirement of  $\eta$-causality, it is not necessary to consider (\ref{du0du1<}). The restriction (\ref{generic-cone_x2}) has to be analyzed in more detail. Considering asymptotic behaviour at $r\goto\infty$ in (\ref{generic-cone_x2}), we are restricted by
\be
\l|f\r|_{r\goto\infty} <~\sim \frac{2r_g}{r},
\m{generic-cone_a}
\ee
which is even stronger than the restriction (\ref{first-f}). From (\ref{generic-cone_x2}) for the domain $r_g < r < \infty$, one has
\be
\l|f\r| \leq  \frac{2{r_g}/{r}}{1 - {r_g}/{r}}.
\m{generic-cone_b}
\ee
This  restriction is in addition to (\ref{f<N}). Finally, for the case $r = r_g$ with the restricted $f$, see (\ref{f<N}), the expression (\ref{generic-cone2}) describing  the event horizon becomes $+\infty$, as it must for the horizon. Thus, for a monotonic, restricted and negative $f,$ the expression (\ref{generic-cone2}) for the outgoing light ray is positive for $r_g\leq r \leq \infty$.

The case $r < r_g$ requires  special attention. The expression (\ref{generic-cone2}) becomes negative, automatically satisfying the requirement (v) with the natural relation between ingoing and outgoing light rays:
\be
f - 1 \geq f + \frac{1+ {r_g}/{r}}{1 - {r_g}/{r}}.
\m{generic-cone_c}
\ee
The equality in (\ref{generic-cone_c}) holds at the true singularity, which is  where  the light cone becomes degenerated. Again, this fact signals  the continuity of the geodesic all the way to the true singularity. Finally, it could be interesting  to require that after finalizing the collapse the test particle approach the true singularity at a finite time $t$.\footnote{As noted above, one might also consider the alternative requirement that the test particle only approach the true singularity as Minkowski background time goes to infinity.  In that case there might be less worry about black hole information loss, which might never occur. }
 Then, it is necessary to add the restriction (\ref{f<N}) by
 \be
  \l.\l|f\r|\r|_{r \goto 0} < N\,.
 \m{f<N_0}
 \ee

Let us turn to the exterior region including the surface of the star $r\leq R(t)$.
To achieve a smooth matching between exterior and interior regions, one has to require that the function $F(r_g/r)$ in (\ref{PG_to_generic}) be differentiable and monotonic on the interval $0< r \leq \infty$. Another requirement for the function $F(r_g/r)$ in (\ref{PG_to_generic}) is formulated as follows. After integrating (\ref{PG_to_generic}) and replacing $r$ by a surface radius $R(t_p)$, one can choose the constant of integration so that if the surface $R(t_p)$ goes to zero monotonically  at $t_p \goto 0$, then $R(t_f) \goto 0$ when $t_f \goto t \goto 0$. Then the final state (\ref{t-tot-00})-(\ref{t-tot-AB}) is achieved at $t \goto 0$  at a finite time $t$. After satisfying these requirements, the model of the continuous collapse (\ref{PG}) plus (\ref{PG_in}) presented in the PG frame is rewritten in the generic frame with the use of the transformations (\ref{PG_to_generic}). Then, all the requirements hold; only the $\eta$-causality problem remains to be addressed.

 After transformation (\ref{PG_to_generic}) and the shift $t_f \goto t$, the metric (\ref{PG_in_tp}) for the interior region acquires the form:
 \bea
 ds^2 = &-&\l(1- \frac{2m}{r}\r)dt^2 + 2 \l[\sqrt{\frac{2m}{r}} - \l(1- {\frac{2m}{r}}\r)F \r] dr dt\nonumber\\ &+& \l[1+2\sqrt{\frac{2m}{r}}F - \l(1- \frac{2m}{r}\r)F^2\r]dr^2 + r^2d\Omega^2\,.
\m{PG_in_tf}
 \eea
 For the sake of simplicity in formulae, here and below, we set $G=c=1$.
A standard calculation gives the expression for the ingoing ray of the light cone
\be
\l.\frac{dt}{dr}\r|_1=  -\frac{1}{1+ \sqrt{2m/r}} - F(r_g/r),
\m{cone_minus}
\ee
whereas the outgoing ray is determined by
\be
\l.\frac{dt}{dr}\r|_2=  \frac{1}{1- \sqrt{2m/r}} - F(r_g/r).
\m{cone_plus}
\ee
The necessary requirement for (\ref{cone_minus}) is that it has to be negative in all the regions. Then the $\eta$-causality condition, $\l.{dt}/{dr}\r|_1 \leq -1$, implies the restriction
\be
F \geq \frac{\sqrt{2m/r}}{1+ \sqrt{2m/r}}.
\m{cone_minus_eta}
\ee

Let us consider three cases for (\ref{cone_plus}), each of which corresponds to a concrete instant $t_p$:

\bit

 \item The first case corresponds to the PG instant of time when the star boundary $R(t_p) > r_g \equiv 2M/r$. Because $m\leq M$ and  $m$ in  (\ref{m_PG_ct}) decreases as $r\goto 0$ at the instant $t_p$, one finds that
\be
\frac{2m}{r} <1.
\m{m<1}
\ee
Next, the outgoing expression (\ref{cone_plus}) has to be positive in order to be matched with the exterior smoothly. Then the $\eta$-causality condition, $\l.{dt}/{dr}\r|_2 \geq 1$, implies the restriction
\be
  \frac{\sqrt{2m/r}}{1- \sqrt{2m/r}} - F \geq 1.
\m{cone_plus_eta}
\ee
Combining (\ref{cone_minus_eta}) and (\ref{cone_plus_eta}) gives a unified restriction on $F(r_g/r)$
\be
\frac{\sqrt{2m/r}}{1+ \sqrt{2m/r}}\leq F \leq \frac{\sqrt{2m/r}}{1- \sqrt{2m/r}}\,.
\m{cone_minus_plus_eta}
\ee

 \item The second case is classified by the position of the star surface at the horizon $R(t_p) = r_g$.  Then (\ref{cone_plus}) gives $\l.{dt}/{dr}\r|_2 = + \infty$. It matches the exterior region continuously, see (\ref{generic-cone2}). For the interior region, where again the condition (\ref{m<1}) holds, the result (\ref{cone_minus_plus_eta}) of the first case is repeated.

 \item The third case that is classified by the position of the star surface, $R(t_p) < r_g$, is more complicated. The interior region is decomposed into the three subregions: a) $2m/r>1$, b) $2m/r = 1$ and c) $2m/r <1$. In the case a) for the outgoing ray defined by (\ref{cone_plus}) one has $\l.{dt}/{dr}\r|_2 < 0$. Then, because the light cone must not to be degenerate, $\l.{dt}/{dr}\r|_1 > \l.{dt}/{dr}\r|_2$, one obtains
\be
- \frac{1}{1+ \sqrt{2m/r}}> \frac{1}{1- \sqrt{2m/r}}
\m{minus}
\ee
so the restriction for case a) holds automatically. Analyzing subregions b) and c), one finds easily that the results correspond exactly to the results of the second case and the first case, respectively. Thus again one obtains only the restriction (\ref{cone_minus_plus_eta}).

\eit

To conclude the description of continuous collapse in the field-theoretic formulation, it is instructive to analyze the matter part $t^m_{\mu\nu}$ of the total energy-momentum. Thus, applying the transformations (\ref{PG_to_generic}) to the metric, $g_{\mu\nu}$, represented by (\ref{PG_in_tp}), we obtain $g^f_{\mu\nu}$ in (\ref{PG_in_tf}); applying the transformations (\ref{PG_to_generic}) to the energy-momentum, $T_{\mu\nu}$, presented in (\ref{T_00}), we obtain $T^f_{\mu\nu}$. Then, formula (\ref{(2.20+)}) takes the form:
 \be
t^{m}_{\mu\nu} = T^f_{\mu\nu} - \half
g^f_{\mu\nu}T^f - \half
\gamma_{\mu\nu}\gamma^{\alpha\beta}\l(T^f_{\alpha\beta} - \half
g^f_{\alpha\beta}T^f\r)\, .
 \m{t-T-f}
 \ee
 Thus we conclude that the interior region defined by the energy-momentum (\ref{t-T-f}) is in fact contracted at $t \goto 0$ to a point-like state described by the $\del$-function.

 Keeping in mind that the Schwarzschild black hole and its collapse to this stage are physical realities as described by Einstein's equations, one can view the description in the field-theoretic formalism as merely an alternative equivalent mathematical language. On the other hand, physically reasonable criteria such as $\eta$-causality make it possible to take the background geometry to be not purely fictitious, but rather to have a qualitative physical meaning that makes sense of quantization techniques that are often used anyway.  Thus our treatment of gravitational collapse can be useful both  for practical calculations and fundamental considerations.

\vspace{0.3cm}


\vspace{0.3cm}
\sect{The field-theoretic method in an arbitrary $D$-dimensional metric theory}

Here, following \cite{GPP,[15]}, see also \cite{Petrov+_2017}, we generalize the Lagrangian-based field-theoretic method in arbitrary $D$-dimensional metric theories.
 Consider a theory with the action:
 \be
 S = \frac{1}{c}\int d^Dx \lag_{\sst D}(g,\Phi) = - \frac{1}{2 \k c}\int d^Dx \lag_{\sst G}(g_{\mu\nu}) +  \frac{1}{c}\int d^Dx \lag^M(g_{\mu\nu},\Phi^A)\,,
 \m{lag-g}
 \ee
 where $\lag_{\sst G}$ is the pure gravitational Lagrangian in an arbitrary metric theory and $\k$ is the $D$-dimensional Einstein's constant. Here, unlike section 1, we consider just the components of $g_{\mu\nu}$ as dynamical variables.  They are more appropriate in this section than are the  $\gog^{\mu \nu }=\sqrt{-g}g^{\mu \nu }$ components used earlier.  For a different choice of dynamic variables from the set
  \be
 g = \l\{g^{\mu \nu },~g_{\mu
\nu },~\sqrt{-g}g^{\mu \nu },~\sqrt{-g}g_{\mu \nu },~(-g)g^{\mu \nu
},~\ldots\r\}\, \m{(1)}
 \ee
 see \cite{Petrov_2019,Petrov+_2017}.
The matter part $\lag^M$ in (\ref{lag-g}) depends on $\Phi^A$ --- generalized matter variables interacting with $g_{\mu\nu}$, the same as  (\ref{(a2.1)}). Varying
(\ref{lag-g}) with respect to $g_{\alf\beta}$, one obtains
  \be
\frac{\delta \lag_{\sst G}}{\delta g_{\alf\beta}} = 2\k \frac{\delta
\lag^{M}}{\delta g_{\alf\beta}}\qquad \goto \qquad \frac{\delta
\lag_{\sst G}}{\delta g^{\mu\nu}} \equiv {\cal C}_{\mu\nu}= \k {\cal T}_{\mu\nu}\, . \m{ddd}
 \ee
The last equality gives the gravitational equations in the standard form obtained after contracting the first one with $\di g_{\alf\beta}/\di g^{\mu\nu}= - g_{\alf(\mu}g_{\nu)\beta}$; see the analogous coefficient in (\ref{g-di_g}).  Variation of (\ref{lag-g}) with
respect to $\Phi^A$ gives corresponding matter equations, the same as (\ref{(a2.3)}). The
background gravitational equations are in the form
 \be
 \frac{\delta
\bar\lag_{\sst G}}{\delta \bar g_{\alf\beta}} = 2\k \frac{\delta \bar
\lag^{M}}{\delta \bar g_{\alf\beta}}\qquad \goto \qquad \frac{\delta
\bar\lag_{\sst G}}{\delta \bar g^{\mu\nu}} \equiv \bar{\cal C}_{\mu\nu} = \k \bar {\cal T}_{\mu\nu}\,,
 \m{bbb}
 \ee
where the background Lagrangian is defined by the barred procedure
in (\ref{lag-g}): $\bar \lag_{\sst D} = \lag_{\sst D}(\bar g_{\alf\beta},\bar\Phi^A)$.
The background matter equations are derived analogously.  It is assumed that background fields $\bar g_{\alf\beta}$ and $\bar\Phi^A$ are specified and satisfy the
background equations.

A physical system described by equations (\ref{ddd}) can be considered as a perturbed one
with respect to a background system with the equations (\ref{bbb}). Just as in (\ref{(a2.4)}) and (\ref{(a2.4m)}), we
decompose metric and matter variables in (\ref{lag-g}) into the
background (barred) parts and the dynamic variables (perturbations)
$\varkappa_{\mu\nu}$ and $\phi^A$:
 \bea
 g_{\mu\nu} &=& \bar g_{\mu\nu} + \varkappa_{\mu\nu}\, ,
 \m{g-Dec}\\
 \Phi^A &=& \bar \Phi^A + \phi^A\, .
  \m{Phi-Dec}
 \eea
We note that $\goh^{\mu\nu}$ in (\ref{(a2.4)}) and $\varkappa_{\mu\nu}$ in (\ref{g-Dec}) one differ  %
(aside from moving indices and densitizing with the background metric) both in trace at lowest order and in second and higher order terms, see  \cite{Petrov_2019,Petrov+_2017}. Now, analogously to the definition  (\ref{(2.10)}), we derive the dynamical Lagrangian:
 \be
\lag^{dyn}(\bar g,\bar\Phi;\varkappa,\,\phi) = \lag_{\sst D} (\bar g+\varkappa,\,\bar
\Phi+\phi ) - \varkappa_{\mu\nu} \frac{\delta \bar \lag_{\sst D}}{\delta \bar g_{\mu\nu}} - \phi^A
\frac{\delta \bar \lag_{\sst D}}{\delta \bar \Phi^A}- \bar \lag_{\sst D} = -\frac{1}{2\k} \lag_g +\lag_m\,.
 \m{lag}
 \ee
Thus, the field-theoretic method applied in an arbitrary metric theory is based on the dynamical Lagrangian (\ref{lag}).

To obtain the gravitational equations
related to the Lagrangian (\ref{lag}), one needs to vary it with respect to $\varkappa_{\alf\beta}$. Using the property ${\delta \lag_{\sst D} (\bar g+\varkappa,\,\bar
\Phi+\phi)}/{\delta \bar g_{\alf\beta}} = {\delta \lag_{\sst D} (\bar g+\varkappa,\,\bar
\Phi+\phi)}/{\delta \varkappa_{\alf\beta}}$,  we present them in the form:
 \be
\frac{\delta \lag^{dyn}}{\delta \varkappa_{\alf\beta}} = \frac{\delta }{\delta \bar g_{\alf\beta}}\l[\lag_{\sst D}(\bar g+\varkappa,\,\bar \Phi+\phi) - \bar \lag_{\sst D}\r] = 0\, .
 \m{PERTeqs1}
 \ee
It is clear that the field equations for $\varkappa_{\alf\beta}$ are {\em equivalent} to the gravitational equations in the standard form
(\ref{ddd}) if the background equations (\ref{bbb}) are satisfied.

Let us define the metric energy-momentum for perturbations defined in (\ref{g-Dec}) and (\ref{Phi-Dec}). Let us rewrite the equation (\ref{PERTeqs1}) as
\be
\frac{\delta \lag^{dyn}}{\delta \bar g_{\alf\beta}} =
 -\frac{\delta }{\delta \bar
g_{\alf\beta}}\l(\varkappa_{\rho\sig} \frac{\delta \bar \lag_{\sst D}}{\delta \bar g_{\rho\sig}} + \phi^A
\frac{\delta \bar \lag_{\sst D}}{\delta \bar \Phi^A}\r)\,.
 \m{PERTcurrent}
 \ee
Note that the background equations should not  be
taken into account before variation of $\lag^{dyn}$ with respect to
$\bar g_{\alf\beta}$ and $\bar \Phi^A$. Then, contracting (\ref{PERTcurrent}) with $2\k\di \bar g_{\alf\beta}/\bar g^{\mu\nu} = - 2\k \bar g_{\alf(\mu}\bar g_{\nu)\beta}$,  one obtains another form of the equations
(\ref{PERTeqs1}):
  \be
{\cal C}^{L}_{\mu\nu} + {\cal F}^{L}_{\mu\nu} = \k{\bm t}^{\rm tot}_{\mu\nu}\,
 \m{PERTmunu}
 \ee
equivalent to the equations (\ref{ddd}) if the background equations (\ref{bbb}) hold. The  linear operators on the
left hand side of (\ref{PERTmunu}) are defined by the expressions:
 \bea
 {\cal C}^{L}_{\mu\nu} =
 \frac{\delta }{\delta \bar g^{\mu\nu}} \varkappa_{\rho\sig} \frac{\delta \bar \lag_{\sst G}}{\delta \bar g_{\rho\sig}}\,,
 \m{GL-q}\\
  {\cal F}^{L}_{\mu\nu} =
 -2\k\frac{\delta }{\delta \bar g^{\mu\nu}}\l(\varkappa_{\rho\sig} \frac{\delta \bar \lag^M}{\delta \bar g_{\rho\sig}} + \phi^A
\frac{\delta \bar \lag^M}{\delta \bar \Phi^A} \r)\,.
 \m{PhiL-q}
 \eea
Finally, the right hand side in (\ref{PERTmunu}) becomes the total symmetric
(metric) energy-momentum tensor density for the fields (perturbations) $\varkappa_{\alf\beta}$ and $\phi^A$ defined as usual,
 \be
 {\bm t}^{\rm tot}_{\mu\nu} = 2\frac{\delta\lag^{dyn}}{\delta \bar
 g^{\mu\nu}} = {\bm t}^g_{\mu\nu} + {\bm t}^m_{\mu\nu}\, ,
 \m{em-q}
 \ee
 see also (\ref{(2.20)}).
Here, ${\bm t}^g_{\mu\nu}$  is the energy-momentum related to a pure gravitational part of the Lagrangian (\ref{lag}); ${\bm t}^m_{\mu\nu}$ is the energy-momentum of matter fields $\phi^A$ in (\ref{lag}) interacting with the gravitational field $\varkappa_{\alf\beta}$.

To conclude this section, let us note the following. As a general rule, a
difference between definitions of metric perturbations is not important for calculating conserved quantities for
static solutions. It does not influence calculations in quantum gravity either \cite{B-Deser}. A difference in the second
order becomes important, however, in a real calculation for radiating isolated systems in 4D GR. It turns out that {\em only} the choice of the metric perturbations
 $
 \goh^{\mu\nu} = \gog^{\mu\nu} - \bar {\gog}^{\mu\nu}$ gives (see \cite{PK}) the standard Bondi-Sachs momentum
\cite{BMS}. All the other decompositions (including the popular $ \varkappa_{\mu\nu} = g_{\mu\nu} - \bar g_{\mu\nu}$, used in this section and, for example, in \cite{AbbottDeser82}) do not lead to the right result.


\vspace{0.3cm}
\sect{Currents and superpotentials in an arbitrary field theory of the Lovelock type}

Here we construct currents and superpotentials for generic theories presented in the field-theoretic description, as in the above section. We restrict ourselves to Lovelock-like theories; see \cite{Petrov_2019} for a detailed treatment. For such constructions we use many results from the Appendix. Before constructing conserved quantities for perturbations in the Lovelock theory, we consider the Lagrangian (\ref{lagQ}) in the Appendix in a more concrete form:
 \be
 \lag = \lag_{\sst G}(\varkappa_{\alf\beta}; g_{\pi\sig};  R^\alf{}_{\mu\beta\nu})\,,
 \m{LagR}
 \ee
although it is still quite abstract. The fields in (\ref{lagQ}) is now $\psi^A = \{\varkappa_{\alf\beta},\, g_{\pi\sig} \}$. The Lagrangian  (\ref{LagR}) is an arbitrary enough smooth algebraic function of $\varkappa_{\alf\beta}$, $g_{\pi\sig}$ and the Riemann tensor $R^\alf{}_{\mu\beta\nu}$. We note especially that derivatives of the metric $g_{\pi\sig}$ are
included only in $R^\alf{}_{\mu\beta\nu}$, not anywhere else.

It is very useful to define the quantity
 \be
 {\bm \omega}^{\rho\lam|\mu\nu}  =   \frac{\di \lag_{\sst G}}{\di
g_{\rho\lam,\mu\nu}}\,.
 \m{NL1}
 \ee
It has the evident symmetries
 \be
 {\bm \omega}^{\rho\lam|\mu\nu}  = {\bm \omega}^{\lam\rho|\mu\nu}  =
 {\bm \omega}^{\rho\lam|\nu\mu}\,.
  \m{omega}
 \ee
Recalling that the Riemannian tensor is linear in second derivatives of
$g_{\pi\sig}$, we conclude that the quantity (\ref{NL1}) is covariant automatically. Turning again to the Appendix and
making use of the definitions for the coefficients (\ref{(+3+)}) - (\ref{(+5+)})
 and (\ref{Uu}) - (\ref{Nn}),
 the identities  (\ref{(+9+4A)}) and  (\ref{(+9+4)}), the
quantity (\ref{NL1}) and its symmetries (\ref{omega}), we can rewrite (\ref{Uu}) - (\ref{Nn}) for the Lagrangian (\ref{LagR}) as
  \bea
{\bm u}_{\sig}{}^{\mu} & =& \lag_{\sst G}\delta_\sigma^\mu +\frac{\delta
\lag_{\sst G}}{\delta \psi^A}\l.\psi^A \r|^\mu_\sigma   - {\bm \omega}^{\lambda\mu|\rho\nu} R_{\lambda\rho\nu\sigma}
\, ,\m{uL}\\
 {\bm m}^{\rho\mu\nu}  &=& 2 \nabla_\lam
 {\bm \omega}^{\rho\nu|\mu\lam}; \qquad   {\bm m}^{\rho\mu\nu} =
 {\bm m}^{\nu\mu\rho}\, ,\m{mL} \\
 {\bm n}^{\rho\lam\mu\nu} & =& {\bm \omega}^{\rho\lam|\mu\nu};\qquad\qquad{\bm \omega}^{\rho\lam|\mu\nu} = {\bm \omega}^{\mu\nu|\rho\lambda}\,.
 \m{nL}
 \eea
The concrete expressions (\ref{uL}) - (\ref{nL}) are covariant, and, of course, they satisfy the identities (\ref{(+9+2)}) -
(\ref{(+9+4)}) of the general form derived in the Appendix.

Keeping in mind the expressions (\ref{uL}) - (\ref{nL}), we rewrite
the current (\ref{(+7+)}) - (\ref{(+8+)}) and the superpotential
(\ref{(+11+)}) of the general form for the Lagrangian (\ref{LagR}) in the concrete form:
  \bea
{\bm i}^\mu &=& -\l(\lag_{\sst G}\xi^\mu + \frac{\delta \lag_{\sst G}}{\delta
\psi^A}\l.\psi^A \r|^\mu_\sigma \xi^\sig + {\bm z}^\mu\r)\,,
 \label{Jmu}\\
  {\bm z}^\mu &= & 2\zeta_{\rho\lam}\nabla_\nu
 {\bm \omega}^{\rho\lam|\mu\nu}- 2{\bm \omega}^{\rho\lam|\mu\nu}
  \nabla_\nu \zeta_{\rho\lam} \,.\label{ZDmu}\\
{\bm i}^{\mu\nu} & =&{\textstyle{4\over 3}}\l(
 2\xi^\sig \nabla_\lam  {\bm \omega}_{\sig}{}^{[\mu|\nu]\lam}   -
{\bm \omega}_{\sig}{}^{[\mu|\nu]\lam}
 \nabla_\lam  \xi^\sig\r)\,.
 \m{(+16+A)}
 \eea
Here $ 2\zeta_{\rho\sigma} \equiv - {\pounds}_\xi g_{\rho\sigma} =
2\nabla_{(\rho}\xi_{\sigma)}\,.$
We note that due to the symmetry in (\ref{mL}), the $m$-term disappears from the
current;  compare (\ref{(+7+)}) with (\ref{Jmu}). Furthermore, the expression for the superpotential
  (\ref{(+16+A)}) only depends on the quantity ${\bm \omega}^{\rho\lam|\mu\nu}$ defined in (\ref{NL1}).

Keeping in mind the abstract definitions (\ref{uL})-(\ref{(+16+A)}) related to the system (\ref{LagR}), we can construct currents and superpotentials in metric theories in the field-theoretic formulation (\ref{lag}). We use the structure of the Lagrangian $\lag^{dyn}$ defined in (\ref{lag}). Consider this  purely gravitational part that is linear in metric perturbations:
\be
\lag_1 = -\frac{1}{2\k}\varkappa_{\alf\beta} \frac{\delta \bar
\lag_{\sst G}}{\delta \bar g_{\alf\beta}}\, .
 \m{Lag-1}
 \ee
It plays a role of an {\em auxiliary} Lagrangian (\ref{LagR}), where we can set $\psi^A = \{\varkappa_{\alf\beta},\,\bar g_{\pi\sig} \}$.
 The index "$_1$" is used because Lagrangian (\ref{Lag-1}) is of the first order in  $\varkappa_{\alf\beta}$ in expansion of $\lag_{\sst G}$. To apply the above technique to the Lagrangian (\ref{Lag-1}), we assume that $\lag_1 = \lag_1(\varkappa_{\alf\beta}; \bar g_{\pi\sig};
\bar R^\alf{}_{\mu\beta\nu})$. Then, because $\lag_1$ in (\ref{Lag-1}) is proportional to the Lagrange derivative of $\bar \lag_{\sst G}$, the gravitational part of the Lagrangian (\ref{lag-g}) can present a theory of the Lovelock type, see, for example, \cite{Petrov_2019}. However, here we consider generic expressions only. Analysing (\ref{Lag-1}) itself, we can obtain only {\em identically} conserved quantities. However, because the Lagrangian (\ref{Lag-1}) induces the construction of the linear operator (\ref{GL-q}) in (\ref{PERTmunu}), making the use of the field equations (\ref{PERTmunu}), we can transform the {\em identically} conserved quantities to {\em physically} conserved quantities.

The explicit expression for the linear operator ${\cal C}^{L}_{\mu\nu}$ in (\ref{PERTmunu}) defined in (\ref{GL-q}) is quite important in numerous applications. Let us derive it. For the Lagrangian (\ref{Lag-1}) that is of the type (\ref{LagR}) and for the expressions (\ref{uL}) - (\ref{nL}), one finds from the related identity (\ref{(+9+2)}) of the Appendix:
 \be
{\cal C}_\sigma^{L\mu} =- \frac{1}{2}\frac{\delta \bar \lag_{\sst G}}{\delta
\bar {g}_{\alf\beta}}\l.\l(\varkappa_{\alf\beta}\delta^\mu_\sig + \varkappa_{\alf\beta}\r|^\mu_\sig\r) +2\k \l(
\bar\nabla_{\rho\lam}  {\bm \omega}_{1\sig}{}^{\mu|\rho\lam} +
 {\bm \omega}_1^{\mu\tau|\rho\lam} \bar R_{\sig\lam\tau\rho} + \frac{1}{3} {\bm \omega}_{1\sig}{}^{\lam|\tau\rho} \bar R^{\mu}{}_{\tau\rho\lam} \r)\, .
 \m{Linear-Gen}
 \ee
 %
Note that this formula can be obtained following the method in \cite{Rund};
however, the use of the identities (\ref{(+9+1)}) - (\ref{(+9+4)}) in the Appendix is more economical.

Now, let us proceed to constructing concrete quantities. Substituting the expression for the Lagrangian (\ref{Lag-1}) into
the current expression (\ref{Jmu}), one obtains
  \be
{\bm i}_1^\mu = {\bm \theta}_\sigma{}^\mu\xi^\sigma  -
{\bm z}^\mu_{1} \,,
 \label{Jmu1}
 \ee
 where the coefficient ${\bm \theta}_\sigma{}^\mu$ of $\xi^\sigma$ is interpreted as the energy-momentum in the standard way:
  \be
{\bm \theta}_\sigma{}^\mu =\frac{1}{\k} \l({\cal C}_\sigma^{L\mu} + \frac{1}{2}\frac{\delta \bar \lag_{\sst G}}{\delta
\bar g_{\alf\beta}}\l(\varkappa_{\alf\beta}\delta^\mu_\sig + \l.\varkappa_{\alf\beta}\r|^\mu_\sig\r)\r) \,,
 \label{EM1}
 \ee
 and
 $z$-term ${\bm z}^\mu_{1}$  has exactly the form
(\ref{ZDmu}) with ${\bm \omega^{\rho\lam|\mu\nu}_1}$ defined in (\ref{NL1}) and related to $\lag_1$.  Combining the last expression with  (\ref{Linear-Gen}), one finds a quite simple formula for the energy-momentum:
 \be
{\bm \theta}_\sigma{}^\mu =2\l(
\bar\nabla_{\rho\lam}  {\bm \omega}_{1\sig}{}^{\mu|\rho\lam} +
 {\bm \omega}_1^{\mu\tau|\rho\lam} \bar R_{\sig\lam\tau\rho} + \frac{1}{3} {\bm \omega}_{1\sig}{}^{\lam|\tau\rho} \bar R^{\mu}{}_{\tau\rho\lam} \r) \,.
 \label{EM1A}
 \ee
Formally the energy-momentum (\ref{EM1A}) is related to the auxiliary and arbitrary Lagrangian (\ref{Lag-1}). We note again the nice property of the expression (\ref{EM1A}) that, being quite general, depends {\em essentially } on the quantity ${\bm \omega}_{1}^{\sig\lam|\tau\rho}$, not on any other quantities.

We recall that the current (\ref{Jmu1}) is identically conserved:
 \be
 \bar\nabla_\mu {\bm i}_1^\mu \equiv \di_\mu {\bm i}_1^\mu
\equiv 0\, .\m{BDi-1}
 \ee
As a consequence, the current in this identity, making use of the Klein-Noether identities, can be rewritten as a divergence of a superpotential,
 \be
 {\bm i}_1^\mu \equiv \bar\nabla_\nu {\bm i}_1^{\mu\nu} \equiv \di_\nu {\bm i}_1^{\mu\nu}
\, ,\m{BDi-sup}
 \ee
 where ${\bm i}_1^{\mu\nu}$ has exactly the form
(\ref{(+16+A)}) with ${\bm \omega^{\rho\lam|\mu\nu}_1}$ related to the Lagrangian $\lag_1$ in (\ref{Lag-1}).

Both (\ref{BDi-1}) and (\ref{BDi-sup}) are merely identities. To make them physically meaningful conservation laws, one has to use the field equations. Substituting the part linear in metric perturbations  from (\ref{PERTmunu}) into the energy-momentum (\ref{EM1}), one obtains
\be
{\bm \theta}_{\mu\nu} \goto {\bm \tau}_{\mu\nu} ={\bm t}^g_{\mu\nu} + {\bm t}^m_{\mu\nu} - \frac{1}{\k} {\cal F}^{L}_{\mu\nu} + \frac{1}{2\k}\frac{\delta \bar \lag_{\sst G}}{\delta
\bar {g}_{\alf\beta}}\l(\varkappa_{\alf\beta}\bar g_{\mu\nu} + \l.\varkappa_{\alf\beta}\r|_\mu^\sig\bar g_{\nu\sig}\r) \,.
 \label{EM1+}
 \ee
 Here, the first term is the energy-momentum for a free gravitational field related to the gravitational part of the Lagrangian (\ref{lag}). The second term is the energy-momentum for matter fields related to the matter part of the Lagrangian (\ref{lag}). The last term describes interaction of the gravitational field $\varkappa_{\alf\beta}$ with a curved background described by the metric $\bar g_{\mu\nu}$. However, the role of the third term is not clear. Let us clarify it. Using the definitions (\ref{PhiL-q}) and (\ref{em-q}), combining the second and third terms, and taking into account (\ref{ddd}) and (\ref{bbb}), one transforms (\ref{EM1+}) to
\be
{\bm \tau}_{\mu\nu} ={\bm t}^g_{\mu\nu} + \delta{\cal T}_{\mu\nu}+ \frac{1}{2\k}\frac{\delta \bar \lag_{\sst G}}{\delta
\bar {g}_{\alf\beta}}\l(\varkappa_{\alf\beta}\bar g_{\mu\nu} + \l.\varkappa_{\alf\beta}\r|_\mu^\sig\bar g_{\nu\sig}\r)   \,,
 \label{EM1++A}
 \ee
 where $\delta {\cal T}_{\mu\nu} = {\cal T}_{\mu\nu} -  \bar{\cal T}_{\mu\nu}$ describes a perturbation of the matter energy-momentum of the gravity theory in (\ref{ddd}) with respect to the background one in (\ref{bbb}). If we examine a concrete solution to the field equations
(\ref{PERTmunu}), we can turn to the energy-momentum (\ref{EM1A}),  instead of (\ref{EM1+}) or (\ref{EM1++A}). Thus, it is
\be
{\bm \tau}^{\mu\nu} = 2\l.\l(
\bar\nabla_{\rho\lam}  {\bm \omega}_{1}^{\mu\nu|\rho\lam} +
 {\bm \omega}_1^{\mu\tau|\rho\lam} \bar R^{\nu}{}_{\lam\tau\rho} + \frac{1}{3} {\bm \omega}_{1}^{\nu\lam|\tau\rho} \bar R^{\mu}{}_{\tau\rho\lam} \r)\r|_{(\ref{PERTmunu})} \,.
\m{EM1+++}
\ee

After that let us consider the current (\ref{Jmu1}) and transform it to
 \be
{\bm i}_1^\mu \goto {\cal I}^\mu  =   {\bm \tau}^{\mu\nu}\xi_\nu - {\bm z}^\mu_{1}\,.
 \m{current}
 \ee
Then the identity (\ref{BDi-1}) becomes the physically sensible conservation law:
 \be
\bar\nabla_\mu  {\cal I}^\mu =\di_\mu  {\cal I}^\mu = 0\,.
 \m{nablacurrent+}
 \ee
 At last, substituting potentials of a concrete solution into the superpotential expression (\ref{BDi-sup}), we denote it as
 \be
{\bm i}_1^{\mu\nu} \goto {\cal I}^{\mu\nu}  \,.
 \m{current++}
 \ee
Then the identity (\ref{BDi-sup}) turns into the physically sensible conservation law:
 \be
 {\cal I}^\mu = \bar\nabla_\nu {\cal I}^{\mu\nu} = \di_\nu {\cal I}^{\mu\nu}
\, .\m{BDi-sup+}
 \ee

All the above has been constructed for arbitrary curved backgrounds, even
non-vacuum ones. However,  the case of a vacuum background,
 \be
 \bar{\lag}_{m} = 0  \goto \qquad \frac{\delta \bar\lag_{\sst G}}{\delta \bar {g}_{\mu\nu}} = 0, \qquad
\bar{\cal T}_{\mu\nu}=0,\qquad  {\cal F}^{L}_{\mu\nu}=0\,,
\m{vacuumB}
 \ee
is of  special interest; we describe this below. The field equations (\ref{PERTmunu}) go over to
\be
{\cal C}^{L}_{\mu\nu} = \k\l({\bm t}^g_{\mu\nu} + {\bm t}^m_{\mu\nu} \r)\,.
\m{EqsVacuum}
\ee
Under the conditions (\ref{vacuumB}) the expression (\ref{Linear-Gen}) becomes
\be
{\cal C}_\sigma^{L\mu} =2\k \l(
\bar\nabla_{\rho\lam}  {\bm \omega}_{1\sig}{}^{\mu|\rho\lam} +
 {\bm \omega}_1^{\mu\tau|\rho\lam} \bar R_{\sig\lam\tau\rho} + \frac{1}{3} {\bm \omega}_{1\sig}{}^{\lam|\tau\rho} \bar R^{\mu}{}_{\tau\rho\lam} \r)\,.
  \m{Linear-Vac}
 \ee
 In this case, the total energy-momentum (\ref{EM1+}) becomes
\be
{\bm \tau}_{\mu\nu} = {\bm t}^g_{\mu\nu} + {\bm t}^m_{\mu\nu} \,.
\m{EM1++B}
\ee
Again we highlight the nice property that the expression (\ref{Linear-Vac})  depends on the quantity ${\bm \omega}_{1}^{\sig\lam|\tau\rho}$ only, defined as in (\ref{NL1}) for the Lagrangian $\lag_1$.
Next, in the case (\ref{vacuumB}), making use of the Killing vectors $\xi^\alf =\bar\xi^\alf$, the current (\ref{current}) transforms to the standard
form:
  \be
{\cal I}^\mu = {\bm \tau}^{\mu\nu}\bar\xi_\nu\, .
 \m{current+1}
 \ee
Assuming {\em arbitrary} Killing vectors $\xi^\alf=\bar\xi^\alf$ in the identity
(\ref{BDi-1}), one obtains the identity
 \be
 \bar\nabla^\mu{\cal C}^{L}_{\mu\nu} \equiv 0
 \m{DGL-q}
 \ee
 under the conditions (\ref{vacuumB}). Then, from the field equations
(\ref{EqsVacuum}) one obtains the differential conservation law for the total energy-momentum tensor density (\ref{EM1++B}):
 \be
 \bar\nabla_\nu {\bm\tau}^{\mu\nu}= 0\,.
 \m{CLfor-t}
 \ee
 %


 \vspace{0.3cm}
\sect{Conserved quantities in the Lovelock theory}

In this section, based on the results of the previous two sections, the Lagrangian based field-theoretic reformulation of Lovelock gravity is provided; for  more detail see \cite{Petrov_2019}. After that we construct conserved currents and superpotentials. Let us concretize Lagrangian (\ref{lag-g}) for the Lovelock theory:
 \be
 \lag_{\sst D}(g,\Phi) = - \frac{1}{2\k}\lag_{\sst L}(g_{\mu\nu}) + \lag^{M}(g_{\mu\nu},\Phi^A)\,,
 \m{lag-ll}
 \ee
where $\k = 2\Omega_{D-2}G_D> 0$  with $G_D$ being the
$D$-dimensional Newton's constant and $\Omega_{D-2}$ being the area  of a $(D-2)$-dimensional sphere with unit radius. Lovelock \cite{Lovelock} has required the following  for the Lagrangian: it must describe  a covariant metric theory in $D$-dimensional space-time and  the field equations from varying $\lag_{\sst L}(g_{\mu\nu})$ must be of  second order only. The unique possibility to satisfy these requirements is  a sum of polynomials in the Riemann tensor in the form:
\begin{equation}
\lag_{\sst L}(g_{\mu\nu})=\sqrt{-g}\sum_{p=0}^{m}\frac{\alf_p}{2^p} \delta _{\lbrack i_{1}i_{2}\cdots
i_{2p}]}^{[j_{1}j_{2}\cdots j_{2p}]}\,{R}_{j_{1}j_{2}}^{i_{1}i_{2}}\cdots {R}_{j_{2p-1}j_{2p}}^{i_{2p-1}i_{2p}}  \,,
\label{Lovelock-lag}
\end{equation}
where $\alf_p$ are coupling constants, $m=\l[(D-1)/2\r]$, and the totally-antisymmetric Kronecker delta is
\begin{equation}
\delta _{\left[ \mu _{1}\cdots \mu _{q}\right] }^{\left[ \nu _{1}\cdots \nu
_{q}\right] }:=\left\vert
\begin{array}{cccc}
\delta _{\mu _{1}}^{\nu _{1}} & \delta _{\mu _{1}}^{\nu _{2}} & \cdots &
\delta _{\mu _{1}}^{\nu _{q}} \\
\delta _{\mu _{2}}^{\nu _{1}} & \delta _{\mu _{2}}^{\nu _{2}} &  & \delta
_{\mu _{2}}^{\nu _{q}} \\
\vdots &  & \ddots &  \\
\delta _{\mu _{q}}^{\nu _{1}} & \delta _{\mu _{q}}^{\nu _{2}} & \cdots &
\delta _{\mu _{q}}^{\nu _{q}}%
\end{array}%
\right\vert \m{determinant}\,.
\end{equation}%
The term of zeroth order, $p=0$, gives a `bare' cosmological constant $\Lambda_0$ with $\alf_0 =-2\Lambda_0$. The  first order term $p=1$ is the Hilbert (Ricci scalar) term with $\alf_1 = 1$.  The second order term $p=2$ is the Gauss-Bonnet term with a coupling constant $\alf_2 $ unspecified.

The Lovelock field equations obtained by varying (\ref{lag-ll}) with respect to $g_{\rho\sig}$ are
\begin{equation}
\frac{\delta\lag_{\sst L}}{\delta g_{\rho\sig}}= \sqrt{-g}g^{\pi\rho}
\sum_{p=0}^{m}\frac{\alf_p}{2^{p+1}} \delta _{\lbrack \pi \nu_{1}\nu_{2}\cdots
\nu_{2p}]}^{[\sig \mu_{1}\mu_{2}\cdots \mu_{2p}]}\,{R}_{\mu_{1}\mu_{2}}^{\nu_{1}\nu_{2}}\cdots {R}_{\mu_{2p-1}\mu_{2p}}^{\nu_{2p-1}\nu_{2p}}  = - \kappa {\cal T}^{\rho\sig} \,.  \label{Lovelock-eqs+}
\end{equation}
It is just an instance of  the equations (\ref{ddd}).

 Now let us derive a Lagrangian linear in metric perturbations (\ref{Lag-1}) for the Lovelock theory. We use the background version of the Lovelock Lagrangian, $\bar\lag_{\sst L}$, and the background version of the Lagrange derivative in (\ref{Lovelock-eqs+}). Then,
the auxiliary Lagrangian (\ref{Lag-1}) becomes
\begin{equation}
\lag_{{\sst L} 1}= - \frac{1}{2\k}\varkappa_{\rho\sig}\frac{\delta\bar\lag_{\sst L}}{\delta \bar g_{\rho\sig}}=- \frac{\sqrt{-\bar g}}{2\k} \varkappa_\sig^\rho
\sum_{p=0}^{m}\frac{\alf_p}{2^{p+1}} \delta _{\lbrack \rho \nu_{1}\nu_{2}\cdots
\nu_{2p}]}^{[\sig \mu_{1}\mu_{2}\cdots \mu_{2p}]}\,{\bar R}_{\mu_{1}\mu_{2}}^{\nu_{1}\nu_{2}}\cdots {\bar R}_{\mu_{2p-1}\mu_{2p}}^{\nu_{2p-1}\nu_{2p}}   \,.  \label{lock-1}
\end{equation}
Another key quantity of the type (\ref{NL1}) calculated for the Lagrangian (\ref{lock-1}) is
\begin{equation}
{\bm \omega}^{\rho\lambda|\mu\nu}_{{\sst L}1}= - \frac{\sqrt{-\bar g}}{2\k} \varkappa^\alf_\beta
\sum_{p=1}^{m}\frac{p\alf_p}{2^{p+1}} \delta _{\lbrack \alf\phi\psi \nu_{3}\nu_{4}\cdots
\nu_{2p}]}^{[\beta\pi\sigma \mu_{3}\mu_{4}\cdots \mu_{2p}]}\,{\bar R}_{\mu_{3}\mu_{4}}^{\nu_{3}\nu_{4}}\cdots {\bar R}_{\mu_{2p-1}\mu_{2p}}^{\nu_{2p-1}\nu_{2p}}\bar g^{\phi\tau}\bar g^{\psi\kappa}D^{\rho\lambda\mu\nu}_{\pi\sigma\tau\kappa}  \,.  \label{NL1-lock}
\end{equation}
The quantity
\be
D^{\rho\lambda\mu\nu}_{\pi\sigma\tau\kappa} = \half \l(\delta^\rho_\pi\delta^\lambda_\k +\delta^\rho_\k\delta^\lambda_\pi \r)\l(\delta^\mu_\sigma\delta^\nu_\tau +\delta^\mu_\tau\delta^\nu_\sigma \r)
\m{D}
\ee
is obtained after differentiating the Riemannian tensor $\bar R_{\pi\sigma\tau\kappa}$ with respect to $\bar g_{\rho\lambda,\mu\nu}$ and using the index symmetry.

The linear operator (\ref{Linear-Gen}) in the Lovelock gravity acquires the form
\be
{\cal C}^{\sigma\mu}_L =- \frac{1}{2}\bar g^{\sigma\mu}\frac{\delta \bar \lag_{\sst L}}{\delta
\bar {g}_{\rho\tau}}\varkappa_{\rho\tau}+\frac{\delta \bar \lag_{\sst L}}{\delta
\bar {g}_{\rho\sig}}\varkappa_{\rho}^{\mu} +2\k \l(
\bar\nabla_{\rho\lam}  {\bm \omega}_{{\sst L} 1}^{\sig\mu|\rho\lam} +
 {\bm \omega}_{{\sst L} 1}^{\mu\tau|\rho\lam} \bar R^{\sig}{}_{\lam\tau\rho} + \frac{1}{3} {\bm \omega}_{{\sst L} 1}^{\sig\lam|\tau\rho} \bar R^{\mu}{}_{\tau\rho\lam} \r)\, .
 \m{Linear-Genl}
 \ee
Keeping in mind the quantity (\ref{NL1-lock}), we derive the conserved quantities in the Lovelock gravity. The conserved current (\ref{current}) becomes
 \be
{\cal I}^\mu_{\sst L}  =   {\bm \tau}^{\sigma\mu}_{\sst L}\xi_\sigma - {\bm z}^\mu_{\sst L}\,,
 \m{current-l}
 \ee
where the energy-momentum (\ref{EM1+++}) has the form:
 \be
 {\bm \tau}^{\sigma\mu}_{\sst L} = \l. 2\l(
\bar\nabla_{\rho\lam}  {\bm \omega}_{{\sst L} 1}^{\sig\mu|\rho\lam} +
 {\bm \omega}_{{\sst L} 1}^{\mu\tau|\rho\lam} \bar R^{\sig}{}_{\lam\tau\rho} + \frac{1}{3} {\bm \omega}_{{\sst L} 1}^{\sig\lam|\tau\rho} \bar R^{\mu}{}_{\tau\rho\lam} \r)\r|_{(\ref{PERTmunu})}\,,
 \m{theta}
 \ee
and $z$-term is
 \be
   {\bm z}^\mu_{\sst L} =  2\bar\zeta_{\rho\lam}\bar\nabla_\nu
 {\bm \omega}_{{\sst L}1}^{\rho\lam|\mu\nu}- 2{\bm \omega}_{{\sst L}1}^{\rho\lam|\mu\nu}
  \bar\nabla_\nu \bar\zeta_{\rho\lam};\qquad 2\bar\zeta_{\rho\lam} \equiv -\Lix\bar g_{\rho\lam} = 2\bar\nabla_{(\rho}\xi_{\lam)}
  \,.\label{ZDmu-l}
 \ee
The superpotential  (\ref{current++}) related to the Lovelock theory is
\be
{\cal I}^{\mu\nu}_{{\sst L}} = {\textstyle{4\over 3}}\l(
 2\xi_\sig \bar\nabla_\lam  {\bm \omega}_{{\sst L}1}^{\sig[\mu|\nu]\lam}  -
{\bm \omega}_{{\sst L} 1}^{\sig[\mu|\nu]\lam}
\bar \nabla_\lam  \xi_\sig\r)\,.
 \m{Super-l}
 \ee
 We again note the remarkable property:
 \bigskip

 \bit

\item For the Lovelock gravity the conserved current (\ref{current-l}) and the superpotential (\ref{Super-l}) constructed for {\em arbitrary} perturbations on {\em arbitrary} curved backgrounds depend on the quantity (\ref{NL1-lock}) {\em only}.
\eit

It is quite important to present conserved quantities for perturbations on arbitrary curved backgrounds. However, let us turn to vacuum backgrounds.  Recall that the linear operator (\ref{Linear-Genl}) in the case of vacuum background depends on (\ref{NL1-lock}) only; the two first terms disappaer. Besides, the energy-momentum (\ref{theta}) is
 conserved for a vacuum background, see (\ref{CLfor-t}).

Among vacuum backgrounds one of the more popular solutions of  Lovelock gravity is the global maximally symmetric space-time with a negative constant curvature - anti-de Sitter (AdS) space. Therefore, it is important to construct conserved quantities for arbitrary perturbations on such backgrounds.
Let us consider the equations (\ref{Lovelock-eqs+}) as background equations under the vacuum condition (\ref{vacuumB}):
\begin{equation}
\frac{\delta\bar\lag_{\ell}}{\delta \bar g_{\rho\sig}}= \sqrt{-\bar g}\bar g^{\pi\rho}
\sum_{p=0}^{m}\frac{\alf_p}{2^{p+1}} \delta _{\lbrack \pi \nu_{1}\nu_{2}\cdots
\nu_{2p}]}^{[\sig \mu_{1}\mu_{2}\cdots \mu_{2p}]}\,{\bar R}_{\mu_{1}\mu_{2}}^{\nu_{1}\nu_{2}}\cdots {\bar R}_{\mu_{2p-1}\mu_{2p}}^{\nu_{2p-1}\nu_{2p}} = 0\,.  \label{Lovelock-eqs++}
\end{equation}
Let the AdS space be with the Riemannian tensor
\be
\bar R^{\rho\lam}_{\mu\nu} = - \frac{1}{\ell^2_{eff}}\delta^{[\rho\lam]}_{[\mu\nu]}\,,
\m{R_AdS}
\ee
where the quantity $\ell_{eff}$ is called the effective AdS radius and defines a length scale. To find $\ell_{eff}$ one has to substitute (\ref{R_AdS}) into (\ref{Lovelock-eqs++}), obtain
\be
\l.V(x)\r|_{x =\ell_{eff}^{-2} } = \sum^{m}_{p=0}\frac{(D-3)!}{(D-2p-1)!}\alf_p(-1)^{p-1}\l(\ell_{eff}^{-2} \r)^p = 0\,
\m{sum}
\ee
and resolve it with respect to $\ell_{eff}$. The effective cosmological constant is defined as usual,
\be
\Lambda_{eff} = -\frac{(D-1)(D-2)}{2\ell^2_{eff}}\,.
\m{Lambda_eff}
\ee
A space-time with the curvature tensor (\ref{R_AdS}) can be described by the metric:
 \be
 d\bar s^2 = -\Bar fdt^2 + \frac{1}{\bar f}dr^2 +
 r^2\sum_{a,b}^{D-2}q_{ab}dx^adx^b\,; \qquad  \bar f(r) \equiv 1+ \frac{r^2}{\ell^2_{eff}}\,,
 \m{AdS_back}
 \ee
 where  the last term describes $(D-2)$-dimensional sphere of the
radius $r$, and $q_{ab}$ depends on coordinates on the sphere only.

As has been emphasized, the key expression in constructing conserved quantities for perturbations $\varkappa_{\mu\nu}$ on vacuum backgrounds in Lovelock gravity is (\ref{NL1-lock}). Let us derive it explicitly making use of the condition (\ref{R_AdS}) for the AdS background. First, we calculate the first term from the sum in (\ref{NL1-lock}) that corresponds to the Hilbert part of the Lovelock Lagrangian, $\alf_1 = 1$,
\bea
&&{\bm \omega}^{\rho\lambda|\mu\nu}_{{\sst H}1}= - \frac{\sqrt{-\bar g}}{8\k} \varkappa^\alf_\beta
\delta _{\lbrack \alf\phi\psi ]}^{[\beta\pi\sigma ]}\bar g^{\phi\tau}\bar g^{\psi\kappa}D^{\rho\lambda\mu\nu}_{\pi\sigma\tau\kappa} = \nonumber\\
&&- \frac{\sqrt{-\bar g}}{4\k}\l[\bar g^{\mu\nu}\varkappa^{\rho\lam} +\bar g^{\rho\lam}\varkappa^{\mu\nu} -\bar g^{\rho(\mu}\varkappa^{\nu)\lam} -\bar g^{\lam(\mu}\varkappa^{\nu)\rho} - \varkappa\l(\bar g^{\mu\nu}\bar g^{\rho\lam} - \bar g^{\rho(\mu}\bar g^{\nu)\lam} \r)\r] \,.  \label{NL1-H}
\eea
Using this expression and the condition (\ref{R_AdS}), and making use of the standard relation,
\begin{equation}
\delta _{\left[ \mu _{1}\cdots \mu _{2k}\mu _{2k+1}\cdots \mu _{2p}\right] }^{\left[ \nu
_{1}\cdots \nu _{2k}\nu _{2k+1}\cdots \nu _{2p}\right] }\,\delta _{\nu _{2k+1}}^{\mu
_{2k+1}}\cdots \delta _{\nu _{2p}}^{\mu _{2p}}=\frac{\left( D-2k\right) !}{%
\left( D-2p\right) !}\,\delta _{\left[ \mu _{1}\cdots \mu _{2k}\right] }^{%
\left[ \nu _{1}\cdots \nu _{2k}\right] }\,,
\m{contract}
\end{equation}
we obtain for (\ref{NL1-lock}):
\be
{\bm \omega}^{\rho\lambda|\mu\nu}_{{\sst L}1} = {\bm \omega}^{\rho\lambda|\mu\nu}_{{\sst H}1} \l[\sum^{m}_{p=1}p\alf_p(-\ell_{eff}^{-2})^{p-1}\frac{(D-3)!}{(D-2p-1)!} \r]\,.
\m{NL1-AdS}
\ee
The expression in square brackets is defined by the differentiation of (\ref{sum})
\be
 V'(\ell_{eff}^{-2}) = \l.\l(\di_x V(x)\r)\r|_{x =\ell_{eff}^{-2} } = \sum^{m}_{p=1}p\alf_p(-\ell_{eff}^{-2})^{p-1}\frac{(D-3)!}{(D-2p-1)!}\,.
\m{NL1-AdS+}
\ee
Thus, the expression (\ref{NL1-AdS}) shows that all the quantities (\ref{Linear-Genl}) - (\ref{Super-l}), if they are constructed for the AdS background, are proportional to the factor (\ref{NL1-AdS+}). The role of the coefficient (\ref{NL1-AdS+}) is discussed in detail in \cite{Rodrigo_2017,Petrov_2019}.

One now finds that the linear operator (\ref{Linear-Genl}) under the condition (\ref{R_AdS}) becomes
\bea
&& {\cal C}^{\mu\nu}_L = \frac{\sqrt{-\bar g}}{2} V'(\ell_{eff}^{-2})\l[\bar \nabla_{\rho}{}^\mu\varkappa^{\nu\rho} + \bar \nabla_{\rho}{}^\nu\varkappa^{\mu\rho} - \bar \nabla_{\rho}{}^\rho\varkappa^{\mu\nu} - \bar g^{\mu\nu}\bar\nabla_{\rho\lam}\varkappa^{\rho\lam}\r.\nonumber\\ &&\l. + \bar g^{\mu\nu}\bar\nabla_{\rho}{}^\rho\varkappa - \bn^{\mu\nu}\varkappa + \bar g^{\mu\nu} \frac{2\Lambda_{eff}}{D-2}\varkappa -  \frac{4\Lambda_{eff}}{D-2}\varkappa^{\mu\nu}\r] \,.  \label{Linear_final}
\eea
The same expression (\ref{Linear_final}) divided by $\k$ is, in fact, the energy-momentum ${\bm \tau}^{\mu\nu}_{\sst L}$ in (\ref{theta}). Of course, it is conserved (see (\ref{CLfor-t}):
\be
\bar\nabla_\nu{\bm \tau}^{\mu\nu}_{\sst L} = 0.
\m{tau_conserve}
\ee
For (\ref{NL1-AdS}) with (\ref{NL1-H}) the conserved current (\ref{current-l}) is calculated by making use of ${\bm \tau}^{\mu\nu}_{\sst L}$ and with $z$-term (\ref{ZDmu-l}) that can be easily found. For (\ref{NL1-AdS}) with (\ref{NL1-H}) the superpotential (\ref{Super-l}) becomes
\be
{\cal I}^{\mu\nu}_{\sst L} = \frac{\sqrt{-\bar g}}{\k} V'(\ell_{eff}^{-2})
 \l[\xi_\rho \bar\nabla^{[\mu}\varkappa^{\nu]\rho} - \xi^{[\mu} \bar\nabla_\rho \varkappa^{\nu]\rho}  + \xi^{[\mu} \bar\nabla^{\nu]}\varkappa
+ \varkappa^{\rho[\mu}\bar\nabla^{\nu]}\xi_\rho + \half \varkappa \bar\nabla^{[\mu} \xi^{\nu]}\r]\, .
 \m{DTsuperpotential}
 \ee

 Recall that here, in the framework of the Lovelock gravity, we apply the {\em Lagrangian-based} method only. To see its advantages, one has to compare our method with others. Possibly the most fruitful and popular method is the approach by Deser and Tekin and their coauthors. They apply the Abbott and Deser procedure in 4D GR \cite{AbbottDeser82} in metric of higher curvature gravity theories in $D$ dimensions; this is called as the ADT approach. Its development and many applications have many very important results; for example, in the framework of any generic $f(Riemann)$, including Lovelock theory, ADT charges have been constructed  \cite{DT6}.  For a broad outline see the recent review \cite{DT7}.
Concerning the Lovelock theory, the ADT method has been developed for AdS backgrounds and using the Killing vectors only. Our approach permits construction of conserved quantities for arbitrary curved backgrounds and arbitrary displacement vectors. It is important to stress that the superpotential (\ref{DTsuperpotential}) constructed for arbitrary displacement vectors coincides with the ADT related superpotential. However the ADT method does not permit one to construct (\ref{DTsuperpotential}) because there is no a possibility to construct the current of the type (\ref{current-l}). Indeed, for Killing vectors $z$-term does not exist, unlike in (\ref{current-l}). For a detailed comparison with the ADT method, see \cite{Petrov_2019}.


\vspace{0.3cm}
\sect{The mass of the Schwarzschild-like black hole and future applications}

In the present section, to apply the above results we calculate the mass for static black holes in the Lovelock gravity. We use the formulae given in the paper \cite{Kofinas_Olea_2007}. Let us derive the Schwarzschild-like metric:
 \be
 d s^2 = -fdt^2 + \frac{1}{f}dr^2 +
 r^2\sum_{a,b}^{D-2}q_{ab}dx^adx^b\, .
 \m{AdS}
 \ee
The function $f$ must satisfy the equation
 \be
 \sum^{m}_{p=0}\frac{\alf_p}{(D-2p-1)!}\l(\frac{1-f}{r^2} \r)^p = \frac{\mu}{(D-3)!\,r^{D-1}}\,.
 \m{EQ_rr}
 \ee
It is a result of integration of the $rr$-component of the Lovelock vacuum equations with the constant of integration $\mu$. For the black hole solution one has to find the event horizon $r_+$ that is the largest solution of the equation $f(r_+) = 0$. We assume that such a solution exists. In \cite{Kofinas_Olea_2007} it is shown that the asymptotic behaviour of $f$ at $r\goto\infty$,
 \be
 f(r)\sim 1 +\frac{r^2}{\ell^2_{eff}} - \frac{1}{V'(\ell_{eff}^{-2})}\frac{\mu}{r^{D-3}}\,,
 \m{f_goto}
 \ee
 occurs. Comparing it with (\ref{AdS_back}), one has
 \be
  \Delta f = f(r) - \bar f(r) \sim - \frac{1}{V'(\ell_{eff}^{-2})}\frac{\mu}{r^{D-3}}\,.
 \m{f_delta}
 \ee
 As a result, one has for the behaviour of perturbations,
 \be
 \varkappa_{00} \sim -\Delta f,\qquad \varkappa_{11} \sim -\Delta f/\bar f^2\,,
 \m{h_00_11}
 \ee
in the necessary order of approximation.

To calculate the mass for the black hole solution (\ref{AdS}) with the AdS asymptotic (\ref{AdS_back}), one has to use the Killing vector $\bar\xi^\alf = \{-1,0,0,0 \}$ and $01$-component of the superpotential (\ref{DTsuperpotential}) with the appropriate order of approximation for the perturbations (\ref{h_00_11}):
\be
{\cal I}^{01}_{\sst L} \sim -\frac{\sqrt{-\bar g}}{\k} V'(\ell_{eff}^{-2}) \frac{D-2}{2r} \Delta f\,.
\m{SUP_01}
\ee
Substituting (\ref{f_delta}) and taking into account $\sqrt{-\bar g} = r^{D-2}\sqrt{\det q_{ij}}$, one obtains
\be
M = \lim_{r \goto \infty} \oint d x^{D-2} {\cal I}^{01}_{\sst L} =  \frac{D-2}{2\k }\mu\oint d x^{D-2}\sqrt{\det q_{ij}}= \frac{D-2}{4 G_D}\mu  \,,
\m{M}
\ee
which is the standard result for the mass obtained by various methods.

In future work, the above-listed advantages (use of arbitrary displacement vectors and arbitrary curved backgrounds), not available using other methods, motivate the study of solutions in Lovelock gravity using a background metric. The Lovelock theory, currently quite popular, is the most natural generalization in higher dimensions. There are also arguments that only a so-called {\em pure} Lovelock gravity leads to acceptable  equations in higher dimensions, see, for example, \cite{Dadhich+_2012,Dadhich+_2013,Dadhich_2016}. Pure Lovelock gravity is characterized by only one  term  from the sum of all the terms in the total Lovelock Lagrangian (\ref{Lovelock-lag}), say, $\alf_{p^*}\neq 0$ with the unique $p^*$ only, whereas all other $\alf_{p}= 0 $, including $\alf_{0}= 0 $ and $\alf_{1}= 0 $.
 Keeping in mind the interest in pure Lovelock gravity, we plan to apply our results to study solutions of this theory obtained in \cite{Dadhich+_2013}. The first one represents collapsing inhomogeneous dust. In our opinion, it is quite important to examine the stability problem for this solution. In other words, one needs to study perturbations and their characteristics and evolution using this solution as a background. Because this background is non-vacuum (with matter), our approach looks very appropriate. On the other hand, approaches constructed for maximally symmetric backgrounds, such as ADT, cannot be used in this case. The second solution in \cite{Dadhich+_2013} represents the Vaidya-type collapsing/radiating model with light-like matter (null dust). To understand this model more deeply, it is important to study densities of conserved quantities measured by a system of observers. Our method is quite appropriate for such a study. Indeed, the field-theoretic method has been elaborated from the start for studying local characteristics; second, in constructing the aforementioned local densities, proper vectors of observers (which are not Killing vectors in general) have to be used---which is just what our formalism permits.

Finally, it has been noted in \cite{Dadhich+_2012,Dadhich+_2013,Dadhich_2016} that pure Lovelock gravity in even dimensions has properties  very close to those in 4-dimensional Einstein theory. Therefore, it could be interesting to represent, for example, 6-dimensional pure Lovelock theory with $\alf_2 \neq 0$ only in the field-theoretic form and to compare it with the field-theoretic reformulation of 4-dimensional Einstein theory that already has been developed in detail \cite[chapter 2]{Petrov+_2017}.


\vspace{0.3cm}
\sect{Modifications of the field-theoretic method, massive gravitons, and spinors}

The field-theoretic method is developed for both in GR and other metric theories.  It also  permits and perhaps suggests the construction of alternative theories of gravity.  From the standpoint of particle physics, in which one routinely thinks of relativistic field theory terms of a taxonomy of particle/field spins and masses (associated with Wigner and others), perhaps the most natural modification of General Relativity might seem to be the introduction of a graviton mass term. An early effort was due to Fierz and Pauli \cite{FierzPauli,PauliFierz,Fierz,Fierz2}.  They recognized that the massless case gives the linear approximation to General Relativity, permitting the identification of Einstein's theory as a theory of interacting massless spin $2$ particles/fields.  They also noted a connection between masslessness and gauge freedom for spins $\geq 1$, found the spin $2$ energy \emph{density} to be gauge dependent though the total energy was gauge invariant (akin to results familiar from GR), noted the mathematical possibility of distinct masses for the expected spin $2$ and perhaps unexpected spin $0$ gravitons included in the formalism, and showed that avoiding negative energy for the spin $0$ (spatial scalar) degree of freedom required tuning the relative coefficients so that the mass of the spin $0$ degree of freedom became infinite.  This occurs not by putting an infinite coefficient in front of the $trace^2$ term (which would seem not to make sense), but by making the derivative terms in the would-be Klein-Gordon equation satisfied by the trace of the gravitational potential disappear due to a vanishing coefficient.  From that point it was often (though not always \cite{OP,FMS}) concluded that only the pure spin $2$ theories were of physical interest, because the negative-energy spin $0$ degrees of freedom would be expected to imply catastrophic instability under quantization:  the conservation of energy-momentum would not prevent the spontaneous development of nothing into something and anti-something.

The development of massive gravity made considerable progress in the 1960s \cite{OP,FMS}.  Unfortunately much of this work was largely forgotten  and hence was reinvented  in the 2010s.  Ogievetsky and Polubarinov considered a spin limitation principle that eliminated wrong-sign spin $1$ and one spin $0$ degree of freedom and inferred Einstein's equations for the massless case and a $2$-parameter family of inequivalent massive generalizations thereof.  In the process they also invented nonlinear group realizations (using non-integral powers of the metric tensor, which have nonlinear coordinate transformation laws; the non-integral powers were defined using a binomial series expansion) and subsumed spinors (almost) into the realm of entities with coordinate transformation properties and no additional local Lorentz gauge freedom \cite{OP,OPspinor,PittsSpinor}.  This supposedly impossible result---which was partly anticipated by Bryce Seligman DeWitt \cite{DeWittDissertation,DeWittSpinor}, who seemed not to grasp the depth of his own work on this point---can be understood in a way that many people continue in effect to reinvent it by the back door, namely, by imposing a symmetric gauge condition on the tetrad, thus fixing the local Lorentz gauge freedom and turning the spinor's coordinate scalar, Lorentz spinor behavior into a nonlinear metric-dependent coordinate spinor transformation rule \cite{vanNGauge,WoodardSymmetricTetrad,BilyalovSpinors,PittsSpinor,DeffayetSymmetricTetrad}.

 As Bilyalov notes in effect,\footnote{Bilyalov actually introduces a matrix $T$ that swaps a pair of coordinates and flips the sign of one of them \cite{BilyalovConservation}. Interpreting $T$ as a coordinate transformation brings the formalism more nearly into the realm of classical geometrical objects.  Flipping the sign of a coordinate seems unnecessary for our purposes.} there are non-perturbative coordinate issues.  Such issues can hardly be noticed if one works perturbatively using $x^4=ict$ as Ogievetsky and Polubarinov do; this fact might tend to vindicate the proposal to put $x^4=ict$ ``to the sword'' \cite[p. 51]{MTW}.  Imaginary time coordinates are clearly not optimal for introducing null coordinates $\sim t \pm x.$ If one works nonperturbatively and introduces the signature matrix $\eta_{AB} =diag(-1,1,1,1),$ most but not all such limitations disappear.
One can isolate those that remain into the transformations that simply swap a time coordinate and a space coordinate along the lines of $\langle t, x, y, z\rangle \leftrightarrow \langle x, t, y, z\rangle$. The nonlinear spinor formalism does not readily permit such transformations, because they tend to create (or destroy!) negative eigenvalues of the matrix $ \sum_{\nu=N} g_{\mu\nu} \eta^{NA}$, negative eigenvalues being the obstruction to taking the principal square root \cite{HighamRoot,HighamRoot87}, which becomes symmetric when an index is moved with $diag(-1,1,1,1).$  If one takes for granted that arbitrary coordinates including $\langle x, t, y, z\rangle$ must be admissible, then this nonperturbative issue looks  significant \cite{Bourguignon}.  Alternatively,  one can  allow the extent of coordinate freedom to be sensitive to the presence or absence of spinors \cite{PittsSpinor}.  Clearly no one would \emph{invent} tensor calculus in order to permit the transformations $\langle t, x, y, z\rangle \leftrightarrow \langle x, t, y, z\rangle$, so a formalism that omits some of this freedom seems satisfactory. The use of a background metric tensor, rather than the matrix $diag(-1,1,1,1)$, permits even coordinates such as $\langle x, t, y, z\rangle.$  Regarding the complications of dealing with spinors and two different metrics, see the discussion early in this review.

The field-theoretic formalism above provides more than one way to treat massive gravity.  One natural way uses a highly symmetric (often flat) background metric.  Another approach makes use of ``clock fields,'' which are the preferred coordinates (often Cartesian) turned formally into dynamical scalar fields through ``parametrization''
 \cite{Kuchar73,DelbourgoSalam,FronsdalMass1,Schmelzer,Arkani,MassiveGravity1,PittsArtificial}.  (The scalars' Euler-Lagrange equations impose no new restrictions.)  This technique proves useful in understanding observables in Hamiltonian General Relativity \cite{ObservablesEquivalentCQG,ObservablesLSEFoP} and in making sense of causality with a physically real and indirectly observable background metric \cite{Pitts_Schive_2004,MassiveGravity1}.\footnote{These works also compare our  notion of $\eta$-causality to the ``causality principle'' imposed by \emph{fiat} in the context of de Donder harmonic gauge fixing in the tradition of the Relativistic Theory of Gravity (\emph{e.g.}, \cite[chapter 6]{LogunovBook}), which came to be  built around the Freund-Maheshwari-Schonberg field equations \cite{FMS}.}

One question rarely considered in the literature but of considerable interest is whether the viciousness of a negative-energy spin $0$ field/particle is a distinctly quantum result, or is it already true in the classical theory?  A realistic answer to this question is likely to require, and is undoubtedly assisted by, numerical simulations, in light of the nonlinearity of the field equations.  In this light the work of  Babak and Grishchuk  \cite{BabakGrishchuk,BabakGrishchuk1} is of considerable value. One also notes work by mathematicians on Hamiltonian field theory that conspicuously fails to exclude negative energies and avoids expecting catastrophe (\emph{e.g.}, \cite{BambusiGrebertTame}.  Issues of resonance are crucial \cite{WeilandWilhelmsson}.  In that regard, the ability to tune the ratio of the scalar and tensor graviton masses \cite{OP,BabakGrishchuk1,PetrovMass,MassiveGravity1,MassiveGravity2,MassiveGravity3} is significant.

As  a preliminary manner, one can recall the state of development of gravitational energy-momentum pseudotensors in the early 1950s.  The Einstein pseudotensor depends on first derivatives of the metric only, but is not symmetric and yields awkward results in non-Cartesian coordinates.  It is awkward, though possible, to define energy-momentum conservation with it \cite{BergmannThomson}.  The Belinfante symmetrized energy-momentum of Papapetrou \cite{Papapetrou} was delayed in its appearance by World War 2 and was not widely known.
An energy-momentum pseudotensor with second derivatives might easily lack the positivity properties that one seeks for the energy density or at least the total energy.  The Landau-Lifshitz pseudotensor, which is symmetric (facilitating conservation of angular momentum), also has no second derivatives; this was progress.  Goldberg noted the possibility of analogs of the Landau-Lifshitz pseudotensor, which, however, all have second derivatives in the symmetric contravariant case; some mixed (contravariant-covariant) entities of arbitrary weight lack second derivatives \cite{GoldbergConservation}. If one wants to avoid the use of $diag(-1,1,1,1)$ or some analogous device (such as a background metric), then a mixed pseudotensor does not readily yield a symmetric contravariant one that is still conserved.  Goldberg discusses some unattractive consequences of the fact that the Landau-Lifshitz pseudotensor is of density weight $2$, not of weight $1$ as one would prefer, under affine coordinate transformations.

One way to  preserve the virtues of symmetry and the absence of second derivatives in an energy-momentum complex is to introduce a background metric and use it to re-weight  and more generally covariantize the Landau-Lifshitz entity.   This was achieved by  Babak and
Grishchuk  \cite{BabakGrishchuk} as a development of the field  approach \cite{GPP}. They consider perturbations in the Minkowski space only, although in arbitrary curved coordinates.
Making use of the definition (\ref{(a2.4)}), let us represent the expression (\ref{DeltaDef}) through the gravitational variables $\goh^{\mu\nu}$:
 \bea \Delta^\lam_{\mu\nu}& \equiv&
\frac{1}{2\sqrt{-g}}\l[ g_{\mu\rho}\bn_\nu\goh^{\lam\rho} +
g_{\nu\rho}\bn_\mu\goh^{\lam\rho} - g_{\mu\alf}g_{\nu\beta}
g^{\lam\rho} \bn_\rho \goh^{\alf\beta} \r. \nonumber\\& +&\l.
\half\l(g_{\alf\beta}\delta^\lam_\mu\bn_\nu\goh^{\alf\beta}  +
g_{\alf\beta}\delta^\lam_\nu\bn_\mu\goh^{\alf\beta} -
g_{\alf\beta} g_{\mu\nu} g^{\lam\rho} \bn_\rho \goh^{\alf\beta}
\r)\r] \m{Delta-l}
 \eea
where $g_{\mu\nu}$, $g^{\mu\nu}$ and $\sqrt{-g}$ are thought as
dependent on the definition (\ref{(a2.4)}). Substituting (\ref{Delta-l}) into  the definition (\ref{(2.20')}) with (\ref{(2.20'')}), we select the part of ${\bm t}^g_{\mu\nu}$ depending on the second derivatives of
$\goh^{\mu\nu}$ explicitly. After making use of the field equations (\ref{(a2.23)}) in Minkowski space
the second derivatives are left anyway, however only minimally, like below
 \bea
 {\bm t}_g^{\mu\nu}& =& {\bm t}_{\rm red}^{\mu\nu} +
Q^{\alf\beta\mu\nu} ({\bm t}^m_{\alf\beta} - \half \bar g_{\alf\beta}
{\bm t}^m_{\rho}{}^\rho) + (2\sqrt{-\bar g})^{-1}
\bn_{\alf\beta}(\goh^{\alf(\mu} \goh^{\nu)\beta}- \goh^{\mu\nu} \goh^{\alf\beta}); \m{Transformed-tg}\\
Q^{\alf\beta\mu\nu}& \equiv & (\sqrt{-\bar g})^{-2}\l[\goh^{\alf(\mu}
\bar{\gog}^{\nu)\beta} + \goh^{\beta(\mu} \bar{\gog}^{\nu)\alf}+ \goh^{\alf(\mu} {\goh}^{\nu)\beta}-  \half \bar{\gog}^{\mu\nu}\goh^{\alf\beta}- \half  \goh^{\mu\nu}\l(\bar{\gog}^{\alf\beta}+\goh^{\alf\beta}\r)\r]. \m{Q}
 \eea
The first term in (\ref{Transformed-tg}) is the reduced part depending on the first derivatives only,
 \bea
\hat t_{\rm red}^{\mu\nu}& =&  \frac{1}{4\k\sqrt{-\bar g} }\l[ 2 \bn_\rho \goh^{\mu\nu} \bn_\sig \goh^{\rho\sig} - 2 \bn_\alf
\goh^{\mu\alf} \bn_\beta \goh^{\nu\beta} + g_{\alf\beta}\l( 2  g^{\rho\sig} \bn_\rho \goh^{\mu\alf}\bn_\sig \goh^{\nu\beta}  +  g^{\mu\nu}
\bn_\sig \goh^{\alf\rho} \bn_\rho\goh^{\beta\sig}\r)\r. \nonumber \\
&-&\l. 4 g_{\beta\rho} g^{\alf(\mu} \bn_\sig \goh^{\nu)\beta} \bn_\alf \goh^{\rho\sig} + {\txt \frac{1}{4}}(2g^{\mu\delta} g^{\nu\omega} -
 g^{\mu\nu}  g^{\omega\delta} )(2 g_{\rho\alf}  g_{\sig\beta} -  g_{\alf\beta}  g_{\rho\sig}) \bn_\delta \goh^{\rho\sig}
 \bn_\omega \goh^{\alf\beta}\r]\, .
\m{Reduced-tg}
 \eea
The matter part in (\ref{Transformed-tg}) has appeared because the field equations (\ref{(a2.23)}) have been used.

Babak and Grishchuk \cite{BabakGrishchuk} have suggested a way to exclude the second derivatives from the energy-momentum
without changing the field equations. Instead of the Lagrangian (\ref{(a2.16)}) they have suggested the modified one,
 \be
 \lag^g_{\rm mod} = \lag^g + {\bm \Lambda}^{\alf\beta\rho\sig}\bar R_{\alf\rho\beta\sig}\, .
 \m{L-mod}
 \ee
Because a background is represented by the Minkowski space, one has to set $\bar R_{\alf\rho\beta\sig} = 0,$ but not before defining the energy-momentum \cite{RosenfeldStress,Kraichnan}. Then in (\ref{L-mod}) the components of ${\bm \Lambda}^{\alf\beta\rho\sig}$ are an undetermined tensor density depending on $\bar g^{\mu\nu}$ and $\goh^{\mu\nu}$ without their derivatives. Besides, ${\bm \Lambda}^{\alf\beta\rho\sig}$ has the symmetries of the Riemannian tensor $\bar R_{\alf\rho\beta\sig}$:  it satisfies ${\bm \Lambda}^{\alf\beta\rho\sig} = -{\bm \Lambda}^{\rho\beta\alf\sig}= -{\bm \Lambda}^{\alf\sig\rho\beta}= {\bm
\Lambda}^{\beta\alf\sig\rho}$. As a result, the field-theoretic equations for perturbations in Minkowski space (\ref{(a2.23)}) do not change. However, in correspondence  with the modified Lagrangian (\ref{L-mod}),  the modified gravitational energy-momentum tensor density is
 \be \k {\bm t}^{\mu\nu}_{\rm gmod} = \k{\bm t}^{\mu\nu}_g -\bn_{\alf\beta}\l({\bm \Lambda}^{\mu\nu\alf\beta} + {\bm \Lambda}^{\nu\mu\alf\beta}\r)
\m{t-mod}
 \ee
instead of (\ref{(2.20')}). Let us define the initially undetermined quantities
${\bm \Lambda}^{\mu\nu\alf\beta}$. We desire to choose them in a way when the remaining second derivatives in (\ref{Transformed-tg}) are compensated. The unique possibility is ${\bm \Lambda}^{\mu\nu\alf\beta} = \l( \goh^{\alf\nu}\goh^{\beta\mu}-\goh^{\alf\beta}\goh^{\mu\nu}\r)/4\sqrt{-\bar g} $.

Thus the equations (\ref{(a2.23)}), being unchanged, are rewritten in another form:
 \bea
 {\cal G}_{\rm mod}^{\mu\nu}& \equiv &
{\cal G}_L^{\mu\nu} - 2\bn_{\alf\beta}{\bm \Lambda}^{(\mu\nu)\alf\beta}  \equiv ({\sqrt{-\bar g}})^{-1}\bn_{\alf\beta}\l[(\bar{\gog}^{\mu\nu} + \goh^{\mu\nu}) (\bar{\gog}^{\alf\beta} + \goh^{\alf\beta})- (\bar{\gog}^{\mu\alf} + \goh^{\mu\alf})(\bar{\gog}^{\nu\beta} + \goh^{\nu\beta})\r]\nonumber\\
&=& \k\l({\bm t}_{\rm gmod}^{\mu\nu} +  {\bm t}_m^{\mu\nu}\r)\equiv \k{\bm t}_{\rm mod}^{\mu\nu}\,.
\m{(a2.23-mod)}
 \eea
We see that the left hand side is no longer linear in $\goh^{\mu\nu}$, but its divergence is identically equal to zero. Then, of course,  $\bn_\nu{\bm t}_{\rm mod}^{\mu\nu} = 0$. Reducing ${\bm t}_{\rm mod}^{\mu\nu}$ by making use of the field equations, we rewrite (\ref{(a2.23-mod)}) as
 \be
 {\cal G}_{\rm mod}^{\mu\nu} = \k \l[{\bm t}_{\rm red}^{\mu\nu} +
Q^{\alf\beta\mu\nu} ({\bm t}^m_{\alf\beta} - \half \bar g_{\alf\beta}
{\bm t}^m_{\rho}{}^\rho)
 +  {\bm t}^m_{\mu\nu}\r]\equiv \k {\bm t}_{\rm new}^{\mu\nu}\,
\m{(a2.23-mod-red)}
 \ee
Thus, finally one can see that the energy-momentum tensor density in (\ref{(a2.23-mod-red)}) has only of first
derivatives of gravitational variables and again $\bn_\nu{\bm t}_{\rm new}^{\mu\nu} = 0$.

Multiplying (\ref{(a2.23-mod-red)}) by $\sqrt{-\bar g}$, and
 using the identification (\ref{(a2.4)}), the definition
(\ref{(2.20+)}) for the flat background and the definition
(\ref{Q}), in the Lorentzian coordinates, one easily gets
 \be
\half\di_{\alf\beta}\l(\gog^{\mu\nu}\gog^{\alf\beta}-
 \gog^{\mu\alf}\gog^{\nu\beta}\r) =
 \k(-g)\l(t_{\sst LL}^{\mu\nu}  +T^{\mu\nu}\r)\, .
 \m{LLT}
 \ee
After substituting  $\k T^{\mu\nu}$ from the Einstein equations (\ref{(a2.2+)}), this equation reduces to the identity.
One finds that $(-g)t_{\sst LL}^{\mu\nu}$ is the Landau-Lifshitz
pseudotensor \cite{LL}. Thus ${\bm t}_{\rm red}^{\mu\nu}$ is the covariantized Landau-Lifshitz's pseudotensor $(-g)t_{\sst LL}^{\mu\nu}/\sqrt{-\bar g}$.

Of course, the gauge transformations (\ref{(5.9)}) and (\ref{(5.9)+}) with $\bar \Phi^A \equiv 0$ are the gauge transformations for the theory with the equations (\ref{(a2.23-mod-red)}). The energy-momentum tensor is gauge invariant up to a divergence, however now in the form:
 \be
  \k{{\bm t}}'^{\rm new}_{\mu\nu} =
  \k{\bm t}^{\rm new}_{\mu\nu}
  + {\cal G}^{\rm mod}_{\mu\nu}(\goh'-\goh).
  \m{tei-gaugeBG+}
  \ee
  on the field equations.
The same as the transformations (\ref{tei-gauge}) and
(\ref{tei-gaugeBG}) the transformations (\ref{tei-gaugeBG+}) express
the non-localization problem of energy and other conserved quantities in GR.

This work on gravitational energy-momentum has been  in the framework of Einstein's equations. However, this reconstruction has been used by Babak and Grishchuk to create a variant of gravity theory with non-zero masses of gravitons \cite{BabakGrishchuk1}. They have assumed that the Lagrangian may also include an additional term similar to the one in (\ref{L-mod}). Let the quantity $\~R_{\alf\rho\beta\sig}$ be the curvature tensor of an abstract space-time with a constant non-zero curvature:
$\widetilde R_{\alf\rho\beta\sig}  = K\l(\widetilde g_{\alf\beta} \widetilde g_{\rho\sig}- \widetilde g_{\alf\sig}\widetilde g_{\rho\beta}\r) $
where the dimensionality of $K$ is $[length]^{-2}$. If one adds ${\bm \Lambda}^{\alf\beta\rho\sig}\widetilde R_{\alf\rho\beta\sig}$
with ${\bm \Lambda}^{\mu\nu\alf\beta} = (4\sqrt{-\widetilde
g})^{-1}\l( \goh^{\alf\nu}\goh^{\beta\mu}-\goh^{\alf\beta}\goh^{\mu\nu}\r) $, changing $\widetilde g^{\mu\nu} \goto \bar g^{\mu\nu}$, then the additional term in the Lagrangian
(\ref{L-mod}) is $ \half  (\sqrt{-\bar g})^{-1}K\l(\goh^{\alf\beta}\goh_{\alf\beta}-\goh^{\alf}{}_{\alf} \goh^{\beta}{}_{\beta}\r)$.
Of course,  the related to such a Lagrangian theory is not GR. However one recognizes in this term the Fierz-Pauli mass-term \cite{FierzPauli} at lowest order. Generalizing it, Babak and Grishchuk present a 2-parametric family of theories with the additional mass terms in the gravitational Lagrangian (\ref{L-mod}):
 \be
 \lag^g_{\rm mass} = \lag^g_{\rm mod}+ (\sqrt{-\bar
g})^{-1}\l(k_1\goh^{\alf\beta}\goh_{\alf\beta}+k_2(\goh^{\alf}{}_{\alf})^{2}\r),
\m{L-mass}
 \ee
where $k_1$ and $k_2$ have a dimensionality  of $[length]^{-2}$.
Studying these theories allows one to ascertain whether negative-energy field degrees of freedom (such as massive gravity almost always implies without deliberate tuning) are vicious already at the classical level.

The additional term in (\ref{L-mass}) gives a contribution both into the right hand side and into the left hand side of (\ref{(a2.23-mod-red)}); the equations of the new gravity theory symbolically can be rewritten as
 \be
{\cal G}_{\rm mass}^{\mu\nu} =
 \k {\bm t}_{\rm mass}^{\mu\nu}\, .
\m{Eqs-mass}
 \ee
These new equations are not gauge invariant under the gauge transformations (\ref{(5.9)}) and (\ref{(5.9)+}) because the background metric $\bar g_{\mu\nu}$ cannot be incorporated in the dynamical metric $g_{\mu\nu}$ totally. As a result, there are no transformations like (\ref{tei-gaugeBG+}). Therefore, there is no problem with the localization of ${\bm t}_{\rm mass}^{\mu\nu}$: that is, energy and other conserved quantities  are not gauge dependent and hence are localized without infinite multiplicity. In accord with Noether's theorem generalized to include fields not varied in the action \cite{TrautmanUspekhi}, the presence of absolute objects \emph{shrinks} the symmetry `group' to the Killing vectors of the fields that are not varied, leaving $10$ symmetries of the Lagrangian rather than infinitely many.

To compare the new theory to GR, it is convenient to present (\ref{Eqs-mass}) in the equivalent form as
 \be
 G_{\mu\nu} + M_{\mu\nu} = \k T_{\mu\nu}\,;
\m{QuasiGeom}
 \ee
here the mass term  is
\be
 M_{\mu\nu} \equiv
\l(2\delta^\alf_\mu\delta^\beta_\nu -
 g^{\alf\beta}g_{\mu\nu}\r) (k_1h_{\alf\beta} + k_2\bar g_{\alf\beta}
 h^\rho{}_\rho)\,.
 \m{MMM}
 \ee
 Recall that $\goh^{\mu\nu} = \sqrt{\bar g}h^{\mu\nu}$ and that the Bianchi identity $\nabla_\nu G^\nu{}_\mu \equiv 0$ is expressed using the effective metric. Matter equations (\ref{(a2.3)}) lead to  $\nabla_\nu T^\nu{}_\mu = 0$, as usual; then after differentiation of (\ref{QuasiGeom}) one obtains $\nabla^\nu M_{\mu\nu} = 0$. Frequently these conditions are more convenient to use instead of some members of the original system (\ref{QuasiGeom}).

We follow the analysis by Ogievetsky and Polubarinov
\cite{OP} and by van Dam and Veltman
\cite{VanDamVeltman} to give a physical interpretation of $k_1$ and $k_2$. Consider the linearization of the (\ref{QuasiGeom}):
 \be
 \half \l({\bn_{\rho}{}^\rho}h_{\mu\nu}
+ {\bar g_{\mu\nu}}{\bn_{\rho\sig}} h^{\rho\sig} - {\bn_{\rho\nu}}h_{\mu}{}^{\rho} - {\bn_{\rho\mu}}h_{\nu}{}^{\rho}\r)+ 2k_1 h_{\mu\nu} -
(k_1+2k_2) \bar g_{\mu\nu}h_{\alf}{}^{\alf}= 0\,.
\m{LinearMass}
 \ee
Let us apply the divergence and obtain
 \be
 \bn^\nu\l[2k_1 h_{\mu\nu} -
(k_1+2k_2) \bar g_{\mu\nu}h_{\alf}{}^{\alf}\r]= 0\,
\m{DivLinearMass}
 \ee
that is the linearized version of the equation $\nabla^\nu M_{\mu\nu} = 0$.

{\em The case with $k_1 \neq - k_2$}. Then the system (\ref{LinearMass}) becomes equivalent to
 \bea
\bar\Box H^{\mu\nu} +\alf^2 H^{\mu\nu}& =& 0,\m{Box-H}\\
\bar\Box h^\alf{}_\alf +\beta^2 h^\alf{}_\alf & =& 0, \m{Box-l}
 \eea
together with (\ref{DivLinearMass}). Here, $\bar\Box \equiv \bar g^{\alf\beta}\bn_{\alf\beta}$,
 \be H^{\mu\nu} \equiv h^{\mu\nu}
-\frac{k_1+k_2}{3k_1} \bar g^{\mu\nu} l^\alf{}_\alf -
\frac{k_1+k_2}{6k_1^2} \bn^{\mu\nu} h^\alf{}_\alf +
\frac{k_1+k_2}{12k_1^2} \bar g^{\mu\nu} \bar\Box h^\alf{}_\alf
\m{H}
 \ee
with $ \bar g_{\mu\nu}H^{\mu\nu} = 0$ and  $\bn_\nu H^{\mu\nu} = 0$. Thus, parameters in the wave-like equations (\ref{Box-H}) and
(\ref{Box-l}) are
 \be \alf^2 = 4k_1\, , \qquad \qquad \beta^2 =
-\frac{2k_1(k_1 +4k_2)}{k_1 +k_2}\, .
\m{AlfBeta}
 \ee
In the standard way, they can be thought as inverse Compton wavelengths of the {\it
spin-2} graviton with the mass $m_2 = \alf\hbar/c$ associated with
the field $H^{\mu\nu}$ and of {\it spin-0} graviton with mass $m_0 =
\beta\hbar/c$ associated with the field $h^\alf{}_\alf$.

Studying weak gravitational waves in the massive gravity, one finds certain modifications of GR. Thus the spin-0 gravitational waves, represented by the trace $h^\alf{}_\alf =
h^{\alf\beta}\eta_{\alf\beta}$, and the polarization state of the
spin-2 graviton represented by the spatial trace $H^{ik}\eta_{ik}$ both, unlike GR, become essential. They provide additional
contributions to the energy-momentum flux carried by the
gravitational wave. However, gravitational wave solutions, their
energy-momentum characteristics, and observational predictions of GR
are fully recovered in the massless limit $\alf \goto 0$ and $\beta
\goto 0$.

{\em The case $k_1 +k_2 = 0$} corresponds to the mass term of Fierz-Pauli type. This means that $\beta^2 \goto \infty$ in (\ref{AlfBeta}), and the
full set of equations (\ref{LinearMass}) is equivalent to
\be
h^\alf{}_\alf = 0, \qquad \bar\Box h^{\mu\nu} +4k_1 h^{\mu\nu} =0, \qquad
\bn_\nu h^{\mu\nu} = 0.
\m{k+k=0}
\ee
This case is interpreted as unacceptable \cite{BabakGrishchuk1} because there is conflict with indirect gravitational-wave
observations of binary pulsars \cite{Taylor}.  Such claims are in tension with recent claims of success for the Vainshstein mechanism as a nonperturbative resolution of the van Dam-Veltman-Zakharov discontinuity and thus might merit further investigation, especially from numerical relativists.

Returning to the case $k_1 \neq - k_2$, the full (total) non-linear equations (\ref{Eqs-mass}), or equivalently (\ref{QuasiGeom}), have been analyzed in \cite{BabakGrishchuk1} searching for black hole and cosmological solutions. The study combined analytical and numerical methods. Thus, static spherically-symmetric vacuum solutions are practically indistinguishable from those of GR for all  $2M \ll r \ll 1/\alf$,  where $r$ and $M$ are
the radial and mass parameters of the Schwarzschild solution, and $\alf$ is infinitesimally small. However, for $r>1/\alf$ the solution is of the Yukawa-type potentials; the event horizon is {\em absent} and at $r=0$ there is a naked (curvature) singularity. Concerning the cosmological (homogeneous and isotropic) solutions, it was shown that the massive solutions have a prolonged time interval where they are practically indistinguishable from the FLRW solution of GR. Only at early times and very late times differences exist. The FLRW expansion is replaced by a regular maximum;  the origin (Big Bang) singularity is replaced by a regular minimum. There are also possibilities of an oscillatory regime.
These studies provide evidence that the negative-energy degree of freedom is not vicious classically.  Hence the traditional worries about negative-energy degrees of freedom \cite{DeserMass} are best seen as arising from the expectation of quantization rather than as features of classical field theory.  The real world is, of course, described quantum field theory (or perhaps some successor) rather than classical field theory, so worries motivated by quantization are  serious.  Concluding that ``ghost'' theories are acceptable classically would therefore hardly vindicate them in a quantum world.

One property of GR \cite{[26],Kraichnan,Deser} that one might wish to have also in a massive theory is  a derivation by universal coupling.  It might appear (\emph{e.g.}, \cite{FMS,Deser}) that universal coupling gives a unique result, the Freund-Maheshwari-Schonberg theory with a negative-energy spin $0$ of the same mass as the spin $2$.  As discussed above, it turns out that universal coupling can accommodate contravariant or covariant gravitational potentials of almost any weight,  as well as tetrad or cotetrad definitions of almost any weight, giving $4$ one-parameter families using linear field redefinitions \cite{MassiveGravity1,MassiveGravity2}.  Introducing nonlinear field (re)definitions, so that the effective metric is a nonlinear function of the gravitational potential and background metric, lets one regard \emph{any} massive gravity theory as universally coupled, as long as a mild condition of invertibility holds \cite{MassiveGravity3}.  Despite the mass terms' taking the form of $potential^2$ rather than $\sqrt{-g} - linear$ previously seen in universally coupled theories, the Babak-Grishchuk theories are included. Hence universal coupling is not very restrictive after all, once nonlinear field (re)definitions are admitted.

As was noted with references in the introduction, massive gravity has become a more lively topic since the 2000s with renewed efforts to address the discontinuous massless limit of pure spin $2$ theories and especially in the 2000s with success in avoiding the nonlinear reappearance of the negative-energy spin $0$ degree of freedom tuned away at the linear level by Pauli and Fierz.  The contingency of scientific history appears in the fact that an exact (nonlinear) argument arriving at a pure spin $2$ theory was already provided \cite{MaheshwariIdentity} prior to the discovery that the nonlinear ghost was typical \cite{TyutinMass,DeserMass}---that is, an exception preceded the rule, though the exception was forgotten and later it and others were reinvented \cite{MassiveGravity3}, as discussed above.  In any event one should note that arguably difficulties remain even for the so-called ghost-free theories   \cite{BabakGrishchuk1,Chamseddine_Mukhanov_2013,Volkov_2014,Deser+3_2014,Deser+3_2015}.


\appendix
\numberwithin{equation}{section}

\sect{ Covariant Klein-Noether identities in an arbitrary field theory}

In this Appendix, we present identities necessary for the goals in the present review. We follow the technique developed in \cite{Petrov_2009,Petrov_2019,Petrov2011,Petrov_Lompay_2013}; see also the book \cite{Petrov+_2017} and related references therein. The conclusions of the Appendix are technical and can be applied to various problems.

Let us consider a theory of arbitrary covariant fields $\psi^A$ with the Lagrangian:
 \be
\lag = \lag (\psi^A; \psi^A{}_{,\alf}; \psi^A{}_{,\alf\beta})\,, \m{lagQ}
 \ee
which depends on partial derivatives up to the second order.\footnote{These results were derived assuming that the fields were tensor densities of some sort.  But the linearity of the transformation law appears to play little or no role in the derivation.  To the degree that that is true,  the  results would also apply using the nonlinear metric-dependent spinor formalism \cite{OPspinor,PittsSpinor}. } Because the Lagrangian
(\ref{lagQ}) is a scalar density of the weight $+1,$ it satisfies the main Noether identity:
 \be
 {\pounds}_\xi {\lag} + (\xi^\alf {\lag})_{,\alf} \equiv 0\,.
 \m{LieLag}
 \ee
After identical transformations it can be represented in the form:
 \be
 \di_\alf \l[{\cal U}_\sig{}^\alf\xi^\sig + {\cal M}_{\sig}{}^{\alf\tau}\di_\tau \xi^\sig + {\cal N}_\sig{}^{\alf\tau\beta}\di_{\beta\tau} \xi^\sig\r] \equiv 0\,.
 \m{(+2+)}
\ee
In (\ref{(+2+)}), the coefficients are defined by the Lagrangian (\ref{lagQ}) without ambiguities in an unique way:
 \bea
&& {\cal U}_\sig{}^\alf  =  \lag
\delta^\alf_\sig +
 {{\delta \lag} \over {\delta \psi_B}} \l.\psi_B\r|^\alf_\sig -
 \l[{{\di \lag} \over {\di \psi_{B,\alf}}} -
 \di_\beta
\l({{\di \lag} \over {\di \psi_{B,\alf\beta}}}\r) \r] \di_\sig
\psi_{B}  -  {{\di \lag} \over {\di \psi_{B,\alf\beta}}}
\di_{\sig\beta} \psi_{B}\, , \m{(+3+)} \\
 && {\cal M}_\sig{}^{\alf\tau}  =
 \l[{{\di \lag} \over {\di \psi_{B,\alf}}} -
 \di_\beta \l({{\di \lag} \over {\di \psi_{B,\alf\beta}}}\r)\r]
 \l.\psi_{B}\r|^\tau_\sig -
 {{\di \lag} \over {\di \psi_{B,\alf\tau}}}
\di_\sig \psi_B +
 {{\di \lag} \over {\di \psi_{B,\alf\beta}}}
 \di_\beta (\l.\psi_{B}\r|^\tau_\sig)\, ,
\m{(+4+)}\\
&& {\cal N}_\sig{}^{\alf\tau\beta} = \half \l[{{\di \lag} \over
{\di\psi_{B,\alf\beta}}}
 \l.\psi_{B}\r|^\tau_\sig +
 {{\di \lag} \over {\di \psi_{B,\alf\tau}}}
 \l.\psi_{B}\r|^\beta_\sig\r].
\m{(+5+)}
 \eea
Because $\di_{\beta\tau}\equiv \di_\beta\di_\tau$ in (\ref{(+2+)}) is symmetrical in $\beta$ and $\tau$, the same symmetry is reflected in coefficients:
${\cal N}_\sig{}^{\alf\tau\beta} = {\cal N}_\sig{}^{\alf\beta\tau}$.

Opening the identity (\ref{(+2+)}), given that $\xi^\sig$, $
\di_{\alf}\xi^\sig$, $ \di_{\beta\alf}\xi^\sig$ and
$\di_{\gamma\beta\alf} \xi^\sig$ are arbitrary at every world point,
we equate to zero the coefficients associated with them and obtain
the system of the Klein-Noether identities:
 \bea
 &{}& \di_\alf  {\cal U} _\sig{}^\alf  \equiv 0, \m{(+9+1A)}\\
&{}&    {\cal U}_\sig{}^\alf + \di_\lam {\cal M}_{\sig}{}^{\lam \alf}
\equiv 0,
 \m{(+9+2A)}\\ &{}&
 {\cal M}_{\sig}{}^{(\alf\beta)}+
\di_\lam  {\cal N}_{\sig}{}^{\lam(\alf\beta)} \equiv 0, \m{(+9+3A)}\\
&{}&
 {\cal N}^{(\alf\beta\gamma)}_\sig \equiv 0.
 \m{(+9+4A)}
 \eea
 These identities are not independent. Indeed, after differentiating (\ref{(+9+2A)}) and using (\ref{(+9+3A)}) and (\ref{(+9+4A)}) one easily finds (\ref{(+9+1A)}).

One can see that expressions (\ref{(+3+)}) - (\ref{(+5+)}) and
the identities (\ref{(+9+1A)}) - (\ref{(+9+4A)}) are not covariant.
On the other hand, the expression in  (\ref{(+2+)}) is  covariant as whole, since it
is a scalar density; the expression under the divergence in
(\ref{(+2+)}) is a vector density. This signals that the above expressions and the identities can be covariantized. We achieve this in the following way (see \cite{Petrov_Lompay_2013}).
Let us replace partial derivatives of $\xi^\sig$ in  (\ref{(+2+)}) by covariant ones, making the use of the expression
$\di_\rho\xi^\sig = \nabla_\rho\xi^\sig - \l.\xi^\sig\r|^\alf_\beta
\Gamma^\beta_{\rho\alf}$. Here, the Christoffel symbols $\Gamma^\beta_{\rho\alf}$ and, consequently, the covariant derivative $\nabla_\rho$ are compatible with $g_{\mu\nu}$. At the moment, $g_{\mu\nu}$ is included in expressions as an external metric only. Note that $g_{\mu\nu}$ can be included in the set $\psi^A$, although this is not necessary.  The identity (\ref{(+2+)}) is now
rewritten as
 \be
 \nabla_\alf\l[{\bm u}_\sig{}^\alf\xi^\sig + {\bm m}_\sig{}^{\alf\tau}\nabla_\tau \xi^\sig + {\bm n}_\sig{}^{\alf\tau\beta} \nabla_{\beta\tau}
\xi^\sig\r]\equiv 0\,,
 \m{secondID}
 \ee
where $\nabla_{\beta\tau}\equiv \nabla_{\beta} \nabla_{\tau}$ and
 \bea
 {\bm u}_\lam{}^\alf &=& {\cal U}_\lam{}^\alf - {\cal M}_\sig{}^{\alf\tau}\Gamma^\sig_{\lam\tau}+ {\cal  N}_\sig{}^{\alf\tau\rho} (\Gamma^\sig_{\tau\pi}\Gamma^\pi_{\lam\rho}
 - \di_\rho\Gamma^\sig_{\lam\tau})\,,\m{Uu}\\
 {\bm m}_\lam{}^{\alf\tau} &=& {\cal M}_\lam{}^{\alf\tau} +
 {\cal N}_\lam{}^{\alf\sig\rho}\Gamma^\tau_{\sig\rho} -
 2{\cal N}_\sig{}^{\alf\tau\rho}\Gamma^\sig_{\lam\rho}\,,
 \m{Mm}\\
 {\bm n}_\lam{}^{\alf\tau\rho} &=& {\cal N}_\lam{}^{\alf\tau\rho}\,.
 \m{Nn}
 \eea
 One can show explicitly that, indeed, these coefficients are covariant \cite{Petrov_Lompay_2013}. Note that in \cite{Petrov_Lompay_2013} we have shown that there are different ways to define coefficients in (\ref{Uu}), (\ref{Mm}) and (\ref{Nn}). Here, however, we use the form  (\ref{Uu}), (\ref{Mm}) and (\ref{Nn}) only.

 The identity (\ref{secondID}) can be rewritten in the form of the
differential conservation law:
 \be
 \nabla_\alf {\bm i}^\alf \equiv
 \di_\alf {\bm i}^\alf \equiv 0\,,
 \m{(+6+)}
 \ee
where the current is rewritten as
 \bea
 {\bm i}^\alf  &=& - \l[({\bm u}_\sig{}^\alf + {\bm n}_\lam{}^{\alf\beta\gamma} R^\lam_{~\beta\gamma\sig})\xi^\sig + {\bm m}^{\rho\alf\beta}\di_{[\beta}\xi_{\rho]} +
 {\bm z}^{\alf}\r]\,,
 \m{(+7+)}\\
  {\bm z}^{\alf} &=& {\bm m}^{\sig\alf\beta}\zeta_{\sig\beta}+ {\bm n}^{\rho\alf\beta\gamma}
\l(2 \nabla_{\gamma}\zeta_{\beta\rho} - \nabla_\rho \zeta_{\beta\gamma}\r); \qquad 2\zeta_{\rho\sigma} \equiv - {\pounds}_\xi g_{\rho\sigma} =
2\nabla_{(\rho}\xi_{\sigma)}\,.
 \m{(+8+)}
 \eea
Thus, the $z$-term disappears, ${\bm z}^{\alf} = 0$, if $\xi^\mu = \bar \xi^\mu$ is a Killing vector of a
 metric $g_{\mu\nu}$. Then the current (\ref{(+7+)}) is determined by the
energy-momentum $(u + nR)$-term and the spin $m$-term only.

Exploring the identity (\ref{(+6+)}) and equating independently
to zero the coefficients at $\xi^\sig$,  $\nabla_{\alf} \xi^\sig$,
$\nabla_{(\beta\alf)} \xi^\sig$ and $\nabla_{(\gamma\beta\alf)} \xi^\sig$,
we get a system of identities that is equivalent to the system (\ref{(+9+1A)}) - (\ref{(+9+4A)}),   reformulated as
 \bea
 &{}& \nabla_\alf  {\bm u}_\sig{}^\alf + \half
 {\bm m}_\lam{}^{\alf\rho} R^{~\lam}_{\sig~\rho\alf}
 +{\textstyle{1\over 3}} {\bm n}_\lam{}^{\alf\rho\gamma}
\nabla_\gamma R^{~\lam}_{\sig~\rho\alf}
  \equiv 0, \m{(+9+1)}\\
&{}&    {\bm u}_\sig{}^\alf + \nabla_\lam {\bm m}_{\sig}{}^{\lam \alf} +
{\bm n}_\lam{}^{\tau\alf\rho} R^{~\lam}_{\sig~\rho\tau} +{\textstyle{2\over 3}} {\bm n}_{\sig}{}^{\lam\tau\rho}R^{\alf}_{~\tau\rho\lam} \equiv 0,
 \m{(+9+2)}\\ &{}&
 {\bm m}_{\sig}{}^{(\alf\beta)}+
\nabla_\lam  {\bm n}_{\sig}{}^{\lam(\alf\beta)} \equiv 0, \m{(+9+3)}\\
&{}&
 {\bm n}^{(\alf\beta\gamma)}_\sig \equiv 0.
 \m{(+9+4)}
 \eea
 These identities are also not independent. After covariantly differentiating (\ref{(+9+2)}) and using (\ref{(+9+3)}) and (\ref{(+9+4)}), one easily finds (\ref{(+9+1)}).
Since (\ref{(+6+)}) is identically satisfied, the current
(\ref{(+7+)}) must be a divergence of a superpotential $ {\bm i}^{\alf\beta}$, an antisymmetric tensor density, for which
$\di_{\beta\alf} {\bm i}^{\alf\beta} \equiv 0$, that is
 \be
 {\bm i}^\alf \equiv \nabla_{\beta} {\bm i}^{\alf\beta} \equiv
\di_{\beta}   {\bm i}^{\alf\beta}.
 \m{(+10+)}
 \ee
Indeed,  substituting ${\bm u}_\sig{}^\alf$  from (\ref{(+9+2)})
 into the current (\ref{(+7+)}), using (\ref{(+9+3)}) and algebraic properties of
${\bm n}_\sig{}^{\alf\beta\gamma}$ and $R^{\alf}_{~\beta\rho\sig}$, we
reconstruct (\ref{(+7+)}) into the form (\ref{(+10+)}), where  the
superpotential acquires the form:
 \be
 {\bm i}^{\alf\beta}  = \l({\textstyle{2\over 3}}
 \nabla_\lam  {\bm n}_{\sig}{}^{[\alf\beta]\lam}  - {\bm m}_{\sig}{}^{[\alf\beta]}\r)\xi^\sig   -
 {\textstyle{4\over 3}} {\bm n}_{\sig}{}^{[\alf\beta]\lam}
 \nabla_\lam  \xi^\sig.
 \m {(+11+)}
 \ee
It is explicitly antisymmetric in $\alf$ and $\beta$;
the differential conservation law
(\ref{(+6+)}) follows from (\ref{(+10+)}).


\received
\noindent
\contactinformationenglish

\noindent{\bf Acknowledgments.} ANP acknowledges the
support from the Program of development of M.V. Lomonosov Moscow State University 2019-2020
(Leading Scientific School ``Physics of stars, relativistic objects and galaxies'').
JBP acknowledges support from the National Science Foundation (USA), Science, Technology and Society, \#1734402.

\end{document}